\documentclass[11pt]{article}
\usepackage{amsfonts,amssymb,bbm}
\usepackage[tbtags]{amsmath}
\usepackage [amsthm,thmmarks,amsmath]{ntheorem} 
\usepackage[arrow, matrix, curve, all, cmtip]{xy}
\usepackage{color,psfrag}
\usepackage[dvips]{graphicx}
\usepackage{slashed}
\usepackage{exscale}
\usepackage{soul}

\usepackage{ulem}

\allowdisplaybreaks[1]

\newtheorem{definition}{Definition}

\newcommand{\D}{\underset{\leftarrow}}
\newcommand{\Dl}{\underset{\longleftarrow}}
\newcommand{\DD}{\underset{\Leftarrow}}
\newcommand{\DDl}{\underset{\Longleftarrow}}

\usepackage{hyperref}

\newtheoremstyle{breakdef}%
  {\item[\rlap{\vbox{\normalfont\bfseries\hbox{\llap{##2}\hskip\labelsep
          ##1:}\hbox{\\[0.1cm]}}}]}%
  {\item[\rlap{\vbox{\normalfont\bfseries\hbox{\llap{##2}\hskip\labelsep
          ##1 (##3):}\hbox{\\[0.1cm]}}}]}
\newtheoremstyle{breaksatz}%
  {\item[\rlap{\vbox{\normalfont\normalsize\bfseries\hbox{\llap{##2}\hskip\labelsep
          ##1:}\hbox{\\[0.1cm]}}}]}%
  {\item[\rlap{\vbox{\normalfont\normalsize\bfseries\hbox{\llap{##2}\hskip\labelsep
          ##1 (##3):}\hbox{\\[0.1cm]}}}]}
\newtheoremstyle{breaklem}%
  {\item[\rlap{\vbox{\normalfont\normalsize\bfseries\hbox{\llap{##2}\hskip\labelsep
          ##1:}\hbox{\\[0.1cm]}}}]}%
  {\item[\rlap{\vbox{\normalfont\normalsize\bfseries\hbox{\llap{##2}\hskip\labelsep
          ##1 (##3):}\hbox{\\[0.1cm]}}}]}
\newtheoremstyle{breakprop}%
  {\item[\rlap{\vbox{\normalfont\normalsize\bfseries\hbox{\llap{##2}\hskip\labelsep
          ##1:}\hbox{\\[0.1cm]}}}]}%
  {\item[\rlap{\vbox{\normalfont\normalsize\bfseries\hbox{\llap{##2}\hskip\labelsep
          ##1 (##3):}\hbox{\\[0.1cm]}}}]}
\newtheoremstyle{breakbem}%
  {\item[\rlap{\vbox{\hbox{\hskip\labelsep\normalfont\bfseries
          ##1 ##2:}\hbox{\\[0.1cm]}}}]}%
  {\item[\rlap{\vbox{\hbox{\hskip\labelsep\normalfont\bfseries
          ##1 ##2 (##3):}\hbox{\\[0.1cm]}}}]}
\newtheoremstyle{breakbsp}%
  {\item[\rlap{\vbox{\hbox{\hskip\labelsep\normalfont\bfseries
          ##1 ##2:}\hbox{\\[0.2cm]}}}]}%
  {\item[\rlap{\vbox{\hbox{\hskip\labelsep\normalfont\bfseries
          ##1 ##2 (##3):}\hbox{\\[0.2cm]}}}]}
\newtheoremstyle{breakkor}%
  {\item[\rlap{\vbox{\hbox{\hskip\labelsep\normalfont\bfseries
          ##1 ##2:}\hbox{\\[0.1cm]}}}]}%
  {\item[\rlap{\vbox{\hbox{\hskip\labelsep\normalfont\bfseries
          ##1 ##2 (##3):}\hbox{\\[0.1cm]}}}]}
\newtheoremstyle{proof}%
  {\item[\rlap{\vbox{\hbox{\hskip\labelsep\normalfont\bfseries
          \underline{##1:}}\hbox{\\[0.1cm]}}}]}%
  {\item[\rlap{\vbox{\hbox{\hskip\labelsep\normalfont\bfseries
          \underline{##1 (##3):}}\hbox{\\[0.1cm]}}}]}
\theoremstyle{breakkor} 			
\theoremstyle{breakkor}

\theoremstyle{breakkor}

\theoremstyle{breakkor}

\theoremstyle{breakkor}

\theoremstyle{breakkor}
\theorembodyfont{\normalfont}

\theoremstyle{proof}

\newcommand{\m}{\mbox{}}
\newcommand{\be}{\begin{eqnarray}}
\newcommand{\ee}{\end{eqnarray}}

\newcommand{\vev}[1]{\left\langle #1 \right\rangle}

\title{{\sf New Variables for Classical and Quantum Gravity in all Dimensions V. Isolated Horizon Boundary Degrees of Freedom. 
} 
}
\author{
{\sf N. Bodendorfer}$^{1,2}$\thanks{{\sf 
norbert.bodendorfer@gravity.fau.de, norbert@gravity.psu.edu}},
{\sf T. Thiemann}$^{1}$\thanks{{\sf 
thomas.thiemann@gravity.fau.de}},
{\sf A. Thurn}$^1$\thanks{{\sf 
andreas.thurn@gravity.fau.de}}\\
\\
{\sf $^1$ Inst. for Theoretical Physics III, FAU Erlangen -- N\"urnberg,}\\
{\sf Staudtstr. 7, 91058 Erlangen, Germany}\\
\\
{\sf $^2$ Institute for
Gravitation and the Cosmos \& Physics
  Department,}\\
{\sf   Penn State, University Park, PA 16802, U.S.A.}\\
}
\date{{\small\sf \today}}
\makeatletter
\@addtoreset{equation}{section}
\makeatother

\begin{document}

\maketitle

{\sf
\begin{abstract}
In this paper, we generalise the treatment of isolated horizons in loop quantum gravity, resulting in a Chern-Simons theory on the boundary in the four-dimensional case, to non-distorted isolated horizons in $2(n+1)$-dimensional spacetimes. 
The key idea is to generalise the four-dimensional isolated horizon boundary condition by using the Euler topological density $E^{(2n)}$ of a spatial slice of the black hole horizon as a measure of distortion. 
The resulting symplectic structure on the horizon coincides with the one of higher-dimensional SO$(2(n+1))$-Chern-Simons theory in terms of a Peldan-type hybrid connection $\Gamma^0$ and resembles closely the usual treatment in $3+1$ dimensions. 
We comment briefly on a possible quantisation of the horizon theory. Here, some subtleties arise since higher-dimensional non-Abelian Chern-Simons theory has local degrees of freedom. 
However, when replacing the natural generalisation to higher dimensions 
of the usual boundary condition by an equally natural stronger one, it is conceivable that the problems originating from the local degrees of freedom are avoided, thus possibly 
resulting in a finite entropy. 

\end{abstract}
}

\newpage

\tableofcontents

\newpage

\section{Introduction}
\label{sec:Introduction}

Black holes in higher dimensions are a subject of great interest in both general relativity and supergravity. Most prominently, the derivation of black hole entropy within string theory was first performed for a five-dimensional black hole \cite{StromingerMicroscopicOriginOf}. Also, no-hair theorems familiar from $d=4$ spacetime dimensions generally fail in higher dimensions, resulting in a large variety of black hole solutions with new (exotic) properties, see \cite{EmparanBlackHolesIn} for a review. 
While this fact has been appreciated in, e.g. string theory, it was not possible so far to perform these calculations in the context of loop quantum gravity, since the Ashtekar-Barbero variables \cite{AshtekarNewVariablesFor, BarberoRealAshtekarVariables} necessary for loop quantisation are restricted to $d=3,4$. On the other hand, the recent extension of this type of connection formulation to higher-dimensional general relativity and supergravity \cite{BTTI, BTTII, BTTIII, BTTIV, BTTVI, BTTVII} opens the window to investigate higher-dimensional black holes also with the methods of loop quantum gravity. 

The treatment of horizons and black hole entropy within loop quantum gravity can be dated back to a remarkable paper by Smolin \cite{SmolinLinkingTopologicalQuantum}, in which it was shown that under some (natural) assumptions, boundaries of spacetime are described by a topological quantum field theory, more precisely SU$(2)$ Chern-Simons theory. This pioneering work already contained many of the ideas which were later necessary to give a rigorous derivation of the black hole entropy within LQG. An entropy associated to a surface which is proportional to the area was first calculated in papers by Krasnov \cite{KrasnovGeometricalEntropyFrom} and Rovelli \cite{RovelliBlackHoleEntropy}, where the important conceptual ingredient was that the punctures of horizon were distinguishable. 

A rigorous technical framework for calculating black hole entropy within loop quantum gravity was derived by Ashtekar and collaborators \cite{AshtekarIsolatedHorizonsThe, AshtekarMechanicsOfIsolated, AshtekarIsolatedHorizonsHamiltonian, AshtekarQuantumGeometryOf}, where the notion of isolated horizon turned out to be crucial in order to have a local description of a black hole horizon. While a classical gauge fixing from SU$(2)$ to U$(1)$ was performed in order to derive the results of \cite{AshtekarIsolatedHorizonsThe, AshtekarMechanicsOfIsolated, AshtekarIsolatedHorizonsHamiltonian, AshtekarQuantumGeometryOf}, it was later shown by Engle, Noui and Perez \cite{EngleBlackHoleEntropy} that the derivation could be extended to an SU$(2)$ invariant framework. 
An error in the state counting for the derivation of black hole entropy in \cite{AshtekarQuantumGeometryOf} was corrected by Domagala and Lewandowski in \cite{DomagalaBlackHoleEntropy}, showing that the dominance of spin $1/2$ representations was incorrect. The detailed state counting has been extensively studied by Barbero and collaborators, see \cite{AgulloDetailedBlackHole} and references therein.
Non-spherical topologies in $3+1$ dimensions were discussed in \cite{DeBenedictisPhaseSpaceAnd, BSTI}. Also, \cite{Diaz-PoloIsolatedHorizonsAnd} provides a recent extensive review of the subject, including a comparison of the U$(1)$ and SU$(2)$ treatments. 

In this paper, we are going to take first steps towards the derivation of higher-dimensional black hole entropy using loop quantum gravity methods by deriving a generalisation of isolated horizon boundary condition $F \propto \Sigma$ first proposed in \cite{SmolinLinkingTopologicalQuantum} and derived rigorously in \cite{AshtekarIsolatedHorizonsThe}. We further show that the canonical transformation to higher-dimensional connection variables induces a higher-dimensional Chern-Simons symplectic structure on the intersection of the spatial slice with the isolated horizon. Also, we shortly comment on the quantisation of the resulting theory on the black hole horizon. The derivations in this paper will be restricted to even spacetime dimensions, since the Euler topological density, which will play a key role in the construction, does not exist otherwise. In even spacetime dimensions, the isolated horizon then is odd-dimensional and a Chern-Simons theory can arise on it.
A corresponding classical higher-dimensional black hole solution (with spherical symmetry) was found by Tangherlini \cite{TangherliniSchwarzschildFieldIn} and generalises the Schwarzschild solution to higher dimensions, see also \cite{EmparanBlackHolesIn} for an overview. Since, in the loop quantum gravity treatment, the notion of isolated horizon is more central than that of a classical black hole solution, we will not go into details about the latter. As the notion of isolated horizon has already been generalised to higher dimensions in \cite{LewandowskiQuasiLocalRotating, KorzynskiMechanicsOfMultidimensional}, we can concentrate on deriving the isolated horizon boundary condition and the symplectic structure in this paper.\\ 
\\
\\
This paper is organised as follows:\\
We start in section \ref{sec:GeneralStrategy} with an outline of the general strategy used in this paper for finding an analogue of the isolated horizon boundary condition in higher dimensions. In order to establish notation for the following calculations, we provide a comprehensive list of the notation used in this paper in section \ref{sec:Notation}.
Next, in section \ref{sec:NewVariables}, we review the canonical transformation to SO$(D+1)$ variables and emphasise the appearance of a boundary term which will later result in the Chern-Simons symplectic structure. In section \ref{sec:HigherDimensionalIsolatedHorizons}, the definition of a higher-dimensional isolated horizon is reviewed and its consequences are given. In the following section \ref{sec:BoundaryCondition}, we derive the isolated horizon boundary condition for the internal gauge group SO$(1,D)$ starting from the Palatini action. Next, in section \ref{sec:Hamiltonianframework}, we will develop the Hamiltonian framework and derive the Chern-Simons symplectic structure on the isolated horizon for the internal gauge group SO$(1,D)$, also starting from the Palatini action. In order to make the connection to SO$(D+1)$ as the internal gauge group, we rederive the isolated horizon boundary condition and the Chern-Simons symplectic structure independently of the internal signature in section \ref{sec:SO(D+1)}, this time purely within the Hamiltonian framework. We shortly comment about the generalisation of the proposed framework to non-distorted horizons in section \ref{sec:NonSymmetric}. Finally, we will discuss a possible quantisation of the boundary degrees of freedom in section \ref{sec:Quantisation} and conclude in section \ref{sec:ConcludingRemarks}. The appendices contain the construction and generalisation of the hybrid connection used in the Chern-Simons symplectic structure, further details on calculations, as well as an overview over higher-dimensional Chern-Simons theory and the higher-dimensional Newman-Penrose formalism. 

\section{General Strategy}
\label{sec:GeneralStrategy}

In this section, we will briefly comment on the general strategy of deriving the isolated horizon boundary condition. It will turn out that there is merely a single reasonable possibility for the general structure of the boundary condition for which a numerical prefactor and an expression for the connection on the horizon have to be fixed by an actual calculation\footnote{Note however that using only the boundary condition discussed in this section will result in local degrees of freedom in the higher-dimensional Chern-Simons theory and that an additional stronger boundary condition will have to be used to determine the Chern-Simons connection by the bulk fields.}. However, the connection used on the boundary is not necessarily unique as already observed in the four dimensional case \cite{EngleBlackHoleEntropyFrom}, where one is free to choose an independent Barbero-Immirzi-type parameter on the black hole horizon.

Let us start with some hints for the boundary condition based on the new connection variables derived in \cite{BTTI, BTTII}.

\begin{itemize}

	\item {\bf Tensorial structure:}\\
	The $3+1$ dimensional SU$(2)$ based boundary condition $F^i_{ab} \propto E^{ci} \epsilon_{abc}$ does not generalise trivially to higher dimensions due to the tensorial structure, i.e. a vector density is dual to a $(D-1)$-form in $D$ spatial dimensions, which is a two-form only for $D=3$. Since, in analogy to the $3+1$ dimensional case, we expect to get a theory which is purely defined in terms of a connection on the horizon, the easiest expression with the correct tensorial structure to write down is 
	\be
		\pi^{aIJ} \propto \epsilon^{a b_1 c_1 \ldots b_n c_n} \epsilon^{IJ K_1 L_1 \ldots K_n L_n} F_{b_1 c_1 K_1 L_1} \ldots F_{b_n c_n K_n L_n} \text{,} \label{eq:StrategyBC}
	\ee
	where $a, b, c$ are spatial tensorial indices and $I,J,K,L$ are fundamental so$(D+1)$ indices, $n=(D-1)/2$, and $\pi^{aIJ}$ is the momentum conjugate to the connection on which the new variables in higher dimensions \cite{BTTI, BTTII} are based. More generally, one could also use a different invariant tensor to intertwine the adjoint so$(D+1)$ representations on the momentum $\pi$ and the field strengths $F$, but the other obvious choice $\delta^{J][K_1} \delta^{L_1 ][K_2} \ldots \delta^{L_{n-1} ][K_n} \delta^{L_n ][I}$ results in a vanishing right hand side for even n and does not allow for the construction performed in this paper in the other cases. The open question at this point is mostly on which connection the field strengths should be based. 
		
	\item {\bf Topological invariants:}\\
	Up to a constant prefactor, the derivation of the boundary condition in three spatial dimensions and spherical symmetry can be easily accomplished by appealing to special properties of curvature tensors in two dimensions. More precisely, the Riemann tensor  $R_{\mu \nu \rho \sigma}$ on a two-dimensional manifold, e.g. a spatial slice of a black hole horizon in a four-dimensional space time, is, due to its symmetries, given by $R^{(2)}_{\mu \nu \rho \sigma} \propto R^{(2)} \, g_{\mu[ \rho} g_{\sigma ]\nu}$. Thus, after obtaining $F^{(4)}_{\mu \nu IJ} = R^{(4)}_{\mu \nu IJ} = R^{(4)}_{\mu \nu \rho \sigma} \Sigma^{\rho \sigma}_{IJ}$ from the field equations and since $R^{(4)}_{\DDl{\mu \nu} \rho \sigma} \Sigma^{\rho \sigma}_{IJ}= \DD{R}\m^{(2)}_{\mu \nu \rho \sigma} \Sigma^{\rho \sigma}_{IJ}$ when using the IH boundary conditions, it directly follows that $\DD{F}^{(4)}_{\mu \nu IJ} \propto R^{(2)} \DD{\Sigma}_{\mu \nu IJ}$, where $\DD{\m}$ denotes the pullback from the spacetime manifold to a spatial slice of the horizon.  In the further discussion of IHs in four-dimensional LQG, it is of importance that in two dimensions, the integral over the Ricci curvature actually is a topological invariant by the Gau{\ss}-Bonnet theorem. The question thus is by which topological invariant that role will be played in higher dimensions. 
	
	From the above calculation, we expect that only the step using $R^{(2)}_{\mu \nu \rho \sigma} \propto R^{(2)} \, g_{\mu[ \rho} g_{\sigma ]\nu}$ does not straight forwardly generalise to higher dimensions. However, this formula is equivalent to $R^{(2)} \propto \epsilon^{\alpha \beta} \epsilon^{IJ} R^{(2)}_{\alpha \beta I J} $, and in this form can be generalised to even dimensions and one is lead to consider the Euler topological density \cite{NakaharaGeometryTopologyAnd} 
	\be
	E^{(D+1)}:= \epsilon^{\mu_1 \nu_1 \ldots \mu_{n+1} \nu_{n+1}} \epsilon^{I_1 J_1 \ldots I_{n+1} J_{n+1}} R_{\mu_1 \nu_1 I_1 J_1} \ldots R_{\mu_{n+1} \nu_{n+1} I_{n+1} J_{n+1}}
	\ee
	as a generalisation. Although this looks already very similar to the above boundary condition (\ref{eq:StrategyBC}), the Euler density would have to be defined on the spatial slices of black hole horizon while the internal gauge group is inherited from the bulk, thus having a representation space which is two dimensions larger than the tangent space of the spatial slice of the horizon. Later in this paper, we will chose a special connection on the boundary, the field strength of which will be inherently ``orthogonal'' on $\pi^{aIJ}$ and thus allowing for a precise implementation of the above idea for a boundary condition based on the Euler topological density. We remark at this point, as also stated in the notation section, that our normalisation of the Euler topological density does not coincide with the standard definition leading to the Euler characteristic.

	\item {\bf Higher-dimensional Chern-Simons theory:}\\
	The notion of $2+1$ dimensional Chern-Simons theory has a straight forward generalisation to higher dimensions, i.e. a higher-dimensional Chern-Simons Lagrangian is defined by $d \mathcal{L}_{\text{CS}} = g_{I_1 J_1 \ldots I_{n+1} J_{n+1}} F^{I_1J_1} \wedge \ldots \wedge F^{I_{n+1} J_{n+1}}$, where $d$ is the exterior derivative and $g$ intertwines $n+1=(D+1)/2$ adjoint representations of so$(D+1)$, see \cite{BanadosExistenceOfLocal}. The right hand side of the previous equation can easily be seen to be the Euler topological density for $g_{I_1 J_1 \ldots I_{n+1} J_{n+1}}  = \epsilon_{I_1 J_1 \ldots I_{n+1} J_{n+1}} $.  The equations of motion derived from this Lagrangian are given by $g_{I_1 J_1 \ldots I_{n+
	1} J_{n+1}} F^{I_2J_2} \wedge \ldots \wedge F^{I_{n+1} J_{n+1}} = 0$, thus fitting nicely in the LQG quantisation scheme for black holes, i.e. the straight forward generalisation of $F^{IJ}=0$ at points of the horizon which are not punctured by spin networks is given by $\epsilon_{I_1 J_1 \ldots I_{n+1} J_{n+1}} F^{I_2J_2} \wedge \ldots \wedge F^{I_{n+1} J_{n+1}} = 0$. 
	
	As the canonical analysis of higher-dimensional Chern-Simons theory reveals \cite{BanadosExistenceOfLocal}, the theory has local degrees of freedom, e.g. $g_{I_1 J_1 \ldots I_{n+1} J_{n+1}} F^{I_2J_2} \wedge \ldots \wedge F^{I_{n+1} J_{n+1}} = 0$ does not imply $F^{IJ}=0$. This tension is discussed in section \ref{sec:Quantisation}.

\end{itemize}

Based on this outline, we will now give a precise derivation of the above proposed generalisation of the isolated horizon boundary condition. The connection used will be a generalisation of Peldan's hybrid connection $\Gamma_{aIJ}^{\text{H}}$ \cite{PeldanActionsForGravity}, which was already used in the construction of the connection variables in higher dimensions \cite{BTTI, BTTII}. We want to stress again that there might be other connections, e.g. a one-parameter family depending on a free parameter unrelated to the Barbero-Immirzi parameter, which satisfy an analogous boundary condition, as observed in \cite{EngleBlackHoleEntropyFrom} in the four-dimensional case.

\section{Notation and Conventions}
\label{sec:Notation}

This chapter gives an overview of the notation and conventions used in this paper. It can be skipped at first reading and should only be consulted as a reference when the notation used is unclear. \\
\\
Let $\mathcal{M}$ denote a $(D+1)$-dimensional (pseudo)-Riemannian manifold $(D \geq 2)$ with metric $g_{\mu \nu}$ of signature $(-,+,...,+)$. We will denote a $D$-dimensional Cauchy surface by $\Sigma$ and a $D$-dimensional null surface by $\Delta$. Equality on $\Delta$ is denoted by $\widehat{=}$.
Throughout this work, we will restrict the topology of $\Delta$ to be $S \times \mathbb{R}$, where $S$ is a $(D-1)$-dimensional compact Riemannian manifold which allows for the isolated horizon boundary conditions defined in section \ref{sec:HigherDimensionalIsolatedHorizons} and has non-zero Euler characteristic. Examples are the $(D-1)$-spheres $S^{(D-1)}$ or hyperbolic spaces $H^{(D-1)}$ divided by a freely acting discrete subgroup $\Gamma$, e.g. handle bodies with genus $g>1$ for $D=3$ (at the level of topology) and the corresponding black hole solutions, given e.g. in \cite{MartinezExactBlackHole}. For notational simplicity, we will (mostly) refer to all these manifolds as spheres in this work but keep in mind that more general topologies are allowed. 
We will in several sections restrict to even spacetime dimensions $D+1 =: 2(n+1)$. This is necessary in the approach taken since (a) there can exist a Chern-Simons theory on the odd ($2n+1$)-dimensional $\Delta$ and (b) the Euler density \cite{NakaharaGeometryTopologyAnd} is defined for the even ($2n$)-dimensional intersections $S \; \widetilde{=} \; S^{D-1}$ of $\Sigma$ and $\Delta$. Here and in the following, we will use index conventions:
\begin{itemize}
	\item tensorial spacetime indices will be denoted by lower Greek letters from the middle of the alphabet: $\mu, \nu, \rho, \hdots \in \{0,\hdots,D\}$.
	\item tensorial spatial indices will be denoted by lower Latin letters: $a, b, c, \hdots \in \{1,\hdots,D\}$.
	\item tensorial indices on $\Delta$ will be denoted by the $\D{\mu}, \D{\nu}, \D{\rho}$ (the pullback arrow will sometimes be omitted if there should be no confusion whether the equation is referring to $\mathcal{M}$ or $\Delta$).
	\item tensorial indices in $(D-1)$-dimensional subspaces $S$ will be denoted by lower Greek letters from the beginning of the alphabet: $\alpha, \beta, \gamma, \hdots \in \{1,\hdots,D-1\}$ or by $\DD{\mu}$.
	\item so$(D+1)$ or so$(1,D)$ Lie algebra indices in the defining representation will be denoted by capital Latin letters: $I,J,K, \hdots \in \{1,\hdots,D+1\}$.
	\item Lower Latin letters $i,j,k,\hdots$ will be used for SO$(D)$ indices (and for labelling normals, only in appendix \ref{app:Connections}).
\end{itemize} 
The spacetime metric will be denoted by $g_{\mu \nu}$, the spatial metric on $\Sigma$ by $q_{ab}$, the (degenerate) metric on $\Delta$ by $h_{\Dl{\mu \nu}}$ and on $(D-1)$ dimensional subspaces $S$ by $h_{\alpha \beta}$. The corresponding Levi-Civita connections will be denoted by $\nabla_{\mu}$, $D_a$, $D_{\D{\mu}}$ and $D_{\alpha}$. For the Riemann tensor, we will use the convention $[\nabla_{\mu}, \nabla_{\nu}] u_{\rho} = R^{(D+1)}_{\mu \nu \rho}\m^{\sigma} u_{\sigma}$ and similar for lower dimensions. The indication of the number of dimensions on the Riemann tensor will be omitted if there is now chance of confusion. If the covariant derivate also acts on internal indices, we denote it by $D^{\Gamma}_\mu u_\nu^I = D_\mu u_\nu^I + \Gamma_\mu \m^I \m_J u_\nu^J $ and its field strength as $R_{\mu \nu IJ}$. We denote by $E^{(D+1)}:= \epsilon^{\mu_1 \nu_1 \ldots \mu_{n+1} \nu_{n+1}} \epsilon^{I_1 J_1 \ldots I_{n+1} J_{n+1}} R_{\mu_1 \nu_1 I_1 J_1} \ldots R_{\mu_{n+1} \nu_{n+1} I_{n+1} J_{n+1}}$ the Euler topological density and remark that it coincides with other definitions in the literature only up to normalisation, i.e. the integral of this density over a closed compact manifold $S$, denoted by $\vev{E^{(2n)}}$, gives a only a {\it multiple} of the Euler characteristic $\chi_S$ of $S$. We choose this definition since it simplifies many formulas. Explicitly, we have
\be
	 \chi_S = \frac{1}{(8\pi)^{n} n!}\int_{S} E^{(2n)}  \text{,}
\ee
which results in $\chi_{S^{(2n)}} = 2$ for spheres spheres $S^{(2n)}$. We will drop integration measures to simplify notation or work directly with differential forms.

The null normal to $\Delta$ will be denoted by $l$ and the vector field normal to the $(D-1)$ -- sphere cross-sections by\footnote{We refrain from using the usual notation $n$ for this normal here, to avoid confusion with the normal to spatial slices, and also to make clear the difference between the hybrid vielbein normal $n^I$ and $k^I = k^{\mu} e_{\mu}\m^{I}$.} $k$, normalised to $l \cdot k = -1$  (cf. sec. \ref{sec:HigherDimensionalIsolatedHorizons}). $k$ can be extended uniquely to a spacetime 1-form at points of $\Delta$ by requiring it to be null. Then, at points of $\Delta$, we can decompose the metric according to $g_{\mu \nu} = h_{\mu \nu} - 2 l_{(\mu} k_{\nu)}$. We will denote the $h$-projected vielbein by $m$, $m_{\mu I} = h_{\mu}^{\nu} e_{\nu I}$, and furthermore use the notation $l^I = l^{\mu} e_{\mu}\m^I$, $k^I = k^{\mu} e_{\mu}\m^I$, and, since $l,k$ are null and normalised, $k^I k^J \eta_{IJ} = 0 = l^I l^J \eta_{IJ}$, $k^I l^J \eta_{IJ} = -1$. We will call $\{l,k,\{m_I\}\}$ a generalised null frame. Elements of higher-dimensional Newman-Penrose formalism in this frame will be introduced in appendix \ref{app:NPFormalism}.

The future pointing timelike unit normal to a spatial slice $\Sigma$ will be denoted by $n_\mu$, $n^2 = -1$. The spacetime metric can be decomposed as follows: $g_{\mu \nu} = q_{\mu \nu} - n_{\mu} n_{\nu}$. We will denote the spatial vielbein as $e_{a I} = (\D{e})_{a I}$ and $n^I = n^{\mu} e_{\mu}\m^I$, $n^I n^J \eta_{IJ} = -1$, $n^{I} e_{a I} = 0$. Furthermore, we will introduce the notation $\bar{\eta}_{IJ} := \eta_{IJ} + n_{I} n_{J} = e_{a I} q^{ab} e_{a J}$, $\bar{\eta}_{IJ} n^J = 0$. 

We will denote with $s$ the spacelike unit normal to the $(D-1)$ - dimensional cross-sections $\Sigma \cap \Delta$, $s^2 = 1, s \cdot n = 0$, pointing outward of $\sigma$. Furthermore, we define the co-normal $\hat{s}_a := \epsilon_{a\alpha_1 \ldots \alpha_{D-1}} \epsilon^{\alpha_1 \ldots \alpha_{D-1}} / \mbox{(D-1)!}$, which is collinear with $s_a$, but normalised appropriately for usage in Stokes theorem later on. When dealing with the Hamiltonian formulation, we will choose the foliation such that $l = \frac{1}{\sqrt{2}} (n-s)$,  $k = \frac{1}{\sqrt{2}} (n+s)$ holds, where $l$ and $k$ are the (representatives of the equivalence class of the) null normals to a given isolated horizon as specified in section \ref{sec:HigherDimensionalIsolatedHorizons}. Furthermore, we will use the notation $s^{I} := s^{\mu} e_{\mu}\m^{I}$ and introduce $\bar{\bar{\eta}}_{IJ} := \eta_{IJ} + n_I n_J - s_I s_J = \eta_{IJ} + 2 l_{(I} k_{J)} = m_{\mu I} m^{\mu}\m_{J}$, $\bar{\bar{\eta}}_{IJ} n^J = \bar{\bar{\eta}}_{IJ} s^J = \bar{\bar{\eta}}_{IJ} l^J = \bar{\bar{\eta}}_{IJ} k^J = 0$. An upper twiddle indicates the density weight of one w.r.t. $h_{\alpha \beta}$, e.g. $\tilde{s}^I := \sqrt{\det{h}} ~ s^I$. 

Finally, a word of caution: In those parts of this work, in which the internal and external signature do not match, several of the above formulas get changed by signs ($n^I$ becomes spacelike, and the $n \; n$ - terms in the definitions of $\bar{\eta}$ and $\bar{\bar{\eta}}$) or even become obsolete (since, to perform the signature switch, we already are in the Hamiltonian framework, $l^I$ and $k^I$ are not null anymore).

\section{Introduction to the New Variables}
\label{sec:NewVariables}

In \cite{BTTI, BTTII}, a new connection formulation for general relativity in any dimensions $D+1 \geq 3$ was introduced, which will be our starting point to extend the results on quantum black holes obtained in LQG. For completeness, we briefly review the construction of the variables from a Hamiltonian perspective, i.e. extend the ADM phase space of general relativity in such a way that we obtain a Poisson self-commuting connection as one of the canonical variables. For a comprehensive treatment including a Lagrangian derivation, see \cite{BTTI, BTTII}. \\
\\
The ADM Hamiltonian formulation of vacuum general relativity in $D+1$ dimensions is based on a phase space with canonical coordinates $(q_{ab},P^{ab})$, where $q_{ab}$ is the (spatial) metric of Euclidean signature on a $D$-dimensional manifold $\sigma$, $a,b,... \in \{1,...,D\}$. The images $\Sigma_t$ of $\sigma$ under one parameter families of embeddings $X_t: \sigma \rightarrow \Sigma_t \subset \mathcal{M}$ into a $(D+1)$-dimensional manifold $\mathcal{M}$ constitute a foliation of $\mathcal{M}$. The conjugate momentum $P^{ab}$ is related to the extrinsic curvature $K_{ab}$ via
\be
	P^{ab} = - s \sqrt{\det{q}} \left[K^{ab} - q^{ab} K_c\m^c\right]~\text,
\ee
where $s = 1$ for a Euclidean and $s=-1$ for a Lorentzian space time manifold. The non vanishing Poisson brackets (we set the gravitational constant to unity for convenience) are given by
\be \label{eq:ADMBrackets}
\{q_{ab}(x),P^{cd}(y)\}=\delta_{(a}^c\;\delta_{b)}^d\; \delta^{(D)}(x-y) \text{,}
\ee 
where $x,y,..$ are coordinates on $\sigma$. Furthermore, we have the following constraints 
\be \label{eq:ADMDiffeo}
\mathcal{H}_a=-2 q_{ac}\; D_b P^{bc} ~\text{,}
\ee
called spatial diffeomorphism constraint, and the Hamiltonian constraint
\be \label{eq:ADMHam}
\mathcal{H}=-\frac{s}{\sqrt{\det(q)}}[q_{ac} q_{bd}-\frac{1}{D-1} q_{ab} q_{cd}]P^{ab} P^{cd}-
\sqrt{\det(q)} R^{(D)} \text{,}
\ee
where $R^{(D)}$ is the Ricci scalar of $q_{ab}$ and $D_a$ denotes the unique torsion free
covariant derivative compatible with $q_{ab}$.\\
\\
To see in detail how the connection formulation is obtained, we split its construction in three steps:
\begin{enumerate}
\item[(1)] Extensions of the ADM phase space to a formulation with a densitised vielbein $\pi^{aIJ}$ ($I,J ... \in \{0,...,D\}$) in the {\it{adjoint}} representation of so$(D+1)$ or so$(1,D)$,
\item[(2)] Constant Weyl rescaling on the extended phase space with a free parameter $\beta$,
\item[(3)] Perform the same extension again but now to a SO$(D+1)$ or SO$(1,D)$ connection formulation.
\end{enumerate}
(1) The new phase space is coordinatised by the canonical pair $(K_{aIJ},\pi^{bKL})$ with non vanishing Poisson brackets
\be \label{eq:Kpi}
	\{K_{aIJ}(x),\pi^{bKL}(y)\} = 2 \delta_a^b \delta^{[K}_I \delta^{L]}_J \delta^{(D)}(x-y)~\text{,}
\ee
subject to the constraints
\be
	G^{IJ} &:=& [K_a,\pi^a]^{IJ} := 2 K_a^{[I}\m_K \pi^{a K|J]}~\text{,} \\
	S^{aIJ~bKL} &:=& \pi^{a[IJ} \pi^{b|KL]} ~\text{,}
\ee
which are called Gau{\ss} and simplicity constraint, respectively, and which form a first class constraint algebra. The symplectic reduction of the extended phase space w.r.t. Gau{\ss} and simplicity constraint leads back to the ADM phase space. More precisely, we can define a map from the extended to the ADM phase space by
\be
	2\zeta \det{q} q^{ab} &:=& \pi^{aIJ} \pi^b\m_{IJ}~\text{,} \label{eq:q} \\
	 P^{ab} &:=& \frac{1}{2} q^{d(a} \;K_{cIJ}~ \pi^{[b)IJ} \delta^{c]}_d ~\text{,} \label{eq:P}
\ee
where $\zeta=+1$ for SO$(D+1)$ and $\zeta = -1$ for SO$(1,D)$. Then, one can easily verify that $q$ and $P$ as in (\ref{eq:q}, \ref{eq:P}) are both, Gau{\ss} and simplicity invariant, i.e. (weak) Dirac observables, and moreover they satisfy the Poisson relations (\ref{eq:ADMBrackets}) on the surface defined by $G = S = 0$. Of course, $\mathcal{H}_a$ and $\mathcal{H}$ become constraints on the extended phase space by replacing $q, P$ in (\ref{eq:ADMDiffeo}, \ref{eq:ADMHam}) via (\ref{eq:q}, \ref{eq:P}) and, by construction, the whole constraint algebra is of first class. We refrain from explicitly writing out the result since it will not be of relevance in the following.\\
\\
(2) The constant scaling transformation 
\be
	\pi^{aIJ} \rightarrow \m^{(\beta)}\pi^{aIJ} := \frac{1}{\beta} \pi^{aIJ} \text{,} ~~~ K_{aIJ} \rightarrow \m^{(\beta)}K_{aIJ} := \beta K_{aIJ}
\ee
for $\beta \in \mathbb{R}^+$ is, of course, canonical.\\
\\
(3) The last step consist of invoking a certain connection constructed from $\pi^{aIJ}$, the {\it{hybrid spin connection $\Gamma_{aIJ}[\pi]$ weakly compatible with $\pi^{aIJ}$}}, and to redo the above extension of the ADM phase space with the role of $\m^{(\beta)}K_{aIJ}$ now played by a connection while the role of $\m^{(\beta)}\pi^{bKL}$ remains unchanged,
\be \label{eq:CanonTrafo}
	(\m^{(\beta)}K_{aIJ},\m^{(\beta)}\pi^{bKL}) \rightarrow (\m^{(\beta)}A_{aIJ} := \Gamma_{aIJ}[\pi] + \m^{(\beta)}K_{aIJ}, \m^{(\beta)}\pi^{bKL})) ~\text{,}
\ee
and non vanishing Poisson brackets given by
\be  \label{eq:Api}
	\{\m^{(\beta)}A_{aIJ}, \m^{(\beta)}\pi^{bKL}\} = 2 \delta_a^b \delta^{[K}_I \delta^{L]}_J \delta^{(D)}(x-y)~\text{.}
\ee
Of course, we still have to define $\Gamma_{aIJ}[\pi]$. To this end, we have to solve the simplicity constraint \cite{FreidelBFDescriptionOf, BTTI}
\be
	S^{aIJ~bKL} = 0 \Leftrightarrow \pi^{aIJ} = 2 n^{[I}E^{a|J]}~\text{,}
\ee
where $E^{aI} = \sqrt{\det{q}}~ e^{aI}$ is a densitised hybrid vielbein (``hybrid" since the dimensions of the internal space $(D+1)$ and the spatial manifold $(D)$ do not match) and $n^I$ is the unique (up to sign) unit normal on the hybrid vielbein, $n^I n_I = \zeta$, $n^I E^a_I = 0$. It has been shown in \cite{PeldanActionsForGravity} that, like in the case of a genuine vielbein, there exists a unique, so-called hybrid spin connection which annihilates the hybrid vielbein if the internal space is, like in our case, one dimension larger than the external one. It is given by (cf. appendix \ref{app:Connections})
\be \label{eq:HybridSpinConnection}
	\Gamma_{aIJ}[e] = e^{b}_{[I} D_a e_{b|J]} + \zeta ~ n_{[I} \partial_a n_{J]}~\text{,}
\ee
where $D_a$ is the torsion-free $q_{ab}$ - compatible covariant derivative. To define $\Gamma_{aIJ}[\pi]$, we demand that on the constraint surface $S=0$, it should be given by (\ref{eq:HybridSpinConnection}), and extend it off this surface. Then, by construction, $\Gamma_{aIJ}[\pi]$ is weakly compatible\footnote{Note that derivatives of the simplicity constraint can always be removed by partial integrations.} with $\pi^{aIJ}$, i.e. it annihilates $\pi^{aIJ}$ on the constraint surface $S=0$, because $\Gamma[\pi]|_{S=0} = \Gamma[e]$ and $\pi|_{S=0} = \pi[e]$. Thus, we can rewrite the Gau{\ss} constraint to obtain its usual form up to terms which vanish if $S=0$,
\be
G^{IJ} &=& [\m^{(\beta)}A_a-\Gamma[\pi]_a, \m^{(\beta)}\pi^a]^{IJ} \approx [\m^{(\beta)}A_a-\Gamma_a[\pi], \m^{(\beta)}\pi^a]^{IJ} + \partial_a \m^{(\beta)}\pi^{aIJ} + [\Gamma_a[\pi],\m^{(\beta)}\pi^a]^{IJ} \nonumber \\
	&=& \partial_a \m^{(\beta)}\pi^{aIJ} + [\m^{(\beta)}A_a, \m^{(\beta)}\pi^a]^{IJ} := \m^{(\beta)}D_a \m^{(\beta)}\pi^{aIJ}~\text{.}
\ee
For $D=3$, there is a simple expression for the weakly $\pi$-compatible hybrid spin connection given by
\be 
	\Gamma_{aIJ}[\pi] = \pi^{b}\m_{[I|}\m^K D_a \pi_{bK|J]} ~ \text{.}
\ee
In higher dimensions, correction terms are necessary, for explicit expression we refer the reader to \cite{BTTI}. To complete step (3), we have to show that again symplectic reduction with respect to $G, S$ leads back to the ADM phase space. The proof is similar to the case (1) above, but it becomes considerably more intricate to show that after the transformation (\ref{eq:CanonTrafo}), the ADM Poisson brackets are still weakly reproduced. 
The key tool in the proving this is the weak integrability of the extension $\Gamma[\pi]$ of $\Gamma[e]$ off the constraint surface, $\Gamma[\pi] \approx \frac{\delta F}{\delta( \m^{(\beta)}\pi^{aIJ})}$. 
The corresponding generating functional $F$ has been constructed in \cite{BTTI} such that
\be
\delta F &=& \int_{\sigma} d^Dx ~ \left( \delta (\m^{(\beta)}\pi^{aIJ}) ~ (\Gamma_{aIJ}[\pi]+S_{aIJ}) + \frac{2}{\beta} n^{[I} E^{a|J]} \delta\Gamma_{aIJ}[e] \right) \nonumber \\
&\approx& \int_{\sigma} d^Dx ~ \left( \delta (\m^{(\beta)}\pi^{aIJ}) ~ \Gamma_{aIJ}[\pi] + \frac{2}{\beta} \partial_a (E^{aI} \delta n_{I})\right) ~ \text{,}
\ee
where $S_{aIJ}$ vanishes on the simplicity constraint surface. The boundary contribution to the symplectic potential can now be read off,
\be
	\frac{1}{\beta}\int_{\sigma} d^Dx~ \partial_a\left(2 E^{aI}\delta n_I\right) = \frac{1}{\beta}\int_S d^{D-1}x ~ 2 \tilde{s}^{I} \delta n_I~~\text{,}
	\label{eq:Boundary}
\ee
where $s_a \in T^*\sigma$ denotes the unit conormal vector to $S$ pointing outward of $\sigma$, $s^I := s_a e^{aI}$ and the twiddle indicates the density weight of one, $\tilde{s}^I := \sqrt{\det{h}} ~ s^I = \hat{s}_a E^{aI}$ (see sec. \ref{sec:Notation} for $\hat{s}_a$). \\
\\
In $3+1$ dimensions, we have the possibility to introduce a Holst - like modification \cite{BTTII}. Repeating the above calculation then yields the modified boundary term
\be
	\frac{1}{\beta}\int_{\sigma} d^3x ~ \partial_a\left(2 E^{aI}\delta n_I - \frac{1}{\gamma} \epsilon^{abc} e_{bM} \delta e_{c}^M \right)= \frac{1}{\beta}\int_S d^{2}x \left( 2 \tilde{s}^{I} \delta n_I - \frac{1}{\gamma} \epsilon^{\alpha \beta} m_{\alpha I} \delta m_{\beta}^I\right)~~\text{.}
\label{eq:3+1Boundary}
\ee
Note that the second term in equation (\ref{eq:3+1Boundary}) corresponds to the boundary term familiar from Ashtekar-Barbero variables. The boundary terms (\ref{eq:Boundary},\ref{eq:3+1Boundary}) will become important in sections \ref{sec:Hamiltonianframework} and \ref{sec:SO(D+1)}.

\section{Higher-Dimensional Isolated Horizons}
\label{sec:HigherDimensionalIsolatedHorizons}
The isolated horizon framework was introduced in a series of seminal papers \cite{AshtekarIsolatedHorizonsThe, AshtekarMechanicsOfIsolated, AshtekarIsolatedHorizonsHamiltonian, AshtekarIsolatedHorizonsAGeneralization} and extended to higher dimensions in \cite{LewandowskiQuasiLocalRotating, KorzynskiMechanicsOfMultidimensional, AshtekarMechanicsOfHigher, LikoIsolatedHorizonsIn}. We will therefore only briefly state the definition of undistorted, non-rotating isolated horizons in higher dimensions which we will be using, and discuss its consequences. The definition is geared towards the goal of the next section, namely to obtain the boundary condition which will lead to a higher-dimensional Chern-Simons theory on the boundary. We will start by giving the weaker definitions of near expanding and weakly isolated horizons and a brief discussion of their consequences in a manner very similar to \cite{AshtekarIsolatedHorizonsHamiltonian}:
\begin{definition}
A sub-manifold $\Delta$ of $(M,g)$ is said to be a non-expanding horizon (NEH) if
\begin{enumerate}
\item $\Delta$ is topologically $ \mathbb{R} \times S^{(D-1)}$ and null.\footnote{As explained in section \ref{sec:Notation}, more general topologies are allowed without modifications of the definitions, but we restrict to spheres for notational simplicity.}
\item Any null normal $l$ of $\Delta$ has vanishing expansion $\theta_l := h^{\mu \nu} \nabla_{\mu} l_{\nu}$\footnote{On $\Delta$, $h^{\mu \nu}$ is any tensor such that $h_{\mu \nu} = h_{\mu \mu'} h^{\mu' \nu'} h_{\nu \nu'}$}.
\item All field equations hold at $\Delta$ and $-T_{\nu}^{\mu} l^{\nu}$ is a future-causal vector for any future directed null normal $l$.
\end{enumerate}
\label{def:NEH}
\end{definition}
We will state the consequences of definition \ref{def:NEH}. For more details on the derivations, we refer the interested reader to the standard literature cited above:\\
\\
(a) {\textit{Properties of $l$}}: Being a null normal to $\Delta$, $l$ is automatically twist free and geodesic. Moreover, using the vanishing of $\theta_l$, the Raychaudhuri equation and the condition on the stress energy tensor, one can show it is additionally shear free and $R_{\mu \nu} l^{\mu} l^{\nu} \; \widehat{=} \; 0$. \\
\\
(b) {\textit{Conditions on the Ricci tensor:}} From the condition on $T_{\mu \nu}$, the field equations and the relation for $R_{\mu\nu}$ in (a) it follows that $R_{\D{\mu} \nu} l^{\nu} \; \widehat{=} \; 0$, or, in Newman-Penrose formalism,
\be
	\Phi_{00} = R_{\mu \nu} l^{\mu} l^{\nu} \; \widehat{=}\; 0 \text{~~~and~~~} \Phi_{0J} = R_{\mu \nu}l^{\mu} m^{\nu}_J \;\widehat{=}\;0\text{.}
\ee
\\
(c) {\textit{Induced Connection on $\Delta$}}: Due to (a), there exists a unique intrinsic derivative operator $D$ on $\Delta$. Its action on vector fields $X \in T\Delta$ and on 1-forms $\eta \in T^*\Delta$ are given by
\be
	D_{\mu} X^{\nu} \;\widehat{=}\; \nabla_{\D{\mu}} \tilde{X}^{\nu} \text{~~~and~~~} D_{\mu} \eta_{\nu} \;\widehat{=}\; \nabla_{\D{\mu}} \tilde{\eta}_{\D{\nu}} \text{~,}
\ee
where $\tilde X$ and $\tilde \eta$ are arbitrary extensions of $X$, $\eta$ to $M$.\\
\\
(d) {\textit{Natural connection 1-form on $\Delta$}}: From the properties of $l$, it follows that there exists a one-form $\omega^l_{\mu}$ such that
\be
	\nabla_{\D{\mu}} l^{\nu} \;\widehat{=}\; \omega^l_{\mu} l^{\nu} \text{,}
\ee
which implies
\be
	\mathcal{L}_l h_{\Dl{\mu \nu}} \;\widehat{=}\;0 \text{.}
\ee
We define the acceleration of $l$ by $l^{\mu} \nabla_{\mu} l^{\nu} = \kappa^l l^{\nu}$. 
We infer $\kappa^l = i_l \omega^{l}$.\\
\\
(e) {\textit{Conditions on the Weyl tensor}}: From the defining equation of the Riemann tensor, it follows that
\be
	2 (D_{[\mu} \omega^l_{\nu]}) l^{\rho} \;\widehat{=} \; - R_{\Dl{\mu \nu \sigma}}\m^{\rho} l^{\sigma}  \;\widehat{=} \; - C_{\Dl{\mu \nu \sigma}}\m^{\rho} l^{\sigma} \text{,} \label{eq:Weyl0}
\ee
where in the last step we used (b). Contracting (\ref{eq:Weyl0}) with $m_{\rho J}$, we find
\be
	\Psi_{0I0J} \;\widehat{=}\; 0 \text{~~~ and ~~~} \Psi_{0IJK} \;\widehat{=}\; 0 \text{,}  \label{eq:Weyl1}
\ee
(see appendix \ref{app:NPFormalism} for notation) and therefore also 
\be
	0 \;\widehat{=}\; \Psi_{010J} = \Psi_{0IJ}\m^I \text{.}
\ee
Using this and (b), we find
\be
	0 \; \widehat{=}\; C_{\Dl{\mu\nu\rho}\sigma} l^{\nu} l^{\rho} k^{\sigma} \;\widehat{=}\; R_{\Dl{\mu \nu \rho}\sigma} l^{\nu} l^{\rho} k^{\sigma}  \;\widehat{=}\; - \mathcal{L}_l \omega^l_{\D{\mu}} + D_{\D{\mu}}\kappa^l \text{.} \label{eq:dkappa}
\ee
\begin{definition}
A pair $(\Delta, [l])$, where $\Delta$ is a NEH and $[l]$ an equivalence class\footnote{Two null normals $l$ and $l'$ are said to belong to the same equivalence class $[l]$ if $l = c l'$ for some positive constant $c$.} of null normals, is said to be a weakly isolated horizon (WIH) if
\begin{enumerate} \item[4.] $ \mathcal{L}_l \omega \; \widehat{=} \: 0$ \end{enumerate}
for any $l \in [l]$.
\label{def:WIH}
\end{definition}
Note that, while $\omega^l$ in general depends on the choice of null normal $l$, it is invariant under constant rescalings of $l$ and therefore depends only on the equivalence class $[l]$ we fixed. Therefore, we will drop the superscript $\m^l$ in the following. We immediately infer from (\ref{eq:dkappa}) that the $0^{\text{th}}$ law holds for WIH,
\be
	\Dl{d\kappa^l} \;\widehat{=}\; 0\text{.}
\ee
In the following, we will slightly strengthen this usual definition of WIHs in a way which is very similar to the definitions given in \cite{AshtekarMechanicsOfIsolated} by introducing some extra structure. Fix a foliation of $\Delta$ by $(D-1)$ - spheres. Denote by $[k]$ an equivalence class of 1-form fields normal to the foliation of $\Delta$ by $(D-1)$ - spheres\footnote{Again, two 1-forms $k,k'$ are called equivalent if $k = c k'$ for some constant $c$.}. We require that any $k \in [k]$ is closed on $\Delta$. We extend them uniquely to spacetime 1-forms on $\mathcal{M}$ by requiring that they be null. Now, we introduce the equivalence class of pairs $[l,k]$ where each pair $(l^{\mu}, k_{\nu})$ satisfies $i_l k = -1$, i.e. we fix $l$ and $k$ up to mutually inverse and constant rescaling. We further\footnote{For spherical topologies, this would already follow from $S^{(D-1)}\times \mathbb{R}$ being simply connected.} demand $k = -dv$ for some function $v$ on $\Delta$, and each leaf $S_v \; \widetilde{=}\; S^{(D-1)}$ of the fixed foliation is characterised by $v = \text{const}$. By {\it{spherically symmetric}}, we will in the following mean {\it{constant on the leaves $S_v$}}, e.g. for a spherically symmetric function $f = f(v)$.
\begin{definition}
A undistorted non-rotating isolated horizon (UDNRIH) is a WIH where to each $l \in [l]$ there is a $k$ like above, such that
\begin{itemize}
	\item[5.] $k$ is shear-free with nowhere vanishing spherically symmetric expansion and vanishing Newman - Penrose coefficients $\pi_J \;\widehat{=}\; l^{\mu}m^{\nu}_J \nabla_{\mu} k_{\nu}$ on $\Delta$.
	\item[6.] The Euler density $E^{(D-1)}$ of the $(D-1)$ -- sphere cross sections obeys $E^{(D-1)} / \sqrt{h} = f(v)$ for some function $f$, i.e. the given ratio is constant on each leaf $S_v$.
\end{itemize}
\label{def:UDNRIH}
\end{definition}
Two remarks are in order: Firstly, in $D=3$, one finds for undistorted non-rotating isolated horizons \cite{AshtekarMechanicsOfIsolated}, instead of the last condition,
\begin{itemize}
	\item[\it{6'.}] $T_{\mu \nu}l^{\mu}k^{\nu}$ \it{is spherically symmetric at $\Delta$.}
\end{itemize}
It is only for $D=3$ that {\it{6.}} and {\it{6'.}} are equivalent. {\it{6'.}} can be shown to be equivalent to demanding that the curvature scalar $R^{(2)}$ of the 2-sphere cross sections be constant. In two dimensions, we have $E^{(2)} = \text{const.} \times R^{(2)} \sqrt{h} = f(v) \sqrt{h}$ for some scalar function $f$. In higher dimensions, condition {\it{6'.}} still is equivalent to demanding that $R^{(D-1)}$ is constant on $S_v$. However, we will see that for our purposes, this condition is unnecessary, but has to be replaced by {\it{6.}} This will be discussed explicitly in section \ref{sec:BoundaryCondition}. Apart from that, compared with \cite{AshtekarMechanicsOfIsolated}, our definition \ref{def:UDNRIH} is slightly stronger (more restrictive) in that \cite{AshtekarMechanicsOfIsolated} does not demand {\it{4.}} Furthermore, whereas we only allow for constant rescaling of $l,k$, in \cite{AshtekarMechanicsOfIsolated} they are fixed up to spherically symmetric and mutually inverse rescaling, but later in that paper, the gauge freedom of rescaling is fixed completely.\\
Secondly, the definition given above is tied to a foliation. The standard definitions of (W)IH are usually foliation independent, though some results rely on the existence of a so called good cuts foliation. Moreover, when going to the Hamiltonian formulation, one usually demands that the spacetime foliation is such that at the boundary, the foliation coincides with this preferred foliation. Note that our fixed foliation is a good cuts foliation. We leave the question if all results obtained here hold in the more general context of weaker definitions of (W)IH or ones without reference to a fixed foliation for further research and continue by stating the consequences of definition \ref{def:UDNRIH}: \\
(f) {\textit{Properties of $k$, $\omega$ and its curvature}}:
By the above requirements, 
we find for vectors $u$ tangential to $\Delta$ using $k^{\mu} \nabla_u k_{\mu} = 0$
\be
	\nabla_{u} k_{\nu} &=& u^{\mu} \left(h_{\nu}^{\nu'} h_{\mu}^{\mu'} \nabla_{\mu'} k_{\nu'} -  k_{\nu} \omega_{\mu}\right) \nonumber \\
	&=& u^{\mu} \left(\frac{1}{D-1} \theta_k h_{\mu \nu} -  k_{\nu} \omega_{\mu}\right) \text{.}
\ee
Furthermore, we have for tangential vectors $u$ and $v$
\be
	0 = u^{\mu} v^{\nu} \nabla_{[\mu} k_{\nu]} = - u^{\mu} v^{\nu} k_{[\nu} \omega_{\mu]} \text{,}
\ee
from which we conclude that $\omega = \hat{f} \, k$ for some function $\hat{f}$. Since $i_l\omega = \kappa^l$, we have $\hat{f} = - \kappa^l$ or 
\be
	\omega = - \kappa^l k \text{.}
\ee
\\
Contraction of (\ref{eq:Weyl0}) with $k_{\rho}$ yields
\be
	2 D_{[\mu} \omega_{\nu]} \;\widehat{=} \; C_{\Dl{\mu\nu\sigma}}\m^{\rho} l^{\sigma} k_{\rho}  \;\widehat{=} \; m_{\D{\mu}}^I m_{\D{\nu}}^J \Psi_{01IJ} \text{,}
\ee
where in the last step we used the trace freeness of the Weyl tensor and (\ref{eq:Weyl1}). We can furthermore conclude that ${d\omega} \;\widehat{=}\; 0$ and $ \Psi_{01IJ} \;\widehat{=}\; 0$, since $\omega \;\widehat{=}\; - \kappa^l \D{k}$ and $\Dl{d\kappa^l} \;\widehat{=}\; 0 \; \widehat{=}\; \Dl{dk}$. This can be traced back to the requirement $\pi_J \;\widehat{=}\; 0$ in the definition of UDNRIHs, and in analogy to the $D=3$ case, this is why we refer to these horizons as non-rotating. (Note that $\Psi_{01IJ}$ is the analog of $\mathfrak{Im}\Psi_2$ in $D=3$).

\section{Boundary Condition}
\label{sec:BoundaryCondition}

In this section, we will derive the boundary condition relating the bulk with the horizon degrees of freedom starting from the Palatini action. This forces us to use SO$(1,D)$ as the internal gauge group as opposed to SO$(D+1)$, which can be used in the Hamiltonian formalism even for Lorentzian signature. In a later chapter, we will rederive the boundary condition independently of the internal signature, thus allowing us to use the loop quantisation based on SO$(D+1)$ connection variables for the bulk degrees of freedom.

Due to {\it{3.}} of definition \ref{def:NEH}, we have at points of $\Delta$
\be
	F_{\mu\nu}\m^{IJ} \;\widehat{=}\; R_{\mu \nu}\m^{IJ} = R^{(D+1)}_{\mu \nu \rho \sigma} ~ e^{\rho I} e^{\sigma J} \text{.}
\ee
In the following, we will use the notation introduced in appendix \ref{app:NPFormalism} for the Weyl tensor also for the Riemann tensor, e.g. $R_{01IJ} = R^{(D+1)}_{\mu \nu \rho \sigma} l^{\mu} k^{\nu} m^{\rho}\m_I m^{\sigma}\m_J$. Note that therefore, the internal indices appearing on $R$ and $\Psi$ are perpendicular to $l^I$ and $k^I$, which will be used in several calculations in this section. Pulling back to $\Delta$, we obtain
\be
	& & F_{\Dl{\mu \nu}}\m^{IJ} = R_{\Dl{\mu \nu}}\m^{IJ} = R^{(D+1)}_{\Dl{\mu \nu}\rho \sigma} e^{\rho I} e^{\sigma J}  \nonumber \\
	&=& \left( \D{h_{\mu}}^{\mu'} \D{h_{\nu}}^{\nu'} R^{(D+1)}_{\mu' \nu'\rho \sigma} - 2 \D{k_{[\mu}} \D{h_{\nu]}}^{\nu'} l^{\mu'} R^{(D+1)}_{\mu' \nu' \rho \sigma} \right) \left(m^{\rho I} m^{\sigma J} - 2 m^{\rho [I} l^{\sigma} k^{J]}  - 2 m^{\rho [I} k^{\sigma} l^{J]}  + 2 l^{[\rho} k^{\sigma]} k^{[I} l^{J]} \right) \nonumber  \\
	&=& \D{h_{\mu}}^{\mu'} \D{h_{\nu}}^{\nu'} R^{(D+1)}_{\mu'\nu'\rho \sigma} m^{\rho I} m^{\sigma J} + \D{m_{\mu}}^K \D{m_{\nu}}^L \left( -2 R_{KL}\m^{[I}\m_0 k^{J]} -2 R_{KL}\m^{[I}\m_1 l^{J]} +2 R_{KL01} k^{[I} l^{J]}\right) \nonumber \\
	& & - 2 \D{k_{[\mu}} \D{m_{\nu]}}^{K} \left(R_{0K}\m^{IJ} -2 R_{0K}\m^{[I}\m_{0} k^{J]} -2 R_{0K}\m^{[I}\m_{1} l^{J]} +2 R_{0K01} k^{[I}l^{J]} \right) \nonumber \\ 
	&=& \D{h_{\mu}}^{\mu'} \D{h_{\nu}}^{\nu'} R^{(D-1)}_{\mu'\nu'\rho \sigma} m^{\rho I} m^{\sigma J} + \D{m_{\mu}}^K \D{m_{\nu}}^L \left( -2 \Psi_{KL}\m^{[I}\m_0 k^{J]} -2 R_{KL}\m^{[I}\m_1 l^{J]} +2 \Psi_{KL01} k^{[I} l^{J]}\right) \nonumber \\
	& & - 2 \D{k_{[\mu}} \D{m_{\nu]}}^{K} \left(\Psi_{0K}\m^{IJ} -2 \Psi_{0K}\m^{[I}\m_{0} k^{J]} -2 R_{0K}\m^{[I}\m_{1} l^{J]} +2 \Psi_{0K01} k^{[I}l^{J]} \right) \nonumber \\
	&=& \D{h_{\mu}}^{\mu'} \D{h_{\nu}}^{\nu'} R^{(D-1)}_{\mu'\nu'\rho \sigma} m^{\rho I} m^{\sigma J} + 4 \D{k_{[\mu}} \D{m_{\nu]}}^{K} R_{0K}\m^{[I}\m_{1} l^{J]} \nonumber \\
	&=& \D{h_{\mu}}^{\mu'} \D{h_{\nu}}^{\nu'} R^{(D-1)}_{\mu'\nu'\rho \sigma} m^{\rho I} m^{\sigma J} + \frac{4}{D-1} \D{k_{[\mu}} \D{m_{\nu]}}^{[I} l^{J]} \left[\nabla_l \theta_k + \kappa^l \theta_k \right] \text{,} \label{eq:Pullback1}
\ee
where in the fourth line, we used that $\Phi_{0J} \;\widehat{=}\; 0$, $\Phi_{00} \;\widehat{=}\; 0$ to replace some Riemann tensor components by the corresponding Weyl tensor components, and in the fifth line we used $0 \;\widehat{=}\; \Psi_{0IJK}  \;\widehat{=}\; \Psi_{01JK}  \;\widehat{=}\; \Psi_{0I0J}  \;\widehat{=}\; \Psi_{010J}$ and furthermore for $u_{\sigma}$ such that $u \cdot l  =0 = u \cdot k$,
\be
	R^{(D-1)}_{\mu \nu \rho} \m^{\sigma} u_{\sigma} 
	&=& [D_{\mu} D_{\nu}] u_{\rho} \nonumber \\ 
	&=& 2 h_{[\mu}^{\mu'} h_{\nu]}^{\nu'} h_{\rho}^{\rho'} \nabla_{\mu'} h_{\nu'}^{\nu''}h_{\rho'}^{\rho''} \nabla_{\nu''} u_{\rho''} \nonumber \\
	&=& h_{\mu}^{\mu'} h_{\nu}^{\nu'} h_{\rho}^{\rho'} h_{\sigma'}^{\sigma} R^{(D+1)}_{\mu' \nu' \rho'}\m^{\sigma'} u_{\sigma} + 2 h_{[\mu}^{\mu'} h_{\nu]}^{\nu'} h_{\rho}^{\rho'} (\nabla_{[\mu'} h_{\nu']}^{\nu''}h_{\rho'}^{\rho''}) \nabla_{\nu''} u_{\rho''} \nonumber \\
	&\widehat{=}& h_{\mu}^{\mu'} h_{\nu}^{\nu'} h_{\rho}^{\rho'} h_{\sigma'}^{\sigma} R^{(D+1)}_{\mu' \nu' \rho'}\m^{\sigma'} u_{\sigma} \text{.} \label{eq:RiemannRiemann}
\ee
The second term in the second to last line vanishes due to
\be
	h_{[\mu}^{\mu'} h_{\nu]}^{\nu'} h_{\rho}^{\rho'} ( \nabla_{[\mu'} h_{\nu']}^{\nu''}h_{\rho'}^{\rho''} ) \nabla_{\nu''} u_{\rho''}
&=& h_{[\mu}^{\mu'} h_{\nu]}^{\nu'} h_{\rho}^{\rho''} \nabla_{[\mu'} (l_{\nu']} k^{\nu''} + k_{\nu']} l^{\nu''})  \nabla_{\nu''} u_{\rho''} \nonumber \\
& & + h_{[\mu}^{\mu'} h_{\nu]}^{\nu''} h_{\rho}^{\rho'} \nabla_{\mu'}( l_{\rho'} k^{\rho''} + k_{\rho'} l^{\rho''} )  \nabla_{\nu''} u_{\rho''}
 \nonumber \\
	&\widehat{=}&  h_{[\mu}^{\mu'} h_{\nu]}^{\nu'} h_{\rho}^{\rho''} ((\nabla_{[\mu'} l_{\nu']}) k^{\nu''} + (\nabla_{[\mu'} k_{\nu']}) l^{\nu''})  \nabla_{\nu''} u_{\rho''}
+ \nonumber \\ 
	& & h_{[\mu}^{\mu'} h_{\nu]}^{\nu''} h_{\rho}^{\rho'} ((\nabla_{\mu'} l_{\rho'}) k^{\rho''} + (\nabla_{\mu'}k_{\rho'}) l^{\rho''} )  \nabla_{\nu''} u_{\rho''}
 \nonumber \\
&\widehat{=}&  h_{[\mu}^{\mu'} h_{\nu]}^{\nu'} h_{\rho}^{\rho''} (\nabla_{[\mu'} k_{\nu']}) l^{\nu''} \nabla_{\nu''} u_{\rho''}
- \nonumber \\ 
	& & h_{[\mu}^{\mu'} h_{\nu]}^{\nu''} h_{\rho}^{\rho'} (\nabla_{\mu'}k_{\rho'}) u^{\rho''}  \nabla_{\nu''} l_{\rho''} \nonumber \\
	&\widehat{=}& 0 ~~ \text{,}
\ee
where in the first line we used $\nabla g = 0$,  in the second line that $h(l,.) = 0 = h(k,.)$, in the third that $\Dl{\nabla_{\mu} l_{\nu}} = 0$ and $l^{\mu} \nabla_{\rho} u_{\mu} = - u^{\mu} \nabla_{\rho} l_{\mu}$, and in the fourth line and $\Dl{dk} = 0$.

Finally, we have to account for the vanishing of $R_{IJK1}$ in (\ref{eq:Pullback1}), which follows from
\be
R_{IJK1} &=& \Psi_{IJK1} + \frac{2}{D-1} \bar{\bar{\eta}}_{K[I} \Phi_{J]1} \nonumber \\
	&=&m^{\mu}_I m^{\nu}_J m^{\rho}_K R_{\mu \nu \rho \sigma}^{(D+1)} k^{\sigma} =m^{\mu}_I m^{\nu}_J m^{\rho}_K \left[\nabla_{\mu},\nabla_{\nu}\right] k_{\rho} \nonumber \\
	&=& 2~m^{\mu}_I m^{\nu}_J m^{\rho}_K  \nabla_{[\mu} \left( \left(h_{\nu]}^{\nu'} - l_{\nu]} k^{\nu'} - k_{\nu]} l^{\nu'} \right) \left(h_{\rho}^{\rho'}-l_{\rho}k^{\rho'}-k_{\rho}l^{\rho'}\right) \nabla_{\nu'} k_{\rho'}\right) \nonumber \\
	&\widehat{=}& 2~m^{\mu}_I m^{\nu}_J m^{\rho}_K  \nabla_{[\mu} \left(h_{\nu]}^{\nu'} \left(h_{\rho}^{\rho'}-k_{\rho}l^{\rho'}\right) \nabla_{\nu'} k_{\rho'}\right) \nonumber \\
	&\widehat{=}& 2~m^{\mu}_{[I} m^{\nu}_{J]} m^{\rho}_K  \nabla_{\mu} \left( \frac{1}{D-1} h_{\nu}^{\nu'} h_{\rho}^{\rho'} h_{\nu' \rho'} \theta_k - h_{\nu}^{\nu'} k_{\rho} \omega_{\nu'}^l \right) \nonumber \\
	&\widehat{=}& \frac{2}{D-1} ~m^{\mu}_{[I} m^{\nu}_{J]} m^{\rho}_K  \left( h_{\nu \rho}  \nabla_{\mu} \theta_k - h_{\mu \rho} \theta_k \omega_{\nu}^l \right) \nonumber \\
	&\widehat{=}& \frac{2}{D-1} ~m^{\mu}\m_{[I} \bar{\bar{\eta}}_{J]K} \left( \nabla_{\mu} \theta_k + \theta_k \omega_{\mu}^l \right) \nonumber \\
	&\widehat{=}& \frac{2}{D-1} ~m^{\mu}\m_{[I} \bar{\bar{\eta}}_{J]K} \left( - (\nabla_l \theta_k) k_{\mu} - \theta_k \kappa^l k_{\mu} \right) ~ \widehat{=} ~ 0 \text{.}
	\label{eq:Weyl3}
\ee
From the third to the fourth line, we dropped the second two summands in the first round bracket because $l$ and $k$ are twist free, and the second summand in the second bracket since $k^{\mu} \D{\nabla} k_{\mu} = 0$. In the fifth line, we used that $k$ is twist and shear free and that $l^{\mu} \D{\nabla} k_{\mu} = \omega^l$. In line 6, we again invoke the twist and shear freeness of $k$. In the last line, we used that $\Dl{d \theta_k} = - k \nabla_l \theta_k$ since it is spherical symmetric by definition \ref{def:UDNRIH} and that $\omega^l = - \kappa^l k$.\footnote{Comparing with the $3+1$ dimensional case, we find $R_{IJK1} = \Psi_{IJK1} + \frac{2}{D-1} \bar{\bar{\eta}}_{K[I} \Phi_{J]1} = 0$ corresponds to $\Psi_3 - \Phi_{21} = 0$, $\Psi_{KLJ0} = 0$ to $\Psi_0 = 0$ and $\Psi_1= 0$, and $\Psi_{KL01}=0$ to the non-rotating condition $\mathfrak{Im} \Psi_2 = 0$.}

In the last line of (\ref{eq:Pullback1}), we furthermore used
\be
	R_{0I1J} &=& C_{0I1J} + \frac{1}{D-1} \left(\bar{\bar{\eta}}_{IJ} \Phi_{01} - \Phi_{IJ} \right) - \frac{1}{D(D+1)} \bar{\bar{\eta}}_{IJ} R^{(D+1)} \nonumber \\
		&=& -\frac{1}{D-1}\bar{\bar{\eta}}_{IJ} \left[\nabla_l \theta_k + \kappa^l \theta_k \right] \text{,} \label{eq:Weyl4}
\ee
which can be shown analogously.

Since the pullback to $H$ of the second summand in (\ref{eq:Pullback1}) is zero ($\DD{k} = 0$), we finally obtain when pulling back once more
\be
	\DD{F}\m_{\mu \nu IJ} &=& \DD{R}\m_{\mu\nu IJ} ^{(D+1)} = \DDl{h_{\mu}}^{\mu'} \DDl{h_{\nu}}^{\nu'} R_{\mu' \nu' \rho \sigma}^{(D+1)} e^{\rho I} e^{\sigma J}  = \DDl{h_{\mu}}^{\mu'} \DDl{h_{\nu}}^{\nu'} R_{\mu' \nu' \rho \sigma}^{(D-1)} m^{\rho I} m^{\sigma J} \text{.} \label{eq:BoundaryCondition1}
\ee
and therefore, for $D-1 = 2n$ even,
\be
	& &\epsilon^{K_1L_1 ... K_{n} L_{n} IJ} \DD{\epsilon}\m^{^{\mu_1 \nu_1... \mu_{n} \nu_{n}}} \DD{F}\m_{\mu_1 \nu_1 K_1 L_1} ... \DD{F}\m_{\mu_{n} \nu_{n} K_{n} L_{n}}   \nonumber \\
	& =&\epsilon^{K_1L_1 ... K_{n} L_{n} IJ} \DD{\epsilon}\m^{^{\mu_1 \nu_1... \mu_{n} \nu_{n}}} R^{(D-1)}_{\mu_1 \nu_1 \rho_1 \sigma_1}  ... R^{(D-1)}_{\mu_{n} \nu_{n} \rho_{n} \sigma_{n}} m^{\rho_1 K_1 } m^{\sigma_{1} L_{1}}... m^{\rho_n K_n }m^{\sigma_{n} L_{n}} \nonumber \\
	& =& \frac{1}{\sqrt{h}}\DD{\epsilon}\m^{\rho_1\sigma_1 ... \rho_{n} \sigma_{n}} \DD{\epsilon}\m^{\mu_1 \nu_1... \mu_{n} \nu_{n}} R^{(D-1)}_{\mu_1 \nu_1 \rho_1 \sigma_1}  ... R^{(D-1)}_{\mu_{n} \nu_{n} \rho_{n} \sigma_{n}} 2 n^{[I} s^{J]} \approx \frac{E^{(2n)}}{\sqrt{h}} \pi^{aIJ} \hat{s}_a~~\text{,}  \label{eq:BoundaryCondition2}
\ee
where $E^{(2n)}$ denotes the Euler density of the $(D-1)$ -- sphere cross sections, $\hat{s}_a$ is the appropriately densitised conormal on $S$, and $\approx$ means equal up to the simplicity constraint. Finally, by {\it{6.}} of definition \ref{def:UDNRIH}, $E^{(2n)} = f(v) \sqrt{h}$. Some comment on the role of the equations (\ref{eq:BoundaryCondition1}, \ref{eq:BoundaryCondition2}) is in order. \\
Firstly, notice that both of these equations are generalisations of the 3+1 dimensional boundary conditions $\DD{F}^{(4)}_{\mu \nu IJ} \propto R^{(2)} \DD{\Sigma}_{\mu \nu IJ}$ known from the U(1) and SU(2) treatments. \eqref{eq:BoundaryCondition1} has the same left hand side, but further manipulation of the right hand side as in the 3+1 dimensional case is not possible, since the Riemann tensor is in general not completely determined by the Ricci scalar in higher dimensions and the Ricci scalar also ceases to play a topological role. \eqref{eq:BoundaryCondition2} generalises the right hand side, the topological role now being played by the Euler density, while the left hand side is more complicated than in the 3+1 dimensional case. \\
Secondly, at the quantum level, we want to work with an independent Chern-Simons connection on the horizon from the onset and demand by constraint that the boundary connection actually is determined by the bulk fields. This constraint is in 3+1 dimensions precisely given by the boundary condition $\DD{F}^{(4)}_{\mu \nu IJ} \propto \DD{\Sigma}_{\mu \nu IJ}$. In higher dimensions, one can easily convince oneself that \eqref{eq:BoundaryCondition2} is insufficient to determine the boundary connection and one has to impose \eqref{eq:BoundaryCondition1} at the quantum level. However, \eqref{eq:BoundaryCondition2} connects the momenta conjugate to the bulk connection with Chern-Simons excitations and therefore is a direct generalisation of what is imposed at the quantum level in the 3+1 dimensional case. It therefore could serve as a consistency requirement additionally to \eqref{eq:BoundaryCondition1}, see the discussion in section \ref{sec:Quantisation}.

One last comment concerning {\it{6'.}}: Assuming this condition to hold, one easily obtains that
\be
	G_{\mu \nu} l^{\mu} k^{\nu} = \Phi_{01} + \frac{D-1}{2(D+1)} R^{(D+1)}
\ee
is spherically symmetric. Moreover, taking the trace of (\ref{eq:Weyl4}), we infer that
\be
	C_{0I1}\m^I + \frac{D-3}{D-1} \Phi_{01} - \frac{D-1}{D(D+1)} R^{(D+1)} = - \nabla_l \theta_k - \kappa^l \theta_k 
\ee
is spherically symmetric since the right hand side is. Finally, from (\ref{eq:RiemannRiemann}),
\begin{align}
	R^{(D-1)} &= R_{IJ}\m^{IJ} = 2 C_{0I1}\m^I + \frac{4(D-2)}{D-1} \Phi_{01} + \frac{(D-2)(D-1)}{D(D+1)} R^{(D+1)}  \nonumber \\
	&= 2 \left(C_{0I1}\m^I + \frac{D-3}{D-1} \Phi_{01} - \frac{D-1}{D(D+1)} R^{(D+1)} \right) + 2 \left( \Phi_{01} + \frac{D-1}{2(D+1)} R^{(D+1)}\right) \text{,}
\end{align}
where Weyl tensor component identities from appendix \ref{app:NPFormalism}  were used. Since both summands in round brackets are spherically symmetric, we find that $R^{(D-1)}$ is also spherically symmetric. As we already remarked at the beginning of section \ref{sec:HigherDimensionalIsolatedHorizons}, this property will not be needed in higher dimensions, but instead {\it{6.}} will be crucial in the next section.

\section{Hamiltonian Framework}
\label{sec:Hamiltonianframework}
In this section, we will show, starting from the Palatini action in $(D+1) = 2(n+1)$ dimensions, how the symplectic structure of $(2n+1)$ - dimensional Chern-Simons theory arises as boundary contribution to the symplectic structure for an internal boundary with UDNRIH conditions. We restrict to a vanishing cosmological constant. Note that the mechanics of higher-dimensional isolated horizons has already been studied in the quasi-local, the asymptotically flat \cite{KorzynskiMechanicsOfMultidimensional} as well as the asymptotically anti-de Sitter \cite{AshtekarMechanicsOfHigher} case. However, in all these treatments, the internal SO$(1,D)$ transformations were (partially) gauge fixed. In view of the boundary term of the generating functional for the canonical transformation to SO$(1,D)$ connection variables which we found in section \ref{sec:NewVariables} and which we expect to be related to the boundary symplectic structure, we are not allowed to fix the internal gauge freedom. In particular, in the usual time gauge $n^I = \delta^I_0$, this boundary term vanishes since it is proportional to $\delta n^I$. Therefore, we will rederive the Hamiltonian framework for IH in higher dimensions for our specific definition of UDNRIH without using any internal gauge fixing. Indeed, the derivation deviates from the usual treatment and we obtain the same boundary contribution to the symplectic structure we derived in \ref{sec:NewVariables}, which a) vanishes in time gauge and b) can be reexpressed as SO$(1,D)$ Chern-Simons symplectic structure.
\\
Consider a region $\mathcal{M}$ in a $(D+1)$ - dimensional Lorentzian spacetime ($\mathcal{M}', g$) bounded by two (partial) Cauchy slices $\Sigma_1$ and $\Sigma_2$, $\Delta$, and possibly an outer boundary. On $\Delta$, we impose the UDNRIH boundary conditions and furthermore require that $\Sigma_1, \Sigma_2$ intersect $\Delta$ in leaves ($(D-1)$ - spheres) of the preferred foliation $S_1, S_2$, respectively. Moreover, as usual in the IH literature, for a given history ($e, A$) the horizon area $A_S$ is constant in time as we will show shortly (below (\ref{eq:DefEpsilonD-1})). We will now furthermore fix the horizon area to be a constant throughout the histories we are considering, $\delta A_S = 0$. The Palatini action\footnote{Note that a well defined action principle can require a boundary term, as e.g. the York-Gibbons-Hawking boundary term \cite{YorkRoleOfConformal, GibbonsActionIntegralsAnd} or its analogue in first order theories \cite{AshtekarActionAndHamiltonians, BNI}. However, such a boundary term does not enter the second variation of the action which will be relevant in this paper for deriving the Chern-Simons symplectic structure. We will thus neglect it for simplicity. 
For a discussion of these issues in higher dimensions, we refer the interested reader to e.g. \cite{AshtekarActionAndHamiltonians, BNI} and, specifically in the IH framework, \cite{KorzynskiMechanicsOfMultidimensional}. } is given by
\begin{align}
	S[A,e] = \int_{\mathcal{M}} \Sigma_{IJ} \wedge F^{IJ} \text{,}
\end{align}
where $F = 1/2 F_{\mu \nu} dx^{\mu}\wedge dx^{\nu}$, $F_{\mu \nu}\m^{IJ} = 2\partial_{[\mu} A_{\nu]}\m^{IJ} + [A_{\mu}, A_{\nu}]^{IJ}$, $\Sigma := -*(e \wedge e)$, or in coordinates $-*(e \wedge e)_{\mu_1 \ldots \mu_{D-1} IJ} = - \frac{1}{(D-1)!} e_{\mu_1}^{K_1} \ldots e_{\mu_{D-1}}^{K_{D-1}} \epsilon_{IJ K_1 \ldots K_{D-1}}$, and as already stated, boundary terms possibly needed for $\mathcal{T}$ are neglected. Variation with respect to $A$ gives rise to a surface term
\begin{align}
	\int_{\Delta} \D{\Sigma}\m_{IJ} \wedge \delta \D{A}^{IJ} \text{,}
\end{align}
which, however, vanishes when imposing the UDNRIH boundary conditions, and therefore, the variation only yields the bulk equations of motion. This is a standard result in the IH literature, but will be derived here without any internal gauge fixing. Using $\D{e}\m_{\mu I} = m_{\mu I} - k_{\mu} l_I$, we immediately find
\be
	\D{\Sigma}\m_{IJ} = -\frac{1}{(D-1)!} \epsilon_{IJK_1 ...K_{D-1}} \left[ m^{K_1} \wedge ... \wedge m^{K_{D-1}} - (D-1) l^{K_1} ~ k \wedge m^{K_2} \wedge ... \wedge m^{K_{D-1}} \right] \text{.} \label{eq:SigmaPullback}
\ee
For the pullback of the spacetime connection $A$ we find analogous to the calculations in section \ref{sec:BoundaryCondition}
\be
	\D{A}\m_{\mu IJ} &=& \D{\Gamma}\m_{\mu IJ} = \Gamma^{0}_{\mu IJ} + \frac{2}{D-1} l_{[I} m_{\mu |J]} \theta_k - 2 \omega_{\mu} l_{[I} k_{J]} \text{,} \\
	\Gamma^{0}_{\mu IJ} &=&  m^{\nu}_{[I} \D{\nabla}\m_{\mu} m_{\nu |J]} - l_{[I} \D{\nabla}\m_{\mu} k_{|J]} - k_{[I} \D{\nabla}\m_{\mu} l_{|J]} \text{,}
\ee
where $\Gamma^0$ here denotes the connection on $\Delta$ which annihilates $m_{\mu K}$, $l_I$ and $k_J$. Here and in the following, we will understand that $m^{\mu I} := h^{\mu \nu} m_{\nu}\m^I$ and $h^{\mu \nu} = g^{\mu \mu'} h_{\mu'\nu'} g^{\nu' \nu}$ such that $h^{\mu \nu} k_{\nu} = 0$. \\
For the variation of $\D{A}$, we find
\begin{alignat}{3}
	\D{\delta A}\m_{\mu IJ} & \; = \; \delta \Gamma^0\m_{\mu IJ} & \; + \; & \frac{2}{D-1} \left[ (\delta l_{[I}) m_{\mu |J]} \theta_k + l_{[I} (\delta m_{\mu |J]}) \theta_k + l_{[I} m_{\mu |J]} (\delta \theta_k) \right] \nonumber \\ 
	& &- \; & 2 \left[(\delta \omega_{\mu}) l_{[I} k_{J]} + \omega_{\mu} \delta( l_{[I} k_{J]}) \right] \text{,} \label{eq:DeltaA}
\end{alignat}
which for the case at hand can be reduced to
\be
	\D{\delta A}\m_{\mu IJ} &=& 2 k_{\mu} l_{[I} k_{J]} l^{K} k^{L} l^{\nu} \D{\delta A}\m_{\nu KL} - 2 k_{[I} \bar{\bar{\eta}}\m_{J]}\m^{L} l^{K} h_{\mu}^{\nu} \D{\delta A}\m_{\nu KL} + \mathcal{R} \nonumber \\ 
	&=& 2 k_{\mu} l_{[I} k_{J]} l^{K} k^{L} l^{\nu} \left[\delta \Gamma^0\m_{\nu KL} - 2 l_{[K} k_{L]} \delta \omega_{\nu}\right] - 2 k_{[I} \bar{\bar{\eta}}\m_{J]}\m^{L} l^{K} h_{\mu}^{\nu} \delta \Gamma^0\m_{\nu KL} + \mathcal{R} \nonumber \\
	&=& 2 k_{\mu} l_{[I} k_{J]}  l^{\nu} \left[ k^{L} D^{\Gamma^0}_{\nu}\delta l_L + \delta \omega_{\nu}\right] - 2 k_{[I} \bar{\bar{\eta}}\m_{J]}\m^{L}  h_{\mu}^{\nu} D^{\Gamma^0}_{\nu} \delta l_L + \mathcal{R}\text{,} \label{eq:ConnectionPullback}
\ee
where in the first line, we made use of the fact that only certain components of $\D{\delta A}$ will appear when contracted with $\D{\Sigma}$ and $\mathcal{R}$ stands for the remaining terms which vanish in this contraction. In the second step, several terms drop out due to $l^{\nu} \delta m_{\nu I} = - m_{\nu I} \delta  l^{\nu} =  - m_{\nu I} c_{\delta} l^{\nu} = 0$ since $l$ is fixed up to constant rescaling on $\Delta$, $l^I \delta l_I = 0$ since $l^2 = 0$ on $\Delta$, and $h_{\mu}^{\nu} \omega_{\nu} \;\widehat{=}\; 0$. Finally, we used that $l^K \delta \Gamma^0_{\mu KL} = -\delta D^{\Gamma^0}_{\mu} l_L + D^{\Gamma^0}_{\mu} \delta l_L = D^{\Gamma^0}_{\mu} \delta l_L$ since $\Gamma^0$ annihilates $l^I$. Putting all together, we recover for the definition of an UDNRIH as given in section \ref{sec:HigherDimensionalIsolatedHorizons} the result that there is no boundary term in symplectic potential for the horizon,
\be
	\int_{\Delta} \D{\Sigma} \wedge \D{\delta A} &=& \int_{\Delta} \D{\Sigma} \wedge \D{\delta \Gamma} \nonumber \\
	&=&  -\frac{1}{(D-1)!} \int_{\Delta} \left( m^{K_1} \wedge ... \wedge m^{K_{D-1}} - (D-1)  l^{K_1} \; k \wedge m^{K_2} \wedge ... \wedge m^{K_{D-1}} \right)  \times \nonumber \\
	& & \qquad \epsilon_{IJK_1 ... K_{D-1}} \wedge \left\{ -2 l^{[I} k^{J]} \left[d(k_M \delta l^M) + \delta \omega \right] + 2 \bar{\bar{\eta}}^{I}_{I'} k_J d_{\Gamma^0} \delta l^{I'}\right\} \nonumber \\
	&=&  -\frac{2}{(D-1)!}  \int_{\Delta}  \epsilon^{D-1} \wedge k \left(\mathcal{L}_l (k_I \delta l^I) \right) + \frac{2}{(D-1)!}  \int_{\Delta} \epsilon^{D-1} \wedge \delta \omega  \nonumber \\
	& & + \frac{2}{(D-2)!}   \int_{\Delta} k \wedge m^{K_2} \wedge ... \wedge m^{K_{D-1}} l^{K_1} \epsilon_{IJK_1 ... K_{D-1}} k^J d_{\Gamma^0} \delta l^{I} \nonumber \\
	&=& 0 \text{,} \label{eq:BoundaryContributionSymplecticPotential}
\ee
where in the second step, we used (\ref{eq:SigmaPullback}) and (\ref{eq:ConnectionPullback}), which results in three terms in the third step, each of which vanishes separately. The first one since we can partially integrate the Lie derivative (boundary terms drop since $\delta l^I = 0$ on $S_1$, $S_2$) and we have $\Dl{\mathcal{L}_l \epsilon}^{D-1} \;\widehat{=}\; 0$ and $\Dl{\mathcal{L}_l k} \;\widehat{=}\; 0$. Note that here, we defined
\be
	\epsilon^{D-1} = \epsilon_{IJK_1 ... K_{D-1}} l^I k^J m^{K_1} \wedge ... \wedge m^{K_{D-1}} \text{.} \label{eq:DefEpsilonD-1}
\ee
To see that it is Lie dragged along $l$, note that
\be
	\Dl{\mathcal{L}_l m}\m_{\mu I} &=&  l^{\nu} \D{\nabla}\m_{\nu} m_{\mu I} + m_{\nu I} \D{\nabla}\m_{\mu} l^{\nu} =  l^{\nu} \D{\nabla}\m_{\nu} m_{\mu I} = - l^{\nu} \Gamma^0\m_{\nu I}\m^J m_{\mu J} \text{,} \\
	\Dl{\mathcal{L}_l l}^I &=& l^{\nu} \D{\nabla}\m_{\nu} l^{I} = - l^{\nu} \Gamma^0\m_{\nu I} \m^J l_J \text{,} \\
	\Dl{\mathcal{L}_l k}^I &=& l^{\nu} \D{\nabla}\m_{\nu} k^{I} = - l^{\nu} \Gamma^0\m_{\nu I} \m^J k_J \text{.}	
\ee
Using this, to prove that $\Dl{\mathcal{L}_l \epsilon}^{D-1} = 0$ we only need to use the invariance of $\epsilon^{I_1...I_{D+1}}$ under (infinitesimal) SO$(1,D)$ transformations. A similar argument shows that
\be 
	\Dl{d \epsilon}^{D-1} = 0 \text{.} 
\ee
The second term in (\ref{eq:BoundaryContributionSymplecticPotential}) is zero since $\delta \omega$ is fixed on $S_1$, $S_2$ and also Lie dragged along $l$, so the whole integrand is Lie dragged an vanishes at the boundary, which implies that the integral vanishes (This argument is e.g. given in \cite{AshtekarIsolatedHorizonsHamiltonian}). The last term vanishes since the derivative $d_{\Gamma^0}$ annihilates the whole expression (note that $\D{dk} \; \widehat{=}\; 0$) and therefore leads only to a boundary contribution which vanishes again due to $\delta l|_{S_1, S_2} = 0$.\\
\\
The second variation of the action yields the symplectic current $\delta_{[1} \Sigma^{IJ} \delta_{2]} A_{IJ}$ which is closed by standard arguments,
\begin{align}
	(\int_{\Sigma_2} - \int_{\Sigma_1} + \int_{\Delta}) \delta_{[1} \Sigma^{IJ} \wedge \delta_{2]} A_{IJ} = 0 \text{.}
\end{align}
Moreover, the contribution at $\Delta$ is a pure surface term, and we will show in the following that
\begin{align}
	\int_{\Delta} \delta_{[1} \D{\Sigma}^{IJ} \wedge \delta_{2]} \D{A}\m_{IJ} = \Omega_{\text{CS}}^{S_2}(\delta_1, \delta_2) -  \Omega_{\text{CS}}^{S_1}(\delta_1, \delta_2) \text{,}
\end{align}
where 
\begin{align}
\Omega_{\text{CS}}^{S} =  \frac{n A_S}{\vev{E^{(2n)}}} \int_{S}  \epsilon^{IJKLM_2N_2 ... M_n N_n} \left(\delta_{[1} \DD{A}\m_{IJ}\right) \wedge \left(\delta_{2]} \DD{A}\m_{KL}\right) \wedge \DD{F}\m_{M_2N_2} \wedge ... \wedge \DD{F}\m_{M_nN_n}  
\end{align}
denotes the Chern-Simons symplectic structure (cf. appendix \ref{sec:ChernSimons}), and therefore, the symplectic structure is given by
\begin{alignat}{3}
\Omega(\delta_1, \delta_2) = &  & \int_{\Sigma} & \delta_{[1} \Sigma^{IJ} \wedge \delta_{2]} A_{IJ} \label{eq:Proof0} \\  \nonumber
 & + \frac{n A_S}{\vev{E^{(2n)}}} & \int_{S} & \epsilon^{IJKLM_2N_2 ... M_n N_n} \left(\delta_{[1} \DD{A}\m_{IJ}\right) \wedge \left(\delta_{2]} \DD{A}\m_{KL}\right) \wedge \DD{F}\m_{M_2N_2} \wedge ... \wedge \DD{F}\m_{M_nN_n}  \text{,}
\end{alignat}
and is independent of the choice of $\Sigma$.
\\
To proof (\ref{eq:Proof0}), we will first show that the contribution to the symplectic structure at $\Delta$ is given by the boundary term we already found in section \ref{sec:NewVariables},
\begin{align}
	\int_{\Delta} \delta_{[1} \D{\Sigma}^{IJ} \wedge \delta_{2]} \D{A}\m_{IJ} = \int_{S_2} 2 (\delta_{[1} \tilde{s}^I)( \delta_{2]} n_I) -  \int_{S_1} 2 (\delta_{[1} \tilde{s}^I)( \delta_{2]} n_I)  \label{eq:Proof1} \text{,}
\end{align}
where $\tilde{s}_I = \sqrt{h} s_I$, and in a second step that the boundary contribution can be rewritten as
 \begin{align}
	\int_{S} 2 (\delta_{[1} \tilde{s}^I)( \delta_{2]} n_I) 
	 &= \frac{A_S}{\vev{E^{(2n)}}} \int_{S}  2\frac{E^{(2n)}}{\sqrt{h}} (\delta_{[1} \tilde{s}^I)( \delta_{2]} n_I)  \nonumber \\
	&=  \frac{n A_S}{\vev{E^{(2n)}}} \int_{S}  \epsilon^{IJKLM_2N_2 ... M_n N_n} \left(\delta_{[1} \DD{A}\m_{IJ}\right) \wedge \left(\delta_{2]} \DD{A}\m_{KL}\right) \wedge \DD{F}\m_{M_2N_2} \wedge ... \wedge \DD{F}\m_{M_nN_n}  \text{.} \label{eq:Proof2}
\end{align}
\\
For the variation of $\D{\Sigma}$, we find using (\ref{eq:SigmaPullback})
\begin{align}
	-(D-1)! \; \delta \D{\Sigma}\m_{IJ} = \epsilon_{IJK_1 ...K_{D-1}} & \left[ (D-1) (\delta m^{K_1}) \wedge m^{K_2} \wedge ... \wedge m^{K_{D-1}} \right. \nonumber \\
	&\left. \; - (D-1)(D-2) l^{K_1} ~ k \wedge (\delta m^{K_2}) \wedge m^{K_3} \wedge ... \wedge m^{K_{D-1}} \right. \nonumber \\
	&\left. \; - (D-1) \left( l^{K_1} (\delta k) + (\delta l^{K_1})k \right) \wedge m^{K_2} \wedge ... \wedge m^{K_{D-1}} \right] \nonumber \\
	= \epsilon_{IJK_1 ...K_{D-1}} & \left[ (D-1) m_L \wedge m^{K_2} \wedge ... \wedge m^{K_{D-1}} (i_{m^L} \delta m^{K_1}) \right. \nonumber \\
	&\left. \; - (D-1)(D-2) l^{K_1} ~ k \wedge m_{L} \wedge m^{K_3} \wedge ... \wedge m^{K_{D-1}} (i_{m^L} \delta m^{K_2}) \right. \nonumber \\
	&\left. \; - (D-1) \left( - l^{K_1} (i_l \delta k) + (\delta l^{K_1}) \right) k \wedge m^{K_2} \wedge ... \wedge m^{K_{D-1}} \right] \label{eq:DeltaSigma}
	\text{,}
\end{align}
where we used
\begin{align}
	\D{\delta m}\m_I &= \D{m}\m_J (i_{m^J} \delta m_I) - \D{k} (i_l \delta m_I) = \D{m}\m_J (i_{m^J} \delta m_I) + \D{k} (i_{m_I} \delta l) \nonumber \\ 
	&=  \D{m}\m_J (i_{m^J} \delta m_I) +  \D{k} c_{\delta} (i_{m_I} l) = \D{m}\m_J (i_{m^J} \delta m_I) \text{,} \\
	\D{\delta k} &= - k (i_l \delta k) \text{.}
\end{align}
In total, after a long calculation explained in appendix \ref{app:StructureOnDelta}, one finds for (\ref{eq:Proof1})
\begin{align} 
	\int_{\Delta} \delta_{[1} \D{\Sigma}^{IJ} \wedge \delta_{2]} \D{A}\m_{IJ} & \nonumber \\
	= \frac{ 2}{(D-1)!} \int_{\Delta} & \left\{ d\left[\delta_{[1}(\epsilon^{D-1} k_I) \delta_{2]} l^I\right] + \delta_{[1} \epsilon^{D-1} \wedge \delta_{2]}\omega^l \right. \nonumber \\
	&\; \left. + (D-1) d\left[\left( c_{\delta} + (k_M \delta_{[1} l^M) \right) k \wedge m^{K_{2}} \wedge ... \wedge m^{K_{D-1}} l^I k^J \epsilon_{IJK_1 ... K_{D-1}} \delta_{2]} l^{K_1}\right] \right. \nonumber \\
	& \; \left. + (D-2) d\left[k\wedge m^M\wedge m^{K_3} \wedge ... \wedge m^{K_{D-1}} l^I k^J \epsilon_{IJK_1...K_{D-1}} (i_{m_M} \delta_{[1}m^{K_2}) \delta_{2]}l^{K_1} \right]\right\} \nonumber \\
	= \frac{ 2}{(D-1)!} \int_{\Delta} & \left\{ d\left[\delta_{[1}(\epsilon^{D-1} k_I) \delta_{2]} l^I\right] + \delta_{[1} \epsilon^{D-1} \wedge \delta_{2]}\omega^l \right\} \nonumber \\
	=  2 \int_{\Delta} & \left\{ d\left[ \delta_{[1} \tilde{s}^I \delta_{2]} n_I \right] + \frac{1}{(D-1)!} \delta_{[1} \epsilon^{D-1} \wedge \delta_{2]}\omega^l \right\} \text{.} \label{eq:DeltaSigmaDeltaA}
\end{align}
We used $\delta k = - c_{\delta} k$ and $\DD{k}=0$. Since we also restricted to constant area $A_{S}$ throughout the phase space region we are considering ($\delta A_S = 0$), we furthermore find
\begin{align}
	\int_{\Delta} \delta_{[1} \epsilon^{D-1} \wedge \delta_{2]}\omega^l &= - \int_{\Delta} \delta_{[1} \epsilon^{D-1} \wedge \delta_{2]} (\kappa^l k) =  \int_{\Delta} \delta_{[1} \epsilon^{D-1} \wedge d \delta_{2]} (\kappa^l v) \nonumber \\
	&=  \left[\delta_{[2} (\kappa^l v)|_{S_2} \int_{S_2} \delta_{1]} \epsilon^{D-1} - \delta_{[2} (\kappa^l v)|_{S_1} \int_{S_1} \delta_{1]} \epsilon^{D-1} \right] \nonumber \\
	&=  \left[\delta_{[2} (\kappa^l v)|_{S_2} \delta_{1]}A_{S_2}- \delta_{[2} (\kappa^l v)|_{S_1} \delta_{1]} A_{S_1} \right]  = 0 \text{.}
\end{align}
Now, since we have $E^{(D-1)} = f(v) \epsilon^{D-1} / (D-1)!$ for a spherically symmetric function $f$ by the conditions for an UDNRIH, and since
\begin{align}
	\int_{S} E^{(2n)} &= (8\pi)^{n} n! \; \chi_S  = : \langle E^{(2n)} \rangle \text{,}\\
	\int_{S} \epsilon^{D-1} &= (D-1)! \; A_S  \text{,}
\end{align}
are both constant in time, we have $f= \frac{ \langle E^{(2n)} \rangle}{A_S}$ where $2n = D-1$. The first line of (\ref{eq:Proof2}) easily follows. In fact, this also shows that $f(v)$ is independent of $v$.
\\
For the second pullback of $A$, we find since $\DD{\omega} = 0$,
\be
	\DD{A}\m_{IJ} = \Gamma^0\m_{IJ} + \frac{2}{D-1}  l_{[I} m_{J]} \theta_k =: \Gamma^0\m_{IJ} + \DD{K}\m_{IJ} \text{.}
\ee
Since $\theta_k$ is constant on the $(D-1)$ - sphere cross sections of the chosen foliation, we have $\DDl{d_{\Gamma^0} K} = 0$. Since also $[\DD{K}, \DD{K}] = 0$, we obtain $\DD{F} = \DD{R}\m^0$ which was already derived in section \ref{sec:BoundaryCondition}. We now want to show that (\ref{eq:Proof2}) holds,
which is shown to be true in (\ref{eq:a3c1}) if the connection would be given by $\Gamma^0$. Therefore, what needs to be checked is if 
\begin{align}
	\epsilon^{IJKLM_2N_2 ... M_n N_n} \left(2 \delta_{[1} \Gamma^0_{IJ} \wedge \delta_{2]} \DD{K}\m_{KL} + \delta_{[1} \DD{K}\m_{ IJ} \wedge \delta_{2]} \DD{K}\m_{KL}\right) \wedge R^0\m_{M_2N_2} \wedge ...  \wedge R^0\m_{M_nN_n}  = 0\text{.} \label{eq:GammaKKK}
\end{align}
Using
\begin{align}
	\delta \DD{K}\m_{IJ} = \frac{2}{D-1} \left[ - l_{[I} k_{J]} l^K  \theta _k \delta m_{K} + l_{[I} \bar{\bar{\eta}}_{J]K} \left( \theta_k \delta m^K - m^K \theta_k k^L \delta l_L + m^K \delta \theta_k \right) + \bar{\bar{\eta}}_{[I}^K m_{J]} \theta_k \delta l_K \right] \label{eq:DeltaK} \text{,}
\end{align}
we find in a first step
\begin{align}
E_{\perp}^{IJKL} \wedge \delta_{[1} \DD{K}\m_{ IJ} \wedge \delta_{2]} \DD{K}\m_{KL} &= - \frac{8}{(D-1)^2} E_{\perp}^{IJKL} \wedge l_I k_J \bar{\bar{\eta}}_{K}^{N} \theta_k^2 l^M \delta_{[1} m_{M} \wedge m_{L} \delta_{2]} l_{N} \nonumber \\
&= \frac{8}{(D-1)^2} E_{\perp}^{IJ[N|L} \wedge l_I k_J \theta_k^2 m^{M]} \wedge m_{L}\delta_{[1} l_M  \delta_{2]} l_{N} = 0\text{.}
\end{align}
$E_{\perp}^{IJKL} = \epsilon^{IJKLM_2N_2 ... M_n N_n} R^0\m_{M_2N_2} \wedge ...  \wedge R^0\m_{M_nN_n} $ in the above formula stands for the terms in (\ref{eq:GammaKKK}) contracted with $\delta K \wedge \delta K$. $\perp$ indicates that fact that $E_{\perp}$ needs to be contracted with $k^I,l^J$ since it vanishes otherwise, therefore only one combination of terms survives when we use (\ref{eq:DeltaK}) in the first step. In the second line, we made use of $l^I \delta m_I = - m_I \delta l^I$ and therefore, the expression is antisymmetric in the index pair $M,N$. Adding terms until all indices of the epsilon symbol in $E_{\perp}$ plus the index $M$ are totally antisymmetric and subtracting the therefore needed terms again, we find that the whole expression vanishes: The total antisymmetrisation since there is no nontrivial rank $D+2$ antisymmetric tensor in $D+1$ dimensions, and the subtracted terms since they are either of the form $l^I m_I=0$ or $k^I m_I = 0$, or $R^0\m_{MN} \wedge m^N$ which vanishes due to the Bianchi identity, or $m^L \wedge m_L = 0$. \\
Furthermore, we have
\begin{align}
E_{\perp}^{IJKL} \wedge \delta_{[1} \Gamma^0\m_{ IJ} & \wedge \delta_{2]} \DD{K}\m_{KL}  \nonumber \\
= \frac{2}{(D-1)} E_{\perp}^{IJKL} \wedge & \left[\;\;- \; \bar{\bar{\eta}}_{[I}^{I'} \bar{\bar{\eta}}_{J]}^{J'} \delta_{[1} \Gamma^{0}\m_{I'J'} \wedge l_{[K} k_{L]} \theta_k l^M \delta_{2]} m_{M} \right. \nonumber \\
  & \quad - 2 k_{[I} l^{I'} \bar{\bar{\eta}}_{J]}^{J'} \delta_{[1} \Gamma^0\m_{I'J'} \wedge  l_{[K} \bar{\bar{\eta}}_{L]M} \left( \theta_k \delta_{2]} m^M - m^M \theta_k k^N \delta_{2]} l_N + m^M \delta_{2]} \theta_k \right) \nonumber \\
  & \quad+ \left. 2 l_{[I} k_{J]} k^{I'} l^{J'} \delta_{[1} \Gamma^0\m_{I'J'} \wedge \bar{\bar{\eta}}_{[K}^M m_{L]} \theta_k \delta_{2]} l_M \right] \nonumber \\
  = \frac{2}{(D-1)} E_{\perp}^{IJKL} \wedge & \left[\;\;- \; \bar{\bar{\eta}}_{[I}^{I'} m_{\alpha |J]} \left(- d_{\Gamma^0} \delta_{[1} m^{\alpha}_{I'} - m^{\beta}_{I'} \delta_{[1} \Gamma_{{\text{\tiny{\textbullet}}} \beta}^{\alpha} \right) \wedge l_{[K} k_{L]} \theta_k l^M \delta_{2]} m_{M} \right. \nonumber \\
  & \quad - 2 k_{[I} \bar{\bar{\eta}}_{J]}^{J'} (d_{\Gamma^0} \delta_{[1} l_{J'}) \wedge  l_{[K} \bar{\bar{\eta}}_{L]M} \left( \theta_k \delta_{2]} m^M - m^M \theta_k k^N \delta_{2]} l_N + m^M \delta_{2]} \theta_k \right) \nonumber \\
  & \quad- \left. 2 l_{[I} k_{J]} [d_{\Gamma^0}(k^{I'}  \delta_{[1} l_{I'} )] \wedge \bar{\bar{\eta}}_{[K}^M m_{L]} \theta_k \delta_{2]} l_M \right] \text{,} \label{eq:DeltaGammaDeltaK}
\end{align}
where we used $\bar{\bar{\eta}}_{IJ} = m^{\alpha}\m_I m_{\alpha J}$ in the last step as well as the fact that $\Gamma^0$ annihilates $m^K, l^I, k^J$ and therefore, e.g. $l^J \delta \Gamma^0\m_{IJ} = \delta (d_{\Gamma^0} l_I) - d_{\Gamma^0} \delta l_I = - d_{\Gamma^0} \delta l_I$. In the last expression, the second summand in the second to last line and the term in the last line together just give a surface term which vanishes since the $(D-1)$ sphere cross sections have no boundary. To see this, one needs to make use of the fact that $d_{\Gamma^0} R^0 = d_{\Gamma^0} m = d_{\Gamma^0} l^I = d_{\Gamma^0}k^J = d_{\Gamma^0} \bar{\bar{\eta}} =  d \theta_K = 0$. Moreover, we also have  $d \delta \theta_K = 0$ since $\delta \theta_K$ has to be constant on the $(D-1)$ - sphere cross sections, and therefore also the last term in the second to last line is a surface term. Using the notation $\delta \Gamma_{{\text{\tiny{\textbullet}}} \beta}^{\alpha}$ to indicate that $\delta \Gamma$ is considered as a form in the index ${\text{\tiny{\textbullet}}}$, the terms in the first line of (\ref{eq:DeltaGammaDeltaK}) give
\begin{align}
 \frac{2\theta_k}{(D-1)} l^K k^L E^{\perp}_{IJKL} \wedge m_{N} \, m_{\alpha}\m^{[J} \wedge & \left[  \left( d_{\Gamma^0} \delta_{[1} m^{\alpha |I]} + m^{\beta |I]} \delta_{[1} \Gamma_{{\text{\tiny{\textbullet}}} \beta}^{\alpha} \right) \right] \delta_{2]} l^N \nonumber \\
= \frac{2\theta_k}{(D-1)} l^K k^L E^{\perp}_{IJKL} \wedge m_M \wedge m_N & \left[ m^{\beta M} m^{\alpha [I} D^{\Gamma^0}_{\beta} \delta_{[1} m_{\alpha}\m^{|J]}  \right .\nonumber \\
& \left. \; +  m^{\beta [J} m^{\alpha |I]} D^{\Gamma^0}_{\beta} \delta_{[1} m_{\alpha}\m^{M} -  m^{\beta [I|} m^{\alpha M} D^{\Gamma^0}_{\beta} \delta_{[1} m_{\alpha}\m^{|J]} \right] \delta_{2]}l^N \nonumber \\
= \frac{2\theta_k}{(D-1)} l^K k^L E^{\perp}_{IJKL} \wedge m_M \wedge m_N & \left[ \frac{1}{3} m^{\beta [M} m^{\alpha I} D^{\Gamma^0}_{\beta} \delta_{[1} m_{\alpha}\m^{J]} + 2 m^{\beta [J} m^{\alpha |I]} D^{\Gamma^0}_{\beta} \delta_{[1} m_{\alpha}\m^{M} \right] \delta_{2]}l^N \nonumber \\
= \frac{4\theta_k}{(D-1)} l^K k^L E^{\perp}_{IJKL} \wedge m_M \wedge m_N & \left[ - m^{\beta [I} m^{\alpha |J]} D^{\Gamma^0}_{\beta} \delta_{[1} m_{\alpha}\m^{M} \right] \delta_{2]}l^N \text{.} \label{eq:klij}
\end{align}
In the third step, the term totally antisymmetric in the indices $M, J, I$ vanishes since
\begin{align}
	l^K k^L E^{\perp}_{[IJ|KL} \wedge m_{|M]} \wedge m_N & =  \epsilon_{[IJ|KLM_2N_2 ... M_n N_n} l^K k^L R_0\m^{M_2N_2} \wedge ...  \wedge R_0\m^{M_nN_n} \wedge m_{|M]} \wedge m_N \nonumber \\
		&= \frac{(D+2)}{3}\epsilon_{[IJKLM_2N_2 ... M_n N_n|} l^K k^L R_0\m^{M_2N_2} \wedge ...  \wedge R_0\m^{M_nN_n} \wedge m_{|M]} \wedge m_N \nonumber \\
		&=0 \text{,}
\end{align}
since $R_0^{KL} \wedge m_L = 0$ due to the Bianchi identity and $m_I l^I = 0 = m_I k^I$, and the antisymmetrisation of $D+2$ indices vanishes. Finally, the first term in the second to last line of (\ref{eq:DeltaGammaDeltaK}) gives
\begin{align}
 & \frac{4 \theta_k}{(D-1)} l^K k^L E^{\perp}_{IJKL} \wedge \left[  (d_{\Gamma^0} \delta_{[1} l^{[J}) \wedge \delta_{2]} m^{I]}  \right] \nonumber \\
 = & \frac{4 \theta_k}{(D-1)} l^K k^L E^{\perp}_{IJKL} \wedge ((d_{\Gamma^0} \delta_{[1} m^{[I}) \delta_{2]} l^{J]}) + d(\ldots) \nonumber \\
 = & \frac{4 \theta_k}{(D-1)} l^K k^L E^{\perp}_{IJKL} \wedge m_M \wedge m_N \left[ m^{\beta [M} m^{\alpha |N]} D^{\Gamma^0}_{\beta} \delta_{[1} m_{\alpha}\m^{[I} \right]\delta_{2]} l^{J]} + d(\ldots) \text{,} \label{eq:ijkl}
\end{align}
up to a boundary term $d(\ldots)$ that vanishes, as above, after integration over $S$, which means that (\ref{eq:klij}) and (\ref{eq:ijkl}) together are of the form
\begin{align}
& l^K k^L E^{\perp}_{IJKL} \wedge m_M \wedge m_N \left[ \alpha^{IJ} \beta^{MN} - \alpha^{MN} \beta^{IJ} \right] \nonumber \\
= & l^K k^L \epsilon_{IJKLM_2N_2 ... M_n N_n} R_0\m^{M_2N_2} \wedge ...  \wedge R_0\m^{M_nN_n}  \wedge m_{M} \wedge m_N \left[ \alpha^{IJ} \beta^{MN} - \alpha^{MN} \beta^{IJ} \right] \nonumber \\
= & (D+2) l^K k^L \epsilon_{[IJKLM_2N_2 ... M_n N_n|} R_0\m^{M_2N_2} \wedge ...  \wedge R_0\m^{M_nN_n}  \wedge m_{|M]} \wedge m_N \left[ \alpha^{IJ} \beta^{MN} - \alpha^{MN} \beta^{IJ} \right] \nonumber \\
 & - 2 l^K k^L \epsilon_{JMKLM_2N_2 ... M_n N_n} R_0\m^{M_2N_2} \wedge ...  \wedge R_0\m^{M_nN_n}  \wedge m_{I} \wedge m_N \left[ \alpha^{IJ} \beta^{MN} - \alpha^{MN} \beta^{IJ} \right] \nonumber \\ 
 = & - 2 l^K k^L \epsilon_{JMKLM_2N_2 ... M_n N_n} R_0\m^{M_2N_2} \wedge ...  \wedge R_0\m^{M_nN_n}  \wedge m_{I} \wedge m_N \left[ \alpha^{NM} \beta^{JI} - \alpha^{MN} \beta^{IJ} \right] \nonumber \\
 = &0 \text{,}
\end{align}
where $\alpha^{IJ}$ and $\beta^{KL}$ are antisymmetric matrices. This furnishes the proof of (\ref{eq:Proof0}).

\section{SO$(D+1)$ as Internal Gauge Group}
\label{sec:SO(D+1)}

In the previous sections, we have derived the isolated horizon boundary condition relating the connection on the horizon with the bulk degrees of freedom, as well as the symplectic structure on the horizon, which coincides with the one of higher-dimensional Chern-Simons theory. Since we started from the space-time covariant Palatini action, the internal gauge group was fixed to SO$(1,D)$. In the light of quantising the bulk degrees of freedom however, it was pointed out in \cite{BTTI} that one can change the internal gauge group to SO$(D+1)$ by a canonical transformation from the ADM phase space. After this reformulation, the quantisation of the bulk degrees of freedom can be performed with standard loop quantum gravity methods as spelled out in \cite{BTTIII}. Thus, we are interested in reformulating the horizon boundary condition and the horizon symplectic structure so that it fits in the SO$(D+1)$ scheme. 

As for the boundary condition, the generalisation to the Euclidean internal group is straight forward, since the construction of the connection $\Gamma^0$ in appendix \ref{app:Connections} works independently of the internal signature. Thus, constructing $\Gamma^0$ such that it annihilates both $n^K$ and $s^K = s_a e^{aK}$ additionally to $m_\alpha^K = \D{e}\m_{\alpha}^K$, the horizon boundary conditions 
\begin{align}
	R^{0, \text{horizon}}_{\alpha \beta IJ} &= R^{0, \text{bulk}}_{\alpha \beta IJ}  \label{eq:BoundaryConditionRiemannCompactGroup} \\
	\frac{1}{\beta} \epsilon^{K_1L_1 ... K_{n} L_{n} IJ} {\epsilon}\m^{^{\alpha_1 \beta_1... \alpha_{n} \beta_{n}}} {R}^{0,\text{horizon}}_{\alpha_1 \beta_1 K_1 L_1} ... {R}^{0,\text{horizon}}_{\alpha_{n} \beta_{n} K_{n} L_{n}}  &= \frac{E^{(2n)}}{\sqrt{h}} \stackrel{(\beta)}{\pi}^{aIJ} \hat{s}_a~~  \label{eq:BoundaryConditionEulerCompactGroup}
\end{align}
follow immediately from the fact that ${R}^0\m_{\alpha \beta KL} n^K = {R}^0\m_{\alpha \beta KL} s^K = 0$. We will drop the superscripts ``bulk" and ``horizon" in what follows.

In order to derive the new symplectic structure, we first perform a symplectic reduction of the theory derived in the previous chapters by solving the Gau{\ss} and simplicity constraint. This leads us to the ADM phase space, from which we can perform further canonical transformations. 
It is shown in section \ref{sec:NewVariables} that the canonical transformation to SO$(D+1)$ connection variables leads to the boundary symplectic structure
\be
	 \Omega^S(\delta_1, \delta_2) = \frac{2}{\beta}\int_S d^{D-1}x ~  \delta_{[1} \tilde{s}_I  \delta_{2]} n^{I} ~~\text{.}
	 \label{eq:SymplecticStructureNS}
\ee
Furthermore, under the non-distortion condition $\delta \frac{E^{(2n)}}{\sqrt{h}} = 0$, i.e. restricting to the part of phase space where $\frac{E^{(2n)}}{\sqrt{h}}$ is constant and thus given by $\frac{\vev{E^{(2n)}}}{A_S}$, it is shown in appendix \ref{app:ChernSimonsSymplecticStructure} that 
\be
	2\frac{E^{(2n)}}{\sqrt{h}} (\delta_{[1} \tilde{s}^I)( \delta_{2]} n_I) &=&  n \epsilon^{IJKLM_1N_1 ... M_{n-1} N_{n-1}} \epsilon^{\alpha \beta \alpha_1\beta _1...\alpha_{n-1}\beta _{n-1}} \nonumber \\
	& & \times \left(\delta_{[1} \Gamma^0_{\alpha IJ}\right) \left(\delta_{2]} \Gamma^0_{\beta  KL}\right) R^0_{\alpha_1\beta _1M_1N_1} ... R^0_{\alpha_{n-1}\beta _{n-1}M_{n-1}N_{n-1}}  \text{,} 
\ee
which results in the Chern-Simons type boundary symplectic structure
\be
	 \Omega_{\text{CS}}^S(\delta_1, \delta_2) &=&  \frac{n A_S}{\beta \vev{E^{(2n)}}} \int_S \epsilon^{IJKLM_1N_1 ... M_{n-1} N_{n-1}}\epsilon^{\alpha \beta \alpha_1\beta _1...\alpha_{n-1}\beta _{n-1}} \nonumber \\ 
	 & & \times \left(\delta_{[1} \Gamma^0_{\alpha IJ}\right) \left(\delta_{2]} \Gamma^0_{\beta  KL}\right) R^0_{\alpha_1\beta _1M_1N_1} ... R^0_{\alpha_{n-1}\beta _{n-1}M_{n-1}N_{n-1}}  ~~\text{.}
\ee

Concluding, we have shown that also for the case of SO$(D+1)$ as an internal gauge group, one arrives at a higher-dimensional Chern-Simons symplectic structure at the isolated horizon boundary of $\sigma$ as well as the boundary conditions \eqref{eq:BoundaryConditionRiemannCompactGroup}, \eqref{eq:BoundaryConditionEulerCompactGroup}. 

A remark concerning the uniqueness of $\Gamma^0$ is in order. In $D=3$, one easily finds that there are more connections which allow for carrying out the whole programme. Exemplarily, we can introduce a constant parameter $\Phi \in \mathbb{R}$ and choose $\Gamma^{\phi}\m_{\alpha}\m^{IJ} = \Gamma^0\m_{\alpha}\m^{IJ} + 2 \Phi n^{[I} m_{\alpha}\m^{J]}$ as connections for the Chern-Simons theory on the boundary. We then find
\begin{align}
	R^{\Phi}\m_{\alpha\beta}\m^{IJ} &= R^0\m_{\alpha\beta}\m^{IJ} -2 \Phi^2 m_{\alpha}\m^{[I} m_\beta\m^{J]} \text{,} \label{eq:RPhi1}
	\\
	\epsilon^{IJKL} \epsilon^{\alpha \beta} R^{\Phi}\m_{\alpha\beta KL} &= \left(\frac{E^{(2)}}{\sqrt{h}} - 4 \Phi^2\right) \pi^{aIJ} \hat{s}_a \text{,} \label{eq:RPhi2}\\
	\frac{A_S}{\langle E^{(2)} \rangle - 4 \Phi^2 A_S} \epsilon^{IJKL} \epsilon^{\alpha \beta} \delta_{[1} \Gamma^{\Phi}\m_{\alpha IJ} \delta_{2]} \Gamma^{\Phi}\m_{\beta KL} &= 2 \delta_{[1} \tilde{s}^I \delta_{2]} n_I \text{.}
\end{align}
A further modification of $\Gamma^0$, which in particular allows for generalisation to distorted horizons, will be introduced in section \ref{sec:BEM}, where a non-constant field $\Psi$ is added to the connection. The introduction of $\Psi$ and $\Phi$ cannot be combined non-trivially, since otherwise there will be terms $\propto n^{[I} m_{\alpha}\m^{J]}$ contributing to $R^{\Phi, \Psi}_{\alpha \beta IJ}$. 

A third possibility to change the connection in $D=3$, which can be combined with both of the above methods, is as follows. As we have already seen at the end of sec. \ref{sec:NewVariables}, if we introduce the Barbero-Immirzi parameter $\gamma$ in $D=3$ \cite{BTTII}, it will appear in the boundary symplectic structure. The boundary condition in this case reads
\be
	\frac{1}{\beta} \epsilon^{\alpha \beta} \left(\epsilon^{IJKL} R^0\m_{\alpha \beta KL} + \frac{1}{\gamma} R^0\m_{\alpha \beta}\m^{IJ}\right)= \frac{E^{(2)}}{\sqrt{h}}\stackrel{(\gamma, \beta)}{\pi}\m^{aIJ} \hat{s}_a \text{,}
\ee
where
\be
	\stackrel{(\gamma, \beta)}{\pi}\m^{aIJ} = \frac{1}{\beta} \left( \pi^{aIJ} + \frac{1}{2\gamma} \epsilon^{IJ}\m_{KL} \pi^{aKL} \right) \text{.}
\ee
To show that the boundary symplectic structure can be rewritten according to
\begin{align}
		\frac{2}{\beta}\int_S d^{2}x \left( \delta_{[1} \tilde{s}^{I} \delta_{2]} n_I - \frac{1}{2 \gamma} \epsilon^{\alpha \beta} \delta_{[1} m_{\alpha I} \delta_{2]} m_{\beta}^I\right) =  \frac{A_S}{\beta \vev{E^{(2)}}} \int_S \epsilon^{\alpha \beta}  &\left(\epsilon^{IJKL} \delta_{[1} \Gamma^0_{\alpha IJ} \delta_{2]} \Gamma^0_{\beta  KL} \right. \nonumber \\
		&\left. \quad + \frac{2}{\gamma} \delta_{[1} \Gamma^0\m_{\alpha IJ} \delta_{2]} \Gamma^0\m_{\beta}\m^{IJ} \right) \text{,}
\end{align}
it remains to verify that
\be
	\frac{E^{(2)}}{\sqrt{h}}  \delta m^I \wedge \delta m_I = -2 \delta \Gamma^{0\, IJ} \wedge \delta \Gamma^0\m_{IJ} \label{eq:ImmirziDeltaDelta} \text{.}
\ee
Since the scalar curvature $R = \frac{E^{(2)}}{2 \sqrt{h}}$ is constant on the  2-spheres, the metric $h$ is fixed up to diffeomorphism. Therefore, $m_{I}, \Gamma^0\m_{IJ}$ are fixed up to diffeomorphism and SO$(D+1)$ rotations, i.e. $\delta m_I = \Lambda_{I}\m^J \delta m_J + \mathcal{L}_{\xi} m_I$ and $\delta \Gamma^0\m_{IJ} = -d_{\Gamma^0} \Lambda_{IJ} + \mathcal{L}_{\xi} \Gamma^0\m_{IJ}$. Using this for the variations, (\ref{eq:ImmirziDeltaDelta}) can be proven straight forwardly using $0 = d_{\Gamma^0}m_I = dm_I + \Gamma^0\m_{IJ} \wedge m^J$, $d \Gamma^0\m_{IJ}  + \frac{1}{2} [\Gamma^0,\Gamma^0]_{IJ} = R^0\m_{IJ}= \frac{1}{2} R \, m_I \wedge m_J$ and the properties of the exterior and Lie derivative.

In higher dimensions, it is less trivial to modify the connection $\Gamma^0$. In particular, the above constructions can at least not be applied trivially. While (\ref{eq:RPhi1}) continues to hold, in (\ref{eq:RPhi2}) mixed terms of the form $R^0 \wedge ... \wedge (\Phi m \wedge m)$ will appear which spoil the construction, and also the introduction of $\gamma$ is tied to $D=3$.

\section{Inclusion of Distortion}
\label{sec:NonSymmetric}

In this section, we are going to comment on the generalisation of the isolated horizon boundary condition derived in the non-distorted case to general isolated horizons. The seminal work on this subject has been a paper by Ashtekar, Engle, and Van Den Broeck \cite{AsthekarQuantumHorizonsAnd}, where treatment was generalised to axi-symmetric horizons. For the generalisation to arbitrary spherical horizons, two methods by Perez and Pranzetti \cite{PerezStaticIsolatedHorizons} and Beetle and Engle \cite{BeetleGenericIsolatedHorizons} exist in four dimensions. We will discuss them in the following and show that an extension of them to higher dimensions is not straight forward. Nevertheless, an extension to distorted isolated horizons seems to exist \cite{BNII}.

\subsection{Beetle-Engle Method}
\label{sec:BEM}

In order to derive the symplectic structure on a spatial slice $S$ of the horizon, it is key to the derivation that $E^{(2n)} / \sqrt{h}$ is a constant on $S$. Otherwise, unwanted terms appear due to the variation of $E^{(2n)} / \sqrt{h}$. Of course, this observation has already been made in the four-dimensional case and a solution of this problem in case of U$(1)$ as the gauge group on $S$ has been proposed by Beetle and Engle \cite{BeetleGenericIsolatedHorizons}. Essentially, they construct a new U$(1)$ connection on $S$ as 
\be
	    \stackrel{\circ}{V}_{\alpha} := \frac{1}{2} \theta_{\alpha} - \epsilon_{\alpha \beta} h^{\beta \gamma} D_{\gamma} \Psi \text{,}
\ee
where $\frac{1}{2} \theta_{\alpha}$ is the U$(1)$ connection used for spherically symmetric isolated horizons and $\Psi$ is a curvature potential defined by the equation
\be
	\Delta \Psi = R - \vev{R} \text{,}
\ee
where $R$ is the intrinsic scalar curvature which is proportional to $E^{(2)} / \sqrt{h}$. Calculating the curvature of $\stackrel{\circ}{V}_{\alpha}$, the terms proportional to $R$ drop out and one gets
\be
	d \stackrel{\circ}{V} = - \frac{\vev{R}}{4} \epsilon = - \frac{2 \pi}{A_S} \Sigma_i s^i \text{.}
\ee
Thus, $\stackrel{\circ}{V}_{\alpha}$ mimics the spin connection of a spherically symmetric horizon, although being defined for {\it any} horizon of spherical topology. 

The method of Beetle and Engle can be generalised to this framework for the case of four dimensions by using the connection
\be
	A_{\alpha IJ} = \Gamma^0_{\alpha IJ} + 2 m_{\alpha [I} m_{\beta |J]} h^{\beta \gamma} (D_{\gamma} \psi) \text{.}
	\label{eq:BeetleEngleConnectionSO(4)}
\ee
Insertion into the boundary condition 
\be
	\epsilon^{\alpha \beta} \epsilon^{IJKL} R_{\alpha \beta KL}(A) = 2 \langle E^{(2)} \rangle n^{[I} \tilde{s}^{J]} 
\ee
yields
\be
\Delta \psi = \frac{1}{4}\left(\frac{E^{(2)}}{\sqrt{h}} - \langle E^{(2)}\rangle\right) \text{.}
\ee
As shown in appendix \ref{app:BeetleEngleConnection}, it follows that 
\begin{align}
	2\langle E^{(2)}\rangle (\delta_{[1} \tilde{s}^I) (\delta_{2]} n_I ) =   \epsilon^{IJKL} \epsilon^{\alpha \beta} \left(\delta_{[1} A_{\alpha IJ}\right) \left(\delta_{2]} A_{\beta KL}\right) \text{.} 
\end{align}
The problem with generalising this method to higher dimensions is that it leads to a non-linear partial differential equation for $\psi$, for which, as opposed to the Laplace operator $\Delta$, a well developed theory ensuring the existence of a solution does not exist. Thus, although a generalisation to higher dimensions seems straight forward, we cannot proceed due to the resulting non-linear partial differential equation.

\subsection{Perez-Pranzetti Method}

The basic idea of Perez and Pranzetti \cite{PerezStaticIsolatedHorizons} in order to solve the problem of a varying scalar curvature on $S$ is to use two Chern-Simons connections on $S$, defined by
\be
	A^i_\gamma = \Gamma^i + \gamma e^i, ~~~~ A^i_\sigma = \Gamma^i + \sigma e^i \text{.}
\ee
For the boundary conditions, it follows that
\be
	F^i(A_\gamma) = \Psi_2 \Sigma^i + \frac{1}{2}(\gamma^2+c) \Sigma^i, ~~~~ F^i(A_\sigma) = \Psi_2 \Sigma^i + \frac{1}{2}(\sigma^2+c) \Sigma^i \text{,}
\ee
where the Newman-Penrose coefficient $\Psi_2$ is proportional to the scalar curvature and $c$ is an extrinsic curvature scalar. Subtracting these two equations, Perez and Pranzetti find
\be
	F^i(A_\gamma) - F^i(A_\sigma) = \frac{1}{2}(\gamma^2-\sigma^2) \Sigma^i \text{,}
\ee
which can be used to derive the symplectic structure of two SU$(2)$ Chern-Simons connections on $S$, since the scalar curvature disappeared from this new boundary condition. Furthermore, they take the additional constraint into account which follows from adding the above two field strengths, which requires to first find a suitable quantisation of the scalar curvature.  

The first steps of this treatment generalise to higher dimensions in a straight forward way: Introduce $N$ Chern-Simons connections of the form
\be
	A_{\alpha IJ}^{(a_i)} = \Gamma_{\alpha IJ} + 2 \sqrt{a_i} s_{[I} m_{\alpha|J]}, ~ i \in \{1,...,N\} \text{.}
\ee
For their field strengths, it follows that
\be
	F_{\alpha \beta IJ}^{(a_i)} = R_{\alpha \beta IJ} - 2 m_{\alpha[I} m_{\beta |J]} a_i \text{.} \label{eq:FieldStrenghtsPP}
\ee
When we insert this in the formula needed for the higher-dimensional boundary condition, we find
\begin{align}
	E_{(a_i)}^{IJ}(A^{(a_i)}) &:=  \epsilon^{\beta_1 \gamma_1 \ldots \beta_n \gamma_n} \epsilon^{IJ K_1 L_1 \ldots K_n L_n} F^{(a_i)}_{\beta_1 \gamma_1 K_1 L_1} \ldots F^{(a_i)}_{\beta_n \gamma_n K_n L_n} \nonumber \\
	&= \sum_{k=0}^n a_i^k X_k \text{,}
\end{align}
where, schematically, $X_k \propto (R^0)^{n-k} \wedge (m\wedge m)^k$. Only the $k=0$ term, being exactly of the form ``$n^{[I} \tilde{s}^{J]} \times \text{const.}$" we need, is allowed to survive when linear combining the $E_{(a_i)}^{IJ}$ with coefficients $b_i \in \mathbb{R}$, $i \in \{1,...,N\}$,
\begin{align}
	\sum_{i=1}^N b_i ~ E_{(a_i)}^{IJ}(A^{(a_i)}) \stackrel{!}{\propto} n^{[I} \tilde{s}^{J]}\text{,}
\end{align}
which leads to the system of equations
\begin{align}
	\sum_{i=1}^N b_i (a_i)^k &= 0\text{,} \hspace{5mm} k \in \{0,...,n-1\} \text{,} \nonumber \\
	\sum_{i=1}^N b_i (a_i)^n & = d \text{,}
\end{align}
for some constant $d \neq 0$. Suppose w.l.o.g. that $a_1 \neq 0$, $b_1\neq 0$. Introducing a new $\tilde{d} = \frac{d}{b_1 (a_1)^n}$, we find that the above $n+1$ equations for fixed $\tilde{d}$, actually only depend on the $2(N-1)$ unknowns $(a_i/a_1)$, $(b_i/b_1)$. Since $N$ is integer and $2n = D-1$, we find that we need at least $N= \lceil\frac{n+1}{2} \rceil + 1= \lceil\frac{D+1}{4} \rceil + 1$ Chern-Simons theories on the boundary, which for $D=3$ reproduces the result of Perez and Pranzetti, namely $N=2$. However, we now have to implement many additional constraints corresponding to \eqref{eq:FieldStrenghtsPP} consistently, which makes a success of this route at the quantum level rather doubtful (see, however, the comments on quantisation in section \ref{sec:Quantisation}).

\section{Comments on Quantisation}
\label{sec:Quantisation}

\subsection{SO$(D+1)$ as Gauge Group}

In order to calculate the entropy associated to a spatial slice of the isolated horizon, we have to quantise the resulting theory on the horizon. In the well known $3+1$-dimensional treatment \cite{AshtekarQuantumGeometryOf}, it is a key result that the field strength on the isolated horizon vanishes almost everywhere due to the isolated horizon boundary condition, except at points where the bulk spin network punctures the isolated horizon. Only at these points, the flux operator, which determines the field strength on $S$ via the isolated horizon boundary condition (\ref{eq:BoundaryConditionEulerCompactGroup}), is non-vanishing. The resulting quantum theory on the horizon is a Chern-Simons theory with topological defects induced by these spin network punctures, which result in a finite-dimensional Hilbert space. 

In higher dimensions, the situation is more complicated and we don't claim to have a satisfactory proposal for a quantisation. In this section, we will comment on how such a quantisation could be performed and where the problems lie. 
In a first attempt, one would expect to obtain a higher-dimensional Chern-Simons theory on the horizon, since the symplectic structure on the isolated horizon is exactly of this type. Due to the distributional nature of the space of generalised connections in loop quantum gravity, see e.g. \cite{ThiemannModernCanonicalQuantum}, one promotes the connection on the isolated horizon to an independent degree of freedom in the quantum theory, here called $A_{IJ}$ with field strength $F_{IJ}=F(A)_{IJ}$. Furthermore, a quantisation of the boundary condition (\ref{eq:BoundaryConditionEulerCompactGroup}) (neglecting for a moment the stronger condition (\ref{eq:BoundaryConditionRiemannCompactGroup}) and thus the fact that the connection on the isolated horizon is given by $\Gamma^0$) yields the quantum equations of motion of a higher-dimensional Chern-Simons theory with punctures exactly as in the $3+1$-dimensional case. The immediate problem with this approach is however that higher-dimensional Chern-Simons theory admits local degrees of freedom (at least at the classical level), since the equations of motion 
\be
	\epsilon_{I_1 J_1 \ldots I_n J_n} F^{I_2J_2} \wedge \ldots \wedge F^{I_n J_n} = 0 \label{eq:ChernSimonsEOM}
\ee
don't constrain the connection to be flat \cite{BanadosExistenceOfLocal}. As a direct consequence, one would expect to obtain an infinite entropy by counting the allowed states in the Hilbert space.

Still, it seems that the objects $\epsilon_{I_1 J_1 \ldots I_n J_n} F^{I_2J_2} \wedge \ldots \wedge F^{I_n J_n}$ constitute an important sub-sector of the higher-dimensional Chern-Simons theory which one should consider for entropy calculations, as we will argue in the following. Using the language of Engle, Noui, Perez, and Pranzetti \cite{EngleBlackHoleEntropyFrom}, the Chern-Simons equations of motion (\ref{eq:ChernSimonsEOM}) are modified by ``particle degrees of freedom'' which are induced by the spin networks puncturing the horizon as
\be
	E^{I_1 J_1}(x) := \epsilon^{I_1 J_1 \ldots I_n J_n} F_{I_2J_2}(x) \wedge \ldots \wedge F_{I_n J_n}(x)  \propto \hat{s}_a  \widehat{\pi}^{a I_1 J_1}(x) \text{,} \label{eq:BoundaryConditionInTermsOfIndepentendConnection}
\ee
where the operator on the right hand side symbolises to the flux operator which acts non-trivially only at points where a spin network punctures the horizon. 
Using the Dirac brackets obtained from solving the second class constraints of the higher-dimensional Chern-Simons theory\footnote{Actually, in order to construct the Dirac brackets of the Chern-Simons theory, we would have to identify the set of second class constraints. This is non-trivial and depends on the choice of invariant tensor, as emphasised in \cite{BanadosExistenceOfLocal}. There, it is also stated that for the epsilon tensor used in this chapter, our choice of second class constraints is correct at least in six spacetime dimensions ($\epsilon^{IJKLMN}$ is ``generic'' in the language of \cite{BanadosExistenceOfLocal}). On the other hand, we can use the horizon boundary condition and the symplectic structure (\ref{eq:SymplecticStructureNS}) to calculate the same algebra, at least under the constraint that the field strength of the Chern-Simons connection is given by $F(\Gamma^0)_{IJ}$.}, we can explicitly calculate the algebra of these ``particle excitations'' as
\be
	\left\{ E^{IJ}(x), E^{KL}(y) \right\} \propto \delta^{(D-1)}(x-y) f^{IJ, KL,} \m_{MN} E^{MN}(x)  \text{,}
\ee
where $f$ are the structure constants of SO$(D+1)$. 
Since a representation of this algebra is just a representation of the Lie algebra so$(D+1)$ for each puncture, the problems which have to be discussed for the quantisation are mainly connected with finding the right subspace of the tensor product of the individual so$(D+1)$ representation spaces which is selected by the criterion of compatibility with the bulk spin networks and the horizon topology. As opposed to the U$(1)$ or SU$(2)$ based constructions in four dimensions, the restrictions imposed by the simplicity constraint will also have to be taken into account properly. While the simplicity constraint is solved on the horizon at the classical level by using the variables $n^I$, $s^I$, and $e^I_{\D{\alpha}}$ to construct $\Gamma_{\alpha IJ}^0$, there might still be non-trivial restrictions coming from imposing the quantum simplicity constraint in the bulk. One of them is to restrict the representations carried by the punctures to be the same as in the bulk, i.e. simple (spherical / class 1) SO$(D+1)$ representations. However, the more interesting question will be if there is a restriction resulting from implementing the vertex simplicity constraints.

Despite these attractive features, we still have to deal with the local degrees of freedom. One point that we overlooked up to now is that the classical analogue of the boundary condition (\ref{eq:BoundaryConditionInTermsOfIndepentendConnection}) does not constrain the Chern-Simons connection $A_{\alpha IJ}$ to be $\Gamma^0_{\alpha IJ}$. In section \ref{sec:SO(D+1)}, it was shown that some modifications of the boundary connection parametrised by constants are allowed. Furthermore, the Beetle-Engle trick from section \ref{sec:BEM} suggests that further modifications are conceivable, possibly an infinite set. Thus, we should introduce a constraint which restricts the degrees of freedom of the higher-dimensional Chern-Simons theory as if the horizon connection would be given by $\Gamma^0$. Since the gauge invariant (local) information of a connection is contained in its field strength, we should introduce the boundary condition (\ref{eq:BoundaryConditionRiemannCompactGroup}) in the form
\be
	F(A)^{\text{horizon}}_{\alpha \beta IJ} = F(\Gamma^0)^{\text{bulk}}_{\alpha \beta IJ}	\label{eq:BoundaryConditionSingleF}
\ee
on $S$. In analogy to the $3+1$-dimensional treatment, we would quantise this boundary condition by promoting the left hand side to an operator in the higher-dimensional Chern-Simons theory and act with a proper quantisation of the right hand side on the bulk spin network (as with a flux operator). Since we would regularise the right hand side by fluxes and commutators involving volume operators as in \cite{ThiemannQSD1, BTTIII}, it would automatically vanish at points where no bulk degrees of freedom are excited\footnote{We would expect that the corresponding operator would even vanish at punctures, since the volume operator annihilates edges. On the other hand, we would demand consistency with (\ref{eq:BoundaryConditionInTermsOfIndepentendConnection}), i.e. we would rather use (\ref{eq:BoundaryConditionInTermsOfIndepentendConnection}) at punctures. This underlines again that the discussion here does not provide a satisfactory answer.}.  
This mechanism could thus get rid of the local degrees of freedom and result in a finite entropy much in the same way as in $3+1$ dimensions. Still, there are many missing and imprecise steps in this argument, e.g. that one would first need an actual quantisation of higher-dimensional Chern-Simons theory before a quantum boundary condition as (\ref{eq:BoundaryConditionSingleF}) could be even imposed. 

This discussion leads us back to reconsider equation \eqref{eq:SymplecticStructureNS} for the following reason: we know that the boundary symplectic structure can be written as  \eqref{eq:SymplecticStructureNS} and the boundary condition can be phrased as $\hat{s}_a \pi^{aIJ} = 2 n^{[I} \tilde{s}^{J]} $. Endowed with the Poisson bracket following from \eqref{eq:SymplecticStructureNS}, the $n^{[I} \tilde{s}^{J]}$ form an so$(D+1)$ Lie algebra in the same way as the fluxes do. On the other hand, we saw that the potentially relevant degrees of freedom $E^{IJ}$ in the Chern-Simons theory are identical to the $n^{[I} \tilde{s}^{J]}$ and obey the same algebra. Thus, one could conclude that the Chern-Simons theory should be avoided already in the beginning. Such a system can be quantised with the same methods as the normal $N^I$ in \cite{BTTVI} resulting from the linear simplicity constraints.
However, we know from the $3+1$-dimensional treatment that valuable insights were gained through the Chern-Simons treatment, e.g. logarithmic corrections resulting from a finite level, see e.g. \cite{EngleBlackHoleEntropyFrom}. 
If neglecting the Chern-Simons theory, these results have to be accounted for differently.

To conclude, we don't have a satisfactory quantisation of the resulting boundary theory and thus also no direct access to full fledged entropy calculations at the moment. The biggest uncertainty certainly is that no quantisation of higher-dimensional Chern-Simons theory with a non-Abelian gauge group is known. A reduction to U$(1)$ as a gauge group would in principle facilitate the problem, but we were not able to perform this reduction as explained in the next subsection. 
It thus seems that to a first approximation, considering the $n^{[I} \tilde{s}^{J]}$ as the boundary degrees of freedom is sensible. A straightforward generalisation of the methods developed in \cite{GhoshAnImprovedEstimate} would then give an entropy proportional to the area to leading order in the same way as in $3+1$ dimensions. One could then fix the free parameter $\beta$ in order to obtain the prefactor $1/4G$ for the entropy. See \cite{JacobsonANoteOn, BNI} for discussions about this issue.

\subsection{Reduction to U(1)}

The above discussion suggests that the local degrees of freedom of the higher-dimensional Chern-Simons theory should be absent after properly implementing the boundary conditions \eqref{eq:BoundaryConditionInTermsOfIndepentendConnection}, \eqref{eq:BoundaryConditionSingleF}. 
Since U(1) Chern-Simons theory on the other hand does not have local degrees of freedom to begin with, this hints that we should investigate the possibility of gauge fixing the SO$(D+1)$ theory down to SO$(2)$, as there is no obvious contradiction due to different degrees of freedom in the quantum theory. 
This question will be pursued in this section, but as we will see, we did not succeed in giving a satisfactory description of the boundary degrees of freedom with this structure group.

Two routes suggest themselves: 1) Gauge fix the SO$(D+1)$ Chern-Simons theory we obtained in the course of this paper down to SO$(2)$, or 2) impose the gauge fixing directly at the level of the boundary symplectic structure and rewrite it in terms of an SO(2) Chern-Simons symplectic structure classically. The first route fails due to the SO$(D+1)$ invariant tensor used to construct the Chern-Simons theory, namely $\epsilon^{I_1 ... I_{D+1}}$, which does not admit this gauge fixing. Therefore, we will follow route 2).

We introduce the gauge fixing $n^{I} = g^{0i} \delta_i^I$, $s^J = g^{1j} \delta_j^J$, where $i,j \in \{0,1\}$ and $g \in$ SO(2). Let us use the usual parametrisation of rotations by an angle $\phi$, $g_{00} = g_{11} = \cos{\phi}$, $g_{01} = - g_{10} = \sin{\phi}$. The boundary contribution to the symplectic structure reads in this gauge
\begin{align}
	\delta_{[1} \tilde{s}^I \delta_{2]} n_I = \delta_{[1} \sqrt{h} ~ \delta_{2]} \phi \text{.}
\end{align}
In the SO$(D+1)$ case, to show that a Chern-Simons symplectic structure arises on the horizon cross sections, it was important that $\sqrt{h}$ and the Euler density are proportional. Introducing an SO(2) connection $A_{\alpha}$, the analogue of this requirement would read 
\be
\sqrt{h} \propto \epsilon^{\alpha_1 ... \alpha_{2n}} F_{\alpha_1 \alpha_2} ... F_{\alpha_{2n-1} \alpha_{2n}} \text{,} \label{eq:U(1)BC}
\ee
where $F_{\alpha \beta} = 2 \partial_{[\alpha} A_{\beta]}$. It follows that $\delta \sqrt{h} \propto 2n \epsilon^{\alpha_1 ... \alpha_{2n}} ~ (\partial_{[\alpha_1} \delta A_{\alpha_2]})~ F_{\alpha_3 \alpha_4} ... F_{\alpha_{2n-1} \alpha_{2n}}$ and therefore (upon partial integration)
\begin{align}
	\delta_{[1} \tilde{s}^I \delta_{2]} n_I \propto 2n \epsilon^{\alpha_1 ... \alpha_{2n}} ~(\delta_{[1} A_{\alpha_1})   ~(\delta_{2]} \partial_{\alpha_2} \phi) ~ F_{\alpha_3 \alpha_4} ... F_{\alpha_{2n-1} \alpha_{2n}} \text{.}
\end{align}
With the additional requirement that $A=d\phi$, this would become the symplectic structure of an SO(2) Chern-Simons theory on the boundary. However, from this requirement we also conclude that $F=0$, which is in contradiction with \eqref{eq:U(1)BC}, and therefore also our second route fails. It thus seems that we have to stick to the SO$(D+1)$ theory on the boundary and one should try to make progress with its quantisation as outlined above.

\subsection{Higher Dimensional versus 4d Entropy} 

As transpires from the discussion so far, the main roadblock towards a 
quantum treatment of the boundary conditions in higher dimensions in analogy to the procedure followed in 3+1 dimensions originates from our lack of understanding of higher dimensional quantum Chern-Simons theory. 
The problem here is in fact two-fold. First, the quantisation of pure Chern-Simons 
theory with punctures by itself is an interesting open problem in mathematical 
physics which is unlikely to be an easy task because the theory has local 
degrees of freedom, gauge freedom, and is self-interacting. The second 
problem which is more relevant in our context is that even if quantum 
Chern-Simons theory was available to us, its Hilbert space (with given puncture data) 
is infinite dimensional due to the presence of the local degrees of freedom 
which suggests that a calculation in analogy to the 3+1 treatment 
would result  in an infinite value of the entropy. 

However, this may not be necessarily the case because the Chern-Simons 
boundary theory that we are actually interested in may in fact only be 
a subsector of pure Chern-Simons theory with punctures. The reason for this is that 
the boundary conditions at the horizon when properly translated into the 
degrees of freedom of Chern-Simons theory may constrain the Chern-Simons connection further 
to such an extent that in fact the resulting theory has no local degrees of freedom. 
Indeed, a phenomenon of this kind must happen if the U$(1)$ point of view sketched 
in the previous section is to be viable at all, because allowed gauge fixings 
cannot change the number of true degrees of freedom, hence it is not 
possible to gauge fix pure non-Abelian Chern-Simons theory in more than $3$ dimensions to 
pure  Abelian Chern-Simons theory (with punctures) unless there is more gauge symmetry available 
than that which is intrinsic to pure Chern-Simons theory (with punctures). 
In our case, this additional gauge symmetry might be available due to the fact that the Chern-Simons theory degrees of freedom that we are interested in are just effective degrees of freedom of a corresponding bulk theory which has more gauge symmetry.  The interesting question is how much of this survives at the boundary.
If this scenario would be valid, it would constitute independent support to the 
new point of view advertised  in this paper to consider the variables $n^{[I} \tilde s^{J]}$ 
as independent boundary degrees of freedom and to base the boundary quantum theory on those. At least naively, this would result in finite entropy as for 
fixed puncture data the allowed quantum states are bounded by the relevant 
highest weights of the SO$(D+1)$ representations. Notice that these are boundary 
degrees of freedom and thus we are not interested in which way the 
highest weights are constituted by bulk edges and representations ending 
in the punctures (for which there are infinitely many possibilities). 

On the other hand, it might also be that the opposite happens and the 
bulk symmetry that survives in the boundary theory is too small to render 
the entropy calculation finite when using the Chern-Simons theory in which 
case the quantisation based on $n^{[I} \tilde s^{J]}$ appears to be an attractive 
option which however must be substantiated by further reasoning.\\ 
\\   
We leave the investigation of these ideas for future research 
and conclude this section with yet another point of view which has the advantage that it makes the contact with the $3+1$ theory more transparent.   
Namely, so far we have talked about the entropy of an observer living 
in $D+1$ dimensions as we used the area of a $D-1$-dimensional surface. 
Whether or not in a $D+1$-dimensional world an actual observer really extends 
in all $D+1$ dimensions, using the usual Kaluza-Klein point of view the 
observer may argue that all but $3+1$ of those dimensions are not accessible 
to him. As such he may argue that what accounts for the entropy of a black 
hole is just a two-dimensional cross section of the 
actual $D-1$-dimensional horizon. This lack of knowledge due to the excess dimensions will result 
in an effective $3+1$-dimensional theory in which the actual $D+1$-dimensional 
pure states are represented by mixed states due to the usual entanglement. 
Ultimately, this could result in a situation identical to the $3+1$ theory with 
the following modifications: 1. The Chern-Simons theory to consider 
is for the gauge group SO$(D+1)$ rather than SO$(3)$ (or SU$(2)$). 2. 
The entropy would be computed not using the 
pure $3+1$-dimensional states but rather the $D+1$-dimensional mixed states. 
The interesting and nontrivial question is of course whether the actual higher-dimensional 
entropy and the effective lower-dimensional one agree with each other at least 
semiclassically. In certain holographic scenarios this could actually be the 
case, see e.g. \cite{SolodukhinEntanglementEntropyOf} and references therein. 

Fundamentally, these mixed states should be computed from the full fledged 
$D+1$ theory. While clean, this approach has the disadvantage that again 
quantum Chern-Simons theory in $D>3$ dimensions would be needed. 
The analysis of \cite{BanadosExistenceOfLocal} reveals that a canonical 
quantisation using techniques of loop quantum gravity is conceivable, 
however, the additional constraints present (beyond gauge invariance and 
diffeomorphism invariance) must be accounted for in a Dirac quantisation. 
Alternatively, a reduced phase space quantisation suggests itself which 
has interesting connections with the WZW conformal field theory about which 
a lot more is known \cite{BanadosTheDynamicalStructure}.        
  
A poor man's version of this which should be viable in the semiclassical sector of the theory (and thus in particular in our context as we are 
using semiclassical reasoning in many places) is to perform a classical 
Kaluza-Klein reduction \cite{NieuwenhuizenSupergravity} 
and to quantise $3+1$ General relativity together 
with matter and the Kaluza-Klein fields in the standard fashion 
\cite{ThiemannQSD5}, \cite{BTTVI, BTTVII}. 
In this approach one simply would have to make sure that the Kaluza-Klein 
modes and in particular their boundary values do not disturb the 
standard reasoning in $3+1$ dimensions, that is, without changing the Barbero-Immirzi parameter as was done explicitly, e.g. for minimally coupled 
Maxwell and scalar fields as well as non-minimally coupled scalar fields, see \cite{AshtekarNonMinimallyCoupled, AshtekarNonMinimalCouplingsQuantum} and references therein.

\section{Concluding Remarks}
\label{sec:ConcludingRemarks}

In this paper, we derived a generalisation of the isolated horizon boundary condition to non-distorted horizons in even dimensional spacetimes and showed that the canonical transformation to SO$(D+1)$ connection variables leads to a higher-dimensional Chern-Simons symplectic structure on the boundary of the spatial slice. While the classical treatment from four spacetime dimensions generalises rather directly, the quantisation of the resulting system is less obvious, since, to the best of the authors' knowledge, there are no known generalisations of the quantisation of $2+1$-dimensional Chern-Simons theory to higher dimensions. 
On the other hand, due to the stronger boundary condition (\ref{eq:BoundaryConditionRiemannCompactGroup}), it could be the case that it is not necessary to quantise ``all of the higher-dimensional Chern-Simons theory'', but just a subalgebra of phase space functions which result as topological defects induced by puncturing the horizon with a spin network, as discussed in the previous section. 
In this line of thought, it suggested itself to forget about the Chern-Simons theory entirely and to use the $n^{[I} \tilde{s}^{J]}$ along with the symplectic structure \eqref{eq:SymplecticStructureNS} as horizon degrees of freedom. While mathematically attractive, this goes against the idea of using a theory based on a connection for both the bulk and the horizon. Also, it is unclear whether one can recover all the insights resulting from a finite level of the Chern-Simons theory in $3+1$ dimensions. As opposed to the Chern-Simons treatment however, the $n^{[I} \tilde{s}^{J]}$ are not restricted to even spacetime dimensions or specific topologies, which makes them very attractive, e.g. to compare with the broad literature on five-dimensional black holes.
As an alternative, we have outlined a possible route along the more standard Kaluza-Klein reduction approach. 
Thus, we underline again that we don't have a fully satisfactory quantisation for the isolated horizon degrees of freedom.

One of the most important questions which should be answered by a suitable quantisation of the theory on the black hole horizon is the treatment of the simplicity constraint. A preliminary analysis shows that the classical simplicity constraint fits nicely into the picture of a Chern-Simons theory with particles as proposed in \cite{EngleBlackHoleEntropyFrom}. While a quantisation of the edge simplicity constraints would just restrict the group representations on the particle defects in the same way as it restricts the edge representations, it might be that a proper quantisation of the horizon degrees of freedom gives us a hint on what the correct implementation of the simplicity constraint on a vertex is. The reason for this comes from the seemingly very effective treatment of a black hole as a single intertwiner, see \cite{KrasnovBlackHolesIn} and more recently also \cite{BianchiEntropyOfNon}. Moreover, this question will have a direct effect on the subleading correction in the entropy formula, which makes it again very interesting to study.

Additionally, it will be interesting to check to what extend the connection on the horizon can be generalised, e.g. as in \cite{EngleBlackHoleEntropyFrom}, where a new free parameter can be associated to the horizon connection which can rescale the entropy. 
The consequences of introducing a two parameter family of connections in the bulk in four dimensions as proposed in \cite{BTTII} should also be investigated. 
From a more general perspective, it is noteworthy that many ingredients of the definition of an isolated horizon were not used in the Hamiltonian treatment, were only the non-distortion condition entered and the fact that there is no boundary term in the ADM symplectic potential. It thus suggests itself to pursue the question of general boundaries of spacetime, not only isolated horizons. This is especially interesting in the context of entropy bounds for general bounded regions of spacetime \cite{BoussoTheHolographicPrinciple}.

\section*{Acknowledgments} 
NB and AT thank the German National Merit Foundation for financial support. We thank Enrique Fern\'andez Borja, I\~{n}aki Garay, Jerzy Lewandowski, Yasha Neiman, and Alexander Stottmeister for stimulating and helpful discussions. During final improvements of this work, NB was supported by the NSF Grant PHY-1205388 and the Eberly research funds of The Pennsylvania State University.

\begin{appendix}


\section{Hybrid Connection and Generalisations}
\label{app:Connections}

In this appendix, we will introduce several connections relevant for the main text, namely Peldan's ``hybrid" spin connection \cite{PeldanActionsForGravity} and extensions thereof to higher-dimensional internal space.
\subsection{Peldan's Hybrid Connection}
It is a well-known fact that, given an SO$(D)$ vielbein $e_a\m^i$ in $D$ dimensions, there exists a unique spin connection $\Gamma_{aij}[e]$ compatible with it, which is obtained by solving
\be
	0 \stackrel{!}{=} D^{\Gamma}_a e_b\m^i = D_a e_b\m^i + \Gamma[e]_{a}\m^i\m_j e_b\m^j
\ee
for $\Gamma[e]_{aij}$, where $D_a$ denotes the torsion free metric compatible covariant derivative. The result is
\be
	\Gamma[e]_{aij} = e^b\m_{[i} D_a e_{b|j]} \text{.}
\ee
Starting from a Lagrangian formulation of general relativity on a $D+1$ dimensional spacetime manifold, the natural gauge group is SO$(1,D)$ or SO$(D+1)$ for the Lorentzian or Euclidean theory, respectively. When passing to the corresponding Hamiltonian system, a $D + 1$ split is performed and we are naturally led to consider a SO$(1,D)$ or SO$(D+1)$ vielbein $e_a\m^J$ on the $D$ dimensional spatial slice, which we will call hybrid vielbein. However, from the Hamiltonian perspective, the signature of the internal space $\zeta$ is not necessarily tied to the spacetime signature $s$, since we can always start with an SO$(D)$ vielbein on the spatial slice and introduce gauge degrees of freedom corresponding either to SO$(1,D)$ or SO$(D+1)$. In the following, we will therefore treat internal and spacetime signature independently. Peldan \cite{PeldanActionsForGravity} investigated if one could define a compatible connection also for this hybrid vielbein. We have
\be
	0 \stackrel{!}{=} D^{\Gamma^{H}}_a e_b\m^J = D_a e_b\m^J + \Gamma^{\text{H}}[e]_{a}\m^J\m_K e_b\m^K \text{,}
\ee
which actually can be solved for the unique ``hybrid" spin connection,
\be
	\Gamma^{\text{H}}[e]_{aIJ} = e^b\m_{[I} D_a e_{b|J]} + \zeta n_{[I} D_a n_{J]} \text{,}
\ee
where $n^I$ is the unique (up to sign) unit normal to the hybrid vielbein, $n^I e_{aI} = 0$, $n^I n^J \eta_{IJ} = \zeta$, and $\zeta$ again denotes the internal signature, $\zeta = -1$ for SO$(1,D)$ and $+1$ for SO$(D+1)$. Note that the sign ambiguity is absent in $\Gamma^{\text{H}}[e]_{aIJ}$ since $n^I$ appears quadratically.

\subsection{Extensions to Higher-Dimensional Internal Space}
\label{sec:GeneralisedHybridConnection}
Now we want to extend this result to a higher-dimensional internal space, which is necessary for black hole applications, since we have to deal with the vielbein on the $D-1$ dimensional inner boundaries of the spatial slice. \\ 
\\
We will start quite general by introducing an $\mathbb{R}^{D+k}$ -- valued vielbein $e_a\m^J$ in $D$ dimensions (only in this section, we will have $I,J,K... = 1,..., D+k$), $e_a\m^I e_b\m^J \eta_{IJ} = q_{ab}$ where $\eta_{IJ} = \text{diag}(\underbrace{-,...,-}_{p},\underbrace{+,...,+}_{D+q})$ and $p+q = k$, and ask for a so$(p,D+q)$ connection $\Gamma^{\text{H}}_{aIJ}$ annihilating $e_a^J$. We have
\be
	0 \stackrel{!}{=} D^{\Gamma^H}_a e_b\m^J = D_a e_b\m^J + \Gamma^{\text{H}}_a\m^J\m_K e_b\m^K \text{,} \label{eq:AnsatzHybridConnection}
\ee
corresponding to $D^2 (D+k)$ equations to determine $\Gamma^{\text{H}}_{aIJ}$. However, these equations are not all independent, since
\be
	0 = e_{(c|}\m^I D^{\Gamma^{\text{H}}}_a e_{|b)I}
\ee 
is identically satisfied due to the antisymmetry of the so$(p,D+q)$ connection and the metric compatibility of $D_a$. The result are 
\be
	D^2(D+k) - D^2(D+1)/2 = D^2((D-1)/2 + k) \label{eq:IndepEquations}
\ee 
independent equations for the 
\be
	D(D+k)(D+k-1)/2 \label{eq:NumberComponents}
\ee
unknowns  $\Gamma^{\text{H}}_{aIJ}$. It is clear that $\Gamma^{\text{H}}_{aIJ}$ cannot be determined uniquely for any $k$, since the number of equations grows, for fixed $D$, linearly with $k$, while the connection components grow quadratically. More precisely, equating both, we obtain  $(\ref{eq:IndepEquations}) = (\ref{eq:NumberComponents}) \Leftrightarrow Dk(k-1)/2 = 0$, i.e. the connection is only uniquely determined for the gauge groups $SO(D)$, corresponding to $k=0$, and $SO(1,D)$ or $SO(D+1)$ for $k=1$. \\
\\
Let us study the indeterminacy for $k > 1$ in more detail. First we ``complete" the vielbein by choosing an orthonormal set of $k$ unit vectors $n_i\m^I$, $i = 1,...,k$, normal to the vielbein, i.e. $n_i\m^I e_{aI} = 0$ $\forall i = 1,...,k$ and $n_i\m^I n_j\m^J \eta_{IJ} = \eta_{ij}$ $\forall i,j = 1,...,k$ where $\eta_{ij} = \text{diag}(\underbrace{-,...,-}_{p},\underbrace{+,...,+}_{q})$\footnote{Actually, we can as well specify $k-1$ vectors, since the last one, $n_k\m^I$, is already determined (up to sign) by the mentioned requirements.}. The indices $i,j,...$ will be raised and lowered using this metric and its inverse $\eta^{ij}$. Then we can decompose $\Gamma^{\text{H}}_{aIJ}$ according to
\be
	\Gamma^{\text{H}}_{aIJ} = \overline{\Gamma}_{aIJ} + 2 n_{i[I|} \overline{\Gamma}^i\m_{a|J]} +  n_{i[I|} n_{j|J]} \Gamma^{ij}\m_a \text{,} \label{eq:ConnectionAnsatz2}
\ee
where summation over repeated indices $i,j$ is understood and $\overline{\Gamma}_{aIJ} n_i\m^J = 0$ $\forall i=1,...,k$, $\overline{\Gamma}^i\m_{aJ} n_{j}\m^{J} = 0$ $\forall i,j = 1,...,k$. Inserting this decomposition of $\Gamma^{\text{H}}_{aIJ}$ into (\ref{eq:AnsatzHybridConnection}), we find that $\Gamma^{ij}\m_a$ simply drops out and therefore cannot be solved for, and the number of its components, $Dk(k-1)/2$ since it is antisymmetric in $i,j$, precisely matches the indeterminacy. For the other components, one obtains
\be
	 \overline{\Gamma}_{aIJ} &=& e^{b}\m_{[I|} \bar{\eta}_{J]K} D_a e_b\m^K \text{,} \\
	 \overline{\Gamma}^i\m_{aJ} &=& \bar{\eta}_{JK} D_a n^{iK} \text{,}
\ee
where $\bar{\eta}_{IJ} := e_{aI} e^a\m_{J}$. Inserting back into (\ref{eq:ConnectionAnsatz2}), we find
\be
	\Gamma^{\text{H}}_{aIJ} = 2 e^b\m_{[I|} D_a e_{b|J]} - e^{b}\m_{[I} \bar{\eta}_{J]K} D_a e_b\m^K +  n_{i[I|} n_{j|J]} \Gamma^{ij}\m_a
\ee
and therefore a $Dk(k-1)/2$ -- parameter family of connections annihilating $e_a\m^I$. To obtain a unique connection, we have to add additional requirements, e.g. we could demand that $\Gamma^{ij}\m_a = 0$ $\forall i,j = 1,...,k$ (these requirements are independent of the choice of ``completion" for the vielbein $\{n_i\m^I\}_{i=1}^{k}$). This connection $\Gamma^1_{aIJ}$ would be special in that it would only depend on $e_a\m^I$,
\be
	\Gamma^{1}_{aIJ} = 2 e^b\m_{[I|} D_a e_{b|J]} - e^{b}\m_{[I} \bar{\eta}_{J]K} D_a e_b\m^K \text{.} \label{eq:Gamma1}
\ee
Having in mind the application to black holes, we will proceed differently. For a fixed extension, the extra conditions we impose are $D^{\Gamma^H}_a n_i\m^I = 0$ $\forall i=1,...,k-1$\footnote{Note that, since $n_k\m^I$ is given by $e_a\m^I$, $n_i\m^J$, $i = 1,...,k-1$, up to sign, it is automatically annihilated by $D_a$ if the latter are.} (these requirements are sensitive to the choice of completion). Again, these conditions are not all independent. We have $e_b\m^I D^{\Gamma^H}_a n_{iI} = 0$ and $n_{(i}\m^I D^{\Gamma^H}_a n_{j)I} = 0$ already satisfied, which results in $D(k-1)(D+k) - (D^2(k-1) + Dk(k-1)/2) = Dk(k-1)/2$ independent equations, which equals the number of undetermined components $\Gamma^{ij}\m_a$. Solving for these, we find
\be
	\Gamma^{ij}\m_a = - n^{[i}\m_I D_a n^{j]I}
\ee
and
\be
	\Gamma^0_{aIJ}[e,n] := e^b\m_{[I|} D_a e_{b|J]} + n^{i}\m_{[I|} D_a n_{i|J]} 
\ee
as the unique connection annihilating the chosen completion of $e_a\m^J$. This connection has several nice properties, e.g. while for all connections of the family, we have
\be
	R^H_{abIJ} e_c\m^I e^{dJ} &=& R_{abc}\m^d \text{,} \\
	R^H_{abIK} n_i\m^{I} \bar{\eta}^{KJ} &=& 0 \text{,}
\ee
which follows from contraction of 
\be
	0 = [D^{\Gamma^H}_a, D^{\Gamma^H}_b] e_{c}\m^I &=& R^H_{ab}\m^I\m_J e_c\m^J + R_{abc}\m^d e_d\m^I \text{,}
\ee
for this connection we additionally have
\be
	R^0_{ab}\m^I\m_J n_i\m^J &=& [D^{\Gamma^0}_a, D^{\Gamma^0}_b] n_i\m^I = 0
\ee
and therefore
\be 
	R^0_{abIJ} &=& R_{abc}\m^d ~e^c\m_I e_{dJ} \text{.} \label{eq:R0identity}
\ee
From the right hand side of (\ref{eq:R0identity}), we see that, while $\Gamma^0_{aIJ}$ depends on the choice of $\{n_i\m^I\}_{i=1}^{k}$, $R^0_{abIJ}$ is independent of $n$, determined completely by $e_a\m^I$ and its first and second derivatives. Explicitly, choosing a different completion $\{\tilde{n}_i\m^I\}_{i=1}^{k}$ of $e_a\m^I$, which is related to $\{n_i\m^I\}_{i=1}^{k}$ by a SO$(p,q)$ transformation $g$ via $\tilde{n}_i = g_i\m^j n_j$, we find
\be
	\Gamma^0_{aIJ}[e,\tilde{n}] = \Gamma^0_{aIJ}[e,n] + K_{aIJ} \text{,} \\
	K_{aIJ} := g^i\m_k n^k\m_{[I|} n_{l|J]} D_a g_i\m^l \text{,}
\ee
and
\be
	R^0_{abIJ}[\Gamma^0[e,\tilde{n}]] = R^0_{abIJ}[\Gamma^0[e,n]] + 2D^{\Gamma^0}[e,n]_{[a} K_{b]IJ} + [K_a,K_b]_{IJ} = ... = R^0_{abIJ}[\Gamma^0[e,n]] \text{.}
\ee
For even dimensions $D = 2n$, it follows from (\ref{eq:R0identity})
\be
	\epsilon^{K_1...K_k I_1J_1 ... I_n J_n}~ \epsilon^{a_1 b_1 ... a_n b_n} ~R^0_{a_1b_1I_1J_1} ... R^0_{a_n b_n I_n J_n} = E^{(D)} \epsilon^{i_1...i_k} n_{i_1}\m^{[K_1} ... n_{i_k}\m^{K_k]} \text{,} 
\ee
the right hand side of which is also manifestly invariant under SO$(p,q)$ rotations and where $E^{(D)}$ denotes the $D$ - dimensional Euler density
\be
	E^{(D)} := \frac{1}{\sqrt{\text{det} \, q_{ab}}} \epsilon^{a_1 b_1 ... a_n b_n} \epsilon^{c_1 d_1 ...c_n d_n} R_{a_1b_1c_1d_1} ... R_{a_nb_nc_nd_n} \text{.}
\ee
Note that $R^0_{abIJ}$ is not the only curvature tensor constructed from $e_a\m^I$ only. Of course, the connection $\Gamma^1_{aIJ}$ we considered earlier, obtained by choosing $\Gamma^{ij}_a = 0$, is constructed solely from $e_a\m^I$ and so is the corresponding curvature tensor, but it fails to satisfy (\ref{eq:R0identity}). More precisely, we find
\be
	R^1_{abIJ} = R^0_{abIJ} + 2 (\eta - \bar{\eta})_{K[I} (\eta - \bar{\eta})_{J]L} q^{cd} (D_{[a|} e_c\m^K)(D_{|b]} e_d\m^L)\text{.}
\ee

\section{Higher-Dimensional Chern-Simons Theory}
\label{sec:ChernSimons}

In this appendix, we will review some facts about Chern-Simons theory in higher dimensions relevant for this work, with focus on the canonical formulation. In particular, we will derive the symplectic structure of the theory. We want to stress that these results are not new, but we state them here for completeness. For a more elaborate canonical treatment of higher-dimensional Chern-Simons theory, we refer the reader to \cite{BanadosTheDynamicalStructure}.\\
\\
The Chern-Simons action is defined for all odd dimensions $2n+1$ and gauge groups $G$ by the equation 
\be
	d\mathcal{L}^{2n+1}_{CS} = i_{A_1 A_2 ... A_{n+1}} F^{A_1} \wedge ... \wedge F^{A_{n+1}}\text{,}
	\label{eq:CSAction1}
\ee
where $F^{A} = dA^{A} + 1/2~ [A,A]^A = dA^A + 1/2~f^{A}\m_{BC} ~ A^B \wedge A^C$ is the field strength of the connection one form $A^B$ valued in the Lie algebra of $G$, $f^A\m_{BC}$ are the structure constants of $G$, $A_j,B,C \in \{1,...,\text{dim}(\mathfrak{g})\}$ are Lie algebra indices and $i_{A_1 ... A_n}$ is a rank $(n+1)$ symmetric tensor invariant under the adjoint action of the group. Explicitly,
\be
	\mathcal{L}^{2n+1}_{CS} &=& i_{A_1 ... A_{n+1}} \sum_{p = 0}^{n} (-1)^p \frac{  {2n+1 \choose n-p} }{{2n+1 \choose n} } ~\times  \nonumber \\& & \underbrace{F^{A_1} \wedge ... \wedge F^{A_{n-p}}}_{n-p} \wedge \underbrace{\left(1/2~[A,A]^{A_{n-p+1}} \right)\wedge ... \wedge \left(1/2 ~[A,A]^{A_{n}}\right)}_{p} \wedge A^{A_{n+1}} \nonumber \\
	&=:& i \cdot \sum_{p = 0}^{n} (-1)^p \frac{  {2n+1 \choose n-p} }{{2n+1 \choose n} } F^{n-p} \wedge (1/2~[A,A])^{p}\wedge A ~~\text{,}
\ee
where the second line defines the short hand notation we will use in the following. For our purposes, it will be sufficient to restrict attention to the groups SO$(1,D)$ or SO$(D+1)$ where $D=2n+1$. It is convenient to label the $\frac{D(D+1)}{2}$ generators of the corresponding Lie algebras by an anti-symmetric combination of two indices in the fundamental representation $I,J = 0,...,D$ (e.g. the connection one form will be denoted by $A^{IJ}$ with $ A^{(IJ)}=0$). We will furthermore restrict the invariant tensor to be the epsilon tensor $\epsilon^{I_1 J_1 ... I_{n+1} J_{n+1}}$, which is the one relevant for our application. However, we want to point out that all results of this section are independent of the choice of gauge group and invariant tensor.\\
\\
In order to obtain the (pre-)symplectic structure, we invoke the covariant canonical formalism \cite{CrnkovicCovariantDescriptionOf, LeeLocalSymmetriesAnd, AshtekarCovariantPhaseSpace}, according to which the presymplectic potential is given by the boundary term of the first variation of the action, while the presymplectic structure is the exterior derivative of the potential. \\
\\
Using the relation
\begin{eqnarray}
\delta \left( \epsilon ~ F^{n-p} \wedge \frac{1}{2}[A,A]^p \wedge A \right) = \epsilon \hspace{-5mm} & &  \left\{(n+p+1)~ \delta A \wedge F^{n-p} \wedge \frac{1}{2}[A,A]^{p}  + \right. \nonumber \\
 											 &&\hspace{3mm}(n-p)~ \delta A \wedge F^{n-p-1} \wedge \frac{1}{2}[A,A]^{p+1} +  \nonumber \\
											 &&\hspace{3mm} \left.  (n-p) ~ d \left[ \delta A \wedge F^{n-p-1} \wedge \frac{1}{2}[A,A]^p\wedge A \right] \right\} \text{,}
\label{eq:CSVar1}
\end{eqnarray}
the first variation of the Chern-Simons action yields
\begin{eqnarray}
\delta S^{2n+1}_{CS} &=& \delta \int_{\mathcal{M}} \mathcal{L}^{2n+1}_{CS} \nonumber \\
&=& \quad \int_{\mathcal{M}} \left[ \epsilon \cdot \sum_{p = 0}^{n} (-1)^p  \frac{{2n+1 \choose n-p} }{{2n+1 \choose n} } (n+p+1) F^{n-p} \wedge \frac{1}{2}\left[A,A\right]^{p} \right] \wedge \delta A  \nonumber \\ 
&&+  \int_{\mathcal{M}} \left[ \epsilon \cdot \sum_{p = 0}^{n-1} (-1)^p  \frac{{2n+1 \choose n-p} }{{2n+1 \choose n} } (n-p) F^{n-p-1} \wedge \frac{1}{2}\left[A,A\right]^{p+1} \right] \wedge \delta A  \nonumber \\ 
&&+  \int_{\mathcal{M}} d \left\{ \delta A \wedge \left[  \epsilon \cdot \sum_{p = 0}^{n-1} (-1)^p  \frac{{2n+1 \choose n-p} }{{2n+1 \choose n} }  (n-p) F^{n-p-1}\wedge \frac{1}{2}\left[A,A\right]^{p}\wedge A \right]\right\} \nonumber \\
&=& \quad  \int_{\mathcal{M}} (n+1) ~ \epsilon \cdot  F^{n} \wedge \delta A  \nonumber \\ 
&&+  \int_{\mathcal{M}} d \left\{ \delta A \wedge \left[  \epsilon \cdot \sum_{p = 0}^{n-1} (-1)^p  \frac{{2n+1 \choose n-p} }{{2n+1 \choose n} }  (n-p) F^{n-p-1}\wedge \frac{1}{2}\left[A,A\right]^{p}\wedge A \right]\right\} ~\text{.}
\end{eqnarray}
Note that the two sums of the bulk contribution cancel each other term by term, and the only term surviving is the $(p=0)$ -- term of the first sum. We obtain the Chern-Simons equations of motion\footnote{Note that the bulk term of the variation can be obtained within two lines by varying \ref{eq:CSAction1}.}
\be
	\epsilon \cdot \underbrace{F \wedge ... \wedge F}_{n \text{ times}} = 0~\text{,}
\ee
which in 2+1 dimensions (which corresponds to $n=1$) reduces to $F=0$. Let $\sigma$ be a $2n$-dimensional Cauchy slice. The presymplectic potential can be read off the boundary term of the first variation and is given by
\begin{eqnarray}
  \theta_{\sigma}(\delta) = \int_{\sigma} \delta A \wedge \left[  \epsilon \sum_{p = 0}^{n-1}  \frac{{2n+1 \choose n-p} }{{2n+1 \choose n} } (-1)^p (n-p) F^{n-p-1}\wedge \frac{1}{2}\left[A,A\right]^{p}\wedge A \right] ~~\text{.}
\end{eqnarray}
For its variation, the equation
\be
\delta_{[2}  \left[ \epsilon \cdot  \delta_{1]} A \wedge F^{n-p-1} \wedge \frac{1}{2}\left[A,A\right]^{p}\wedge A \right] &=& \epsilon \cdot \left[ \frac{1}{2}(n+p+1)~  \delta_{[1} A \wedge \delta_{2]} A \wedge F^{n-p-1} \wedge  \frac{1}{2}\left[A,A\right]^{p} \right.  \\ 
& &\textcolor{white}{\epsilon \cdot} \left. + \frac{1}{2}(n-p-1) ~  \delta_{[1} A \wedge \delta_{2]} A \wedge F^{n-p-2} \wedge \frac{1}{2}\left[A,A\right]^{p+1} \right]~\text{,} \nonumber
\label{eq:CSVar2}
\ee
 is useful.
Actually, in the above result, a boundary term was dropped, but in defining the symplectic current, we are allowed to drop this term since we will integrate the symplectic current we want to derive in this step over the boundary of the spacetime region we are interested in. We find for the symplectic current
 \begin{alignat}{5}
d \theta_{\sigma}(\delta_{2},\delta_{1}) &= \frac{1}{2{2n+1 \choose n} } ~ \epsilon \cdot \delta_{[1} A \wedge \delta_{2]} A \wedge \left[\sum_{p=0}^{n-1} {2n+1 \choose n-p} (-1)^p (n-p) (n+p+1)~  F^{n-p-1} \wedge  \frac{1}{2}\left[A,A\right]^{p}\right. \nonumber \\
&  \textcolor{white}{\frac{1}{2{2n+1 \choose n}} \epsilon \cdot \delta_{[1} A \wedge \delta_{2]} A \wedge} + \left. \sum_{p=0}^{n-2} {2n+1 \choose n-p}(-1)^p (n-p) (n-p-1)~  F^{n-p-2} \wedge  \frac{1}{2}\left[A,A\right]^{p+1}\right] \nonumber \\
&= \frac{n(n+1)}{2} ~\epsilon \cdot \delta A \wedge \delta A \wedge F^{n-1} ~ \text{,} \label{eq:CSSymplCurrent}
\end{alignat}
where again the terms in the two sums cancel each other out, with only the $(p=0)$ -- term in the first sum remaining. Therefore, the presymplectic structure is given by
\be
\Omega_{\sigma}(\delta_{2}, \delta_{1}) = \frac{n(n+1)}{2}  \int_{\sigma} ~\epsilon \cdot \delta_{[1} A \wedge \delta_{2]} A \wedge F^{n-1} ~ \text{.}
\ee
Usually, in order to have a meaningful phase space description, one now imposes suitable boundary conditions and checks if the presymplectic structure is independent of the choice of the Cauchy slice $\sigma$ and, for noncompact $\sigma$, if the integral is finite. However, in this paper we are only interested in a spacetime with internal isolated horizon boundary on which the Chern-Simons symplectic structure arises and we only have to answer this questions for the full spacetime.
\\
From \ref{eq:CSSymplCurrent}, we can also read off that the Dirac matrix of Chern-Simons theory is given, up to numerical factors, by $\epsilon \cdot F^{n-1}$, which coincides with the result in \cite[eq. (2.7)]{BanadosTheDynamicalStructure}.

\section{Details on Calculations}
\label{app:Calculations}

\subsection{Symplectic Structure via the Palatini Action}

\label{app:StructureOnDelta}

In this appendix, we provide calculational details for showing (\ref{eq:DeltaSigmaDeltaA}), 
\begin{align} 
	\int_{\Delta} \delta_{[1} \D{\Sigma}^{IJ} \delta_{2]} \D{A}\m_{IJ}  =  2 \int_{\Delta} & \left\{ d\left[ \delta_{[1} \tilde{s}^I \delta_{2]} n_I \right] + \frac{1}{(D-1)!} \delta_{[1} \epsilon^{D-1} \wedge \delta_{2]}\omega^l \right\} \text{.} 
\end{align}

We will contract any of the three lines of (\ref{eq:DeltaSigma}) separately with (\ref{eq:DeltaA}) and multiply them by $\frac{-1}{(D-1)!}$. For the first line, we find
\begin{align}
	& (D-1)  \epsilon_{IJK_1 ...K_{D-1}} \left[ m_L \wedge m^{K_2} \wedge ... \wedge m^{K_{D-1}} (i_{m^L} \delta_{[1} m^{K_1}) \right] \wedge \left[\delta_{2]} \Gamma^{0IJ} - 2 (\delta_{2]} \omega) l^{[I} k^{J]} - 2 \omega \delta_{2]}(l^{[I} k^{J]})\right] \nonumber \\
	= \; &(D-1)  \epsilon_{IJK_1 ...K_{D-1}} l^I k^J \left[ m_L \wedge m^{K_2} \wedge ... \wedge m^{K_{D-1}} (i_{m^L} \bar{\bar{\eta}}^{K_1 M} \delta_{[1} m_{M}) \right] \wedge \left[ - 2 k_{I'} d_{\Gamma^0} \delta_{2]} l^{I'} - 2 \delta_{2]} \omega \right] \nonumber \\
	 +\; & (D-1) \epsilon_{IJK_1 ...K_{D-1}} \left[ m_L \wedge m^{K_2} \wedge ... \wedge m^{K_{D-1}} ( - i_{m^L} l^{K_1} k^{M} \delta_{[1} m_{M}) \right] \wedge \left[ 2 k^{J}  d_{\Gamma^0} \delta_{2]} l^{I} - 2 k^{J} \omega (\delta_{2]} l^{I}) \right] \nonumber \\
	 +\; & (D-1) \epsilon_{IJK_1 ...K_{D-1}} \left[ m_L \wedge m^{K_2} \wedge ... \wedge m^{K_{D-1}} ( - i_{m^L} k^{K_1} l^{M} \delta_{[1} m_{M}) \right] \wedge \left[ 2 l^{J}  d_{\Gamma^0} \delta_{2]} k^{I} + 2 l^{J} \omega (\delta_{2]} k^{I}) \right] \nonumber \\
	 = \; & -2 (D-1) \epsilon_{IJK_1 ...K_{D-1}} l^I k^J \delta_{[1} m^{K_1} \wedge m^{K_2} \wedge ... \wedge m^{K_{D-1}} \wedge \left[  k_{I'} d_{\Gamma^0} \delta_{2]} l^{I'} + \delta_{2]} \omega \right] \nonumber \\
	&-2 (D-1) \epsilon_{IJK_1 ...K_{D-1}} l^I k^J m_M \wedge m^{K_2} \wedge ... \wedge m^{K_{D-1}}  \wedge \left[  \delta_{[1} k^{M} d_{\Gamma^0} \delta_{2]} l^{K_1} - \omega (\delta_{[1} k^{M}) (\delta_{2]} l^{K_1}) \right]\nonumber \\
	& -2 (D-1) \epsilon_{IJK_1 ...K_{D-1}} l^I k^J m_M \wedge m^{K_2} \wedge ... \wedge m^{K_{D-1}} \wedge \left[  \delta_{[2} l^{M}  d_{\Gamma^0} \delta_{1]} k^{K_1} + \omega (\delta_{[2} l^{M}) (\delta_{1]} k^{K_1}) \right] \nonumber \\
	= \; & -2 \delta_{[1}\epsilon^{D-1} \wedge \left[  d( k_{I'}  \delta_{2]} l^{I'}) + \delta_{2]} \omega \right] \nonumber \\
	&-2 (D-1)^2 \epsilon_{IJK_1[K_2 ...K_{D-1}|} l^I k^J m_{|M]} \wedge m^{K_2} \wedge ... \wedge m^{K_{D-1}} \wedge \left[  \delta_{[1} k^{M} d_{\Gamma^0} \delta_{2]} l^{K_1} - \omega (\delta_{[1} k^{M}) (\delta_{2]} l^{K_1}) \right]\nonumber \\
	& -2 (D-1) \epsilon_{IJK_1 ...K_{D-1}} l^I k^J m_M \wedge m^{K_2} \wedge ... \wedge m^{K_{D-1}} \wedge \left[  \delta_{[2} l^{M}  d_{\Gamma^0} \delta_{1]} k^{K_1} + \omega (\delta_{[2} l^{M}) (\delta_{1]} k^{K_1}) \right] \nonumber \\
	= \; & -2 \delta_{[1}\epsilon^{D-1} \wedge \left[  d( k_{I'}  \delta_{2]} l^{I'}) + \delta_{2]} \omega \right] \nonumber \\
	&-2 (D-1) \epsilon_{IJMK_2 ...K_{D-1}} l^I k^J m_{K_1} \wedge m^{K_2} \wedge ... \wedge m^{K_{D-1}} \wedge \left[  \delta_{[1} k^{M} d_{\Gamma^0} \delta_{2]} l^{K_1} - \omega (\delta_{[1} k^{M}) (\delta_{2]} l^{K_1}) \right]\nonumber \\
	& -2 (D-1) \epsilon_{IJK_1 ...K_{D-1}} l^I k^J m_M \wedge m^{K_2} \wedge ... \wedge m^{K_{D-1}}  \wedge \left[  \delta_{[2} l^{M}  d_{\Gamma^0} \delta_{1]} k^{K_1} + \omega (\delta_{[2} l^{M}) (\delta_{1]} k^{K_1}) \right] \nonumber \\
	= \; & -2 \delta_{[1}\epsilon^{D-1} \wedge \left[  d( k_{I'}  \delta_{2]} l^{I'}) + \delta_{2]} \omega \right] \nonumber \\
	&-2 (D-1) \epsilon_{IJK_1K_2 ...K_{D-1}} l^I k^J m_M \wedge m^{K_2} \wedge ... \wedge m^{K_{D-1}} \nonumber \\
	& \qquad \wedge \left[  \delta_{[1} k^{K_1} d_{\Gamma^0} \delta_{2]} l^{M} + \delta_{[2} l^{M}  d_{\Gamma^0} \delta_{1]} k^{K_1} - \omega (\delta_{[1} k^{K_1}) (\delta_{2]} l^{M})  + \omega (\delta_{[2} l^{M}) (\delta_{1]} k^{K_1}) \right] \nonumber \\
	= \; & -2 \delta_{[1}\epsilon^{D-1} \wedge \left[  d( k_{I'}  \delta_{2]} l^{I'}) + \delta_{2]} \omega \right] \nonumber \\
	&-2 (D-1) \epsilon_{IJK_1[K_2 ...K_{D-1}|} l^I k^J m^M \wedge m^{K_2} \wedge ... \wedge m^{K_{D-1}}  \wedge d_{\Gamma^0} (\delta_{[1} k^{K_1}  \delta_{2]} l_{|M]})  \nonumber \\
	= \; & -2 \delta_{[1}\epsilon^{D-1} \wedge \left[  d( k_{I'}  \delta_{2]} l^{I'}) + \delta_{2]} \omega \right] \nonumber \\
	&+4 \epsilon_{JK_1K_2 ...K_{D-1}M} l^{[I} k^{J]} m^M \wedge m^{K_2} \wedge ... \wedge m^{K_{D-1}}  \wedge d_{\Gamma^0} (\delta_{[1} k^{K_1}  \delta_{2]} l_{I})  \nonumber \\
	&-2 \epsilon_{IJMK_2 ...K_{D-1}} l^I k^J m^M \wedge m^{K_2} \wedge ... \wedge m^{K_{D-1}}  \wedge d(\delta_{[1} k^{K_1}  \delta_{2]} l_{K_1})  \nonumber \\
	= \; & -2 \delta_{[1}\epsilon^{D-1} \wedge \left[  d( k_{I'}  \delta_{2]} l^{I'}) + \delta_{2]} \omega \right] \nonumber \\
	&-2 \epsilon_{JK_1MK_2 ...K_{D-1}} l^{J} k^{K_1} m^M \wedge m^{K_2} \wedge ... \wedge m^{K_{D-1}}  \wedge d_{\Gamma^0} (l_{N} \delta_{[1} k^{N} k^{I} \delta_{2]} l_{I})  \nonumber \\
	&-2 \epsilon^{D-1} \wedge d (\delta_{[1} k^{K_1}  \delta_{2]} l_{K_1})  \nonumber \\
	= \; & -2 \delta_{[1}\epsilon^{D-1} \wedge \left[  d( k_{I'}  \delta_{2]} l^{I'}) + \delta_{2]} \omega \right] +2 \epsilon^{D-1}  \wedge d_{\Gamma^0} ( (k^{N} \delta_{[1} l_{N}) (k^{I} \delta_{2]} l_{I})) -2 \epsilon^{D-1} \wedge d (\delta_{[1} k^{K_1}  \delta_{2]} l_{K_1}) \nonumber \\
	= \; & -2 d\left[ (\delta_{[1}\epsilon^{D-1}  k_{I'}) ( \delta_{2]} l^{I'}) \right] - 2 \delta_{[1}\epsilon^{D-1} \wedge\delta_{2]}\omega \text{.}
\end{align}
Similar calculations of the same length show that for the second and third line of (\ref{eq:DeltaSigma}) contracted with (\ref{eq:DeltaA}), we obtain
\begin{align}
	& - (D-1)(D-2) \epsilon_{IJK_1 ...K_{D-1}}  l^{K_1} ~ k \wedge m_L \wedge m^{K_3} \wedge ... \wedge m^{K_{D-1}} (i_{m^L} \delta_{[1} m^{K_2}) \nonumber \\
	& \qquad \wedge \left[ \delta_{2]} \Gamma^{0IJ} + \frac{2}{D-1} (\delta_{2]} l^{[I}) m^{|J]} \theta_k  \right]\nonumber \\
	= & -2(D-2) \epsilon_{IJK_1...K_{D-1}} d\left[ l^I k^J k \wedge m^M \wedge m^{K_3} \wedge ... \wedge m^{K_{D-1}} \left( (i_{m_M} \delta_{[1} m^{K_2}) \delta_{2]}l^{K_1} \right) \right] \text{,}
\end{align}
and
\begin{align} 
 & - (D-1) \epsilon_{IJK_1 ...K_{D-1}}  \left( - l^{K_1} (i_l \delta_{[1} k) + (\delta_{[1} l^{K_1}) \right) k \wedge m^{K_2} \wedge ... \wedge m^{K_{D-1}} \nonumber \\ 
 & \qquad \wedge \left[\delta \Gamma^{0IJ} + \frac{2}{D-1} \left( (\delta l^{[I}) m^{|J]} \theta_k + l^{[I} (\delta m^{|J]}) \theta_k  \right) \right] \nonumber \\
 = & -2(D-1) \epsilon_{IJK_1 ... K_{D-1}} d\left[ l^{I} k^J k \wedge m^{K_2} \wedge ... \wedge m^{K_{D-1}} \left( i_l \delta_{[1} k + k^M \delta_{[1} l_M \right) \delta_{2]} l^{K_1} \right] \text{,}
\end{align}
respectively. Summing up the three lines, we arrive at (\ref{eq:DeltaSigmaDeltaA}) rescaled by the factor $\frac{-1}{(D-1)!}$ introduced before.

\subsection{Symplectic Structure Independent of the Internal Signature}

\label{app:ChernSimonsSymplecticStructure}

In this appendix, we provide calculational details for showing that under the assumption\footnote{Note that this requirement for an UDNRIH is equivalent to restricting to histories with a fixed value of the horizon area, $\delta A_S = 0$, which can be seen as follows: Since $E^{(2n)} = f(v) \sqrt{h}$, by integrating both sides over $S$ we obtain $f(v) = f = \frac{\langle E^{(2n)}  \rangle}{A_S}$ actually is independent of $v$ since both, $A_S$ and $\langle E^{(2n)}  \rangle$ are. Therefore, we have $\delta \frac{E^{(2n)}}{\sqrt{h}} = \delta \frac{\langle E^{(2n)}  \rangle}{A_S} =  -\frac{\langle E^{(2n)} \rangle}{A_S^2} \delta A_S$, where we used that the topology of S is fixed.}  $\delta \frac{E^{(2n)}}{\sqrt{h}} = 0$ ($2n = D-1$), we have
\begin{align}
	2\frac{E^{(2n)}}{\sqrt{h}} (\delta_{[1} \tilde{s}^I)( \delta_{2]} n_I)  \qquad &  \label{eq:a3c1} \\
	=  n \epsilon^{IJKLM_1N_1 ... M_{n-1} N_{n-1}} &\epsilon^{\alpha \beta \alpha_1 \beta_1...\alpha_{n-1} \beta_{n-1}} \left(\delta_{[1} \Gamma^0_{\alpha IJ}\right) \left(\delta_{2]} \Gamma^0_{\beta KL}\right) R^0_{\alpha_1\beta_1M_1N_1} ... R^0_{\alpha_{n-1}\beta_{n-1}M_{n-1}N_{n-1}}  \text{,} \nonumber
\end{align}
where $\Gamma^0_{\alpha IJ}$ is the generalised hybrid connection and $R^{0}_{\alpha \beta IJ}$ the corresponding curvature tensor which are given in appendix \ref{sec:GeneralisedHybridConnection}.

Starting with (\ref{eq:a3c1}), we first calculate
\begin{align}	
\delta \left( \frac{E^{(2n)}}{\sqrt{h}} \right) &= \delta \left( \frac{1}{h} \epsilon^{\alpha_1 \beta_1 ... \alpha_n \beta_n} \epsilon^{\gamma_1\delta_1 ... \gamma_n \delta_n} R_{\alpha_1 \beta_1 \gamma_1 \delta_1} ... R_{\alpha_n \beta_n \gamma_n \delta_n} \right) \nonumber \\
	&=- (\delta \log h)   \frac{E^{(2n)}}{\sqrt{h}} \nonumber  \\
	&\quad + \frac{n}{h} \epsilon^{\alpha_1 \beta_1 ... \alpha_n \beta_n} \epsilon^{\gamma_1 \delta_1 ... \gamma_n \delta_n} \left(-2 h_{\delta_1 \epsilon_1} D_{\alpha_1} \delta \Gamma_{\beta_1 \gamma_1}^{\epsilon_1} + R_{\alpha_1 \beta_1 \gamma_1}\m^{\epsilon_1} \delta h_{\delta_1\epsilon_1} \right) R_{\alpha_2 \beta_2 \gamma_2 \delta_2} ... R_{\alpha_n \beta_n \gamma_n \delta_n} \nonumber \\
	&=- (\delta \log h)   \frac{E^{(2n)}}{\sqrt{h}} \nonumber  \\
	&\quad - \frac{2n}{h} \epsilon^{\alpha_1 \beta_1 ... \alpha_n \beta_n} \epsilon^{\gamma_1\delta_1 ... \gamma_n \delta_n} \left( D_{\alpha_1} D_{\gamma_1} \delta h_{\beta_1 \delta_1} \right) R_{\alpha_2 \beta_2 \gamma_2 \delta_2} ... R_{\alpha_n \beta_n \gamma_n \delta_n} \nonumber \\
	&\quad + \frac{n}{h} \epsilon^{\alpha_1 \beta_1 ...\alpha_n \beta_n} \epsilon^{\gamma_1 \delta_1 ... \gamma_n \delta_n} R_{\alpha_1 \beta_1 \gamma_1}\m^{\epsilon_1} \left(\delta h_{\delta_1\epsilon_1}\right) R_{\alpha_2 \beta_2 \gamma_2 \delta_2} ... R_{\alpha_n \beta_n \gamma_n \delta_n} \nonumber \\
	&=- (\delta \log h)   \frac{E^{(2n)}}{2\sqrt{h}} \nonumber  \\
	&\quad - \frac{2n}{h} \epsilon^{\alpha_1 \beta_1 ...\alpha_n \beta_n} \epsilon^{\gamma_1 \delta_1 ... \gamma_n \delta_n} \left( D_{\alpha_1} D_{\gamma_1} \delta h_{\beta_1 \delta_1} \right) R_{\alpha_2 \beta_2 \gamma_2 \delta_2} ... R_{\alpha_n \beta_n \gamma_n \delta_n}\text{.} \label{eq:VarEuler}
\end{align}
In the second line, we just explicitly wrote down all variations appearing using $\delta R_{\alpha \beta \gamma}\m^{\delta} = - 2 D_{[\alpha} \delta \Gamma_{\beta]\gamma}^{\delta}$. In the third, we used $\delta \Gamma_{\alpha\beta}^{\gamma} = \frac{1}{2} h^{\gamma\delta} \left(2 D_{(\alpha}\delta h_{\beta)\delta} - D_{\delta} \delta h_{\alpha\beta} \right)$ and in the last step, we used
\begin{align}
\frac{n}{h} \epsilon^{\alpha_1 \beta_1 ... \alpha_n \beta_n} \epsilon^{\gamma_1\delta_1 ... \gamma_n \delta_n}R_{\alpha_1 \beta_1 \gamma_1}\m^{\epsilon_1} \left(\delta h_{\delta_1\epsilon_1}\right) R_{\alpha_2 \beta_2 \gamma_2 \delta_2} ... R_{\alpha_n \beta_n \gamma_n \delta_n} = \frac{E^{(2n)}}{2\sqrt{h}} \left(\delta \log h\right) \text{.} \label{eq:a3z2}
\end{align}
This last identity can be verified as follows:
\begin{align}
 & \frac{n}{h}  \epsilon^{\alpha_1 \beta_1 ... \alpha_n \beta_n} \epsilon^{\gamma_1\delta_1 ... \gamma_n \delta_n}R_{\alpha_1 \beta_1 \gamma_1}\m^{\epsilon_1} \left(\delta h_{\delta_1\epsilon_1}\right) R_{\alpha_2 \beta_2 \gamma_2 \delta_2} ... R_{\alpha_n \beta_n \gamma_n \delta_n}  \nonumber \\
= & -\frac{n}{h} \epsilon^{\alpha_1 \beta_1 ... \alpha_n \beta_n} \epsilon^{[\gamma_1|\delta_1 ... \gamma_n \delta_n} \left(\delta h^{\epsilon_1|\zeta_1]}\right) h_{\delta_1\epsilon_1} R_{\alpha_1 \beta_1 \gamma_1 \zeta_1} R_{\alpha_2 \beta_2 \gamma_2 \delta_2} ... R_{\alpha_n \beta_n \gamma_n \delta_n} \nonumber \\
= & \frac{n}{2h} \epsilon^{\alpha_1 \beta_1 ... \alpha_n \beta_n} \left[ \left(\delta h^{\epsilon_1\delta_1}\right) \epsilon^{\gamma_2\delta_2 ... \delta_n \zeta_1 \gamma_1} + 2(n-1) \left(\delta h^{\epsilon_1\gamma_2}\right) \epsilon^{\delta_2 ... \delta_n \zeta_1 \gamma_1\delta_1}\right] \nonumber \\
 & \hspace{8cm} \times h_{\delta_1\epsilon_1}  R_{\alpha_1 \beta_1 \gamma_1 \zeta_1} R_{\alpha_2 \beta_2 \gamma_2 \delta_2} ... R_{\alpha_n \beta_n \gamma_n \delta_n} \nonumber \\
=& \frac{n E^{(2n)}}{2 \sqrt{h}} \left(\delta \log h\right) -\frac{n(n-1)}{h} \epsilon^{\alpha_1 \beta_1 ...\alpha_n \beta_n}  \epsilon^{\gamma_1 \delta_1... \gamma_n \delta_n} R_{\alpha_1 \beta_1 \gamma_1}\m^{\epsilon_1} \left(\delta h_{\delta_1 \epsilon_1}\right) R_{\alpha_2 \beta_2 \gamma_2 \delta_2} ... R_{\alpha_n \beta_n \gamma_n \delta_n} \text{,} & \label{eq:a3z1}
\end{align}
where in the first step, we used $h^{\epsilon \zeta}\delta h_{\delta \epsilon} = - h_{\delta \epsilon} \delta h^{\epsilon \zeta}$, then we added zero by adding all terms necessary that the expression in the second line becomes antisymmetric in $\gamma_1,\delta_1,...,\gamma_n,\delta_n, \zeta_1$ and immediately subtracting them again. Since these are $D$ indices in dimension $D-1$, the antisymmetrisation vanishes and we are left with the subtracted terms. The first of these gives, using $h_{\delta \epsilon} \delta h^{\delta \epsilon} = - \delta \log h$, the first term in the fourth line, while the remaining ones, after renaming indices, reproduce up to numerical factors the expression we started with. Comparing the first and the last line of (\ref{eq:a3z1}), one easily infers (\ref{eq:a3z2}).

Next, we will calculate $\delta \Gamma^0_{\alpha IJ}$:
\begin{align}
\delta \Gamma^0_{\alpha IJ} &= \left(\delta \Gamma^0_{\alpha KL}\right) \eta^K\m_I\eta^L\m_J \nonumber 
\\ &=  \left(\delta \Gamma^0_{\alpha KL}\right) \left(\bar{\bar{\eta}}^K\m_I + \zeta n^K n_I+ s^K s_I\right) \left(m^{\beta L} m_{\beta J}+ \zeta n^L n_J+ s^L s_J\right) \nonumber
\\ &= \bar{\bar{\eta}}^K\m_{[I} \Big[\left(\left(\delta D^{\Gamma^0}_{\alpha} m^{\beta}\m_K\right) - \left(D^{\Gamma^0}_{\alpha} \delta m^{\beta}\m_K\right) - \left(\delta \Gamma_{\alpha \gamma}^{\beta} \right) m^{\gamma}_K \right) m_{\beta |J]} \nonumber 
\\ &\qquad \qquad +2\zeta \left(\left(\delta D^{\Gamma^0}_{\alpha} n_K\right) - \left(D^{\Gamma^0}_{\alpha} \delta n_K\right)\right)n_{|J]} +2 \left(\left(\delta D^{\Gamma^0}_{\alpha} s_K\right) - \left(D^{\Gamma^0}_{\alpha} \delta s_K\right)\right)s_{|J]} \Big] \nonumber
\\ &\quad +\zeta n_{[I} s_{J]} \left[ n^K \left(\left(\delta D^{\Gamma^0}_{\alpha} s_K\right) - \left(D^{\Gamma^0}_{\alpha} \delta s_K\right)\right) - s^K \left(\left(\delta D^{\Gamma^0}_{\alpha} n_K\right) - \left(D^{\Gamma^0}_{\alpha} \delta n_K\right)\right)\right] \nonumber
\\ &= \bar{\bar{\eta}}^K\m_{[I} m_{\beta |J]} \left[- \left(D^{\Gamma^0}_{\alpha} \delta m^{\beta}\m_K\right) - \left(\delta \Gamma_{\alpha \gamma}^{\beta} \right) m^{\gamma}_K \right] -2\zeta \bar{\bar{\eta}}^K\m_{[I} n_{J]} \left(D^{\Gamma^0}_{\alpha} \delta n_K\right) -2 \bar{\bar{\eta}}^K\m_{[I} s_{J]}\left(D^{\Gamma^0}_{\alpha} \delta s_K\right) \nonumber
\\ &\quad -2\zeta n_{[I} s_{J]} \left(D^{\Gamma^0}_{\alpha} n^K \delta s_K\right) \label{eq:a3delta1}
\end{align}
where in the second step we used $\eta_{IJ} = \bar{\bar{\eta}}_{IJ} + \zeta n_I n_J + s_I s_J$ and $\bar{\bar{\eta}}_{IJ} = m^{\beta}\m_I m_{\beta J}$, in the third that $(\delta \Gamma^0_{\alpha IJ}) m^{\beta J} = \left(\delta D^{\Gamma^0}_{\alpha} m^{\beta}\m_I\right) - \left(D^{\Gamma^0}_{\alpha} \delta m^{\beta}\m_I\right) - \left(\delta \Gamma_{\alpha \gamma}^{\beta} \right) m^{\gamma}_I$ and corresponding equations for $n,s$, and finally in the fourth step we used that $\Gamma^0_{\alpha IJ}$ annihilates the hybrid vielbein and $n,s$. This way of expressing $\delta \Gamma^0_{ \alpha IJ}$ is convenient for several reasons. First of all, we explicitly separated the {\it (bar bar)}, {\it (bar $n$)}, {\it (bar $s$)} and {\it ($n$ $s$)} terms. Since the two variations of $\Gamma^0_{\alpha IJ}$ in (\ref{eq:a3c1}) are contracted with an $\epsilon$, which is {\it bar} projected on all other indices (remember $R^0_{\alpha \beta IJ} = \bar{R}^0_{\alpha \beta IJ}$, cf. \ref{app:Connections}), the only contributions will come from {\it (bar bar)} $\cdot$ {\it ($n$ $s$)} and {\it (bar $n$)} $\cdot $ {\it (bar $s$)} terms. Secondly, many of the terms are such that covariant derivates $D^{\Gamma^0}_{\alpha}$ appear explicitly. This simplifies further manipulations like partial integrations, since almost all appearing objects are annihilated by $D^{\Gamma^0}_{\alpha}$. Furthermore, since $S$ already is a boundary, no boundary terms appear when partially integrating. Using (\ref{eq:a3delta1}), we thus find
\begin{align}
 & n \epsilon^{IJKLM_2N_2 ... M_n N_n} \epsilon^{\alpha \beta \alpha_2 \beta_2...\alpha_n \beta_n} \left(\delta_{[1} \Gamma^0_{\alpha IJ}\right) \left(\delta_{2]} \Gamma^0_{\beta KL}\right) R^0_{\alpha_2\beta_2M_2N_2} ... R^0_{\alpha_n\beta_nM_nN_n} \nonumber \\
= & n \epsilon^{IJKLM_2N_2 ... M_n N_n} \epsilon^{\alpha \beta \alpha_2\beta_2...\alpha_n \beta_n} R^0_{\alpha_2\beta_2M_2N_2} ... R^0_{\alpha_n\beta_nM_nN_n} \nonumber \\
 & \times \left[8\zeta n_I \bar{\bar{\eta}}_{JJ'} \left(D^{\Gamma^0}_{\alpha} \delta_{[1} n^{J'}\right) s_K \bar{\bar{\eta}}_{LL'} \left(D^{\Gamma^0}_{\beta}\delta_{2]}s^{L'}\right)\right. \nonumber \\
 &  \qquad \left. + 4 \zeta \bar{\bar{\eta}}^{I'}\m_{I} m_{\delta J} \left(\left(D^{\Gamma^0}_{\alpha} \delta_{[1} m^{\delta}\m_{I'}\right) + \left(\delta_{[1} \Gamma_{\alpha \gamma}^{\delta} \right) m^{\gamma}_{I'} \right) n_K s_L \left(D^{\Gamma^0}_{\beta} n^P \delta_{2]} s_P\right) \right] \nonumber \\
= & -\frac{4n}{\sqrt{h}} \epsilon^{\gamma \delta \gamma_2\delta_2...\gamma_n\delta_n} \epsilon^{\alpha \beta \alpha_2\beta_2...\alpha_n\beta_n} R_{\alpha_2\beta_2\gamma_2\delta_2} ... R_{\alpha_n\beta_n\gamma_n\delta_n} \nonumber \\
& \times \left[2 m_{\gamma J}\left(D^{\Gamma^0}_{\alpha} \delta_{[1} n^{J}\right) m_{\delta L} \left(D^{\Gamma^0}_{\beta} \delta_{2]}s^{L}\right) - \left(D^{\Gamma^0}_{\gamma} \delta_{[1} h_{\alpha \delta}\right) \left(D_{\beta} n^P \delta_{2]} s_P\right) \right] \nonumber \\
= & -\frac{4n}{h} \epsilon^{\gamma \delta \gamma_2\delta_2...\gamma_n\delta_n} \epsilon^{\alpha \beta \alpha_2\beta_2...\alpha_n\beta_n} R_{\alpha_2\beta_2\gamma_2\delta_2} ... R_{\alpha_n\beta_n\gamma_n\delta_n} \nonumber \\
& \times \left[2 m_{\gamma J}\left(D^{\Gamma^0}_{\alpha} \delta_{[1} n^{J}\right) m_{\delta L} \left(D^{\Gamma^0}_{\beta} \delta_{2]}\tilde{s}^{L}\right) -  \left(D^{\Gamma^0}_{\alpha} D^{\Gamma^0}_{\gamma} \delta_{[1} h_{\beta \delta}\right) n^P \left(\delta_{2]} \tilde{s}_P\right) \right] \nonumber \\
= & -\frac{8n}{h} \epsilon^{\gamma \delta\gamma_2\delta_2...\gamma_n\delta_n} \epsilon^{\alpha \beta\alpha_2\beta_2...\alpha_n\beta_n} R_{\alpha_2\beta_2\gamma_2\delta_2} ... R_{\alpha_n\beta_n\gamma_n\delta_n} m_{\gamma J}\left(D^{\Gamma^0}_{\alpha} \delta_{[1} n^{J}\right) m_{\delta L} \left(D^{\Gamma^0}_{\beta}\delta_{2]}\tilde{s}^{L}\right) \nonumber \\
 & - \left[ 2 \left(\delta_{[1} \frac{E^{(2n)}}{\sqrt{h}}\right) + \frac{E^{(2n)}}{\sqrt{h}} \left(\delta_{[1} \log h\right) \right] n^P \left(\delta_{2]} \tilde{s}_P\right) \text{.} \label{eq:VarGammaGamma}
\end{align}
In the third line, note that the term containing $D^{\Gamma^0}_{\alpha} \delta m^{\beta}_{I}$ vanishes, since when partially integrating, we obtain a term of the form $\left(D^{\Gamma^0}_{[\alpha} D^{\Gamma^0}_{\beta]} n^P \delta s_P\right)$, which vanishes due to torsion-freeness. In the second step, we used $\epsilon^{IJM_1N_1 ... M_nN_n} n_I s_J m^{\gamma_1}\m_{M_1} m^{\delta_1}\m_{N_1} ... m^{\gamma_n}\m_{M_n} m^{\delta_n}\m_{N_n} = \frac{\zeta}{\sqrt{h}} \epsilon^{\gamma_1\delta_1 ... \gamma_n \delta_n}$ and again $\delta \Gamma_{\alpha \beta}^{\gamma} = \frac{1}{2} g^{\gamma \delta} \left(2 D_{(\alpha}\delta h_{\beta)\delta} - D_{\delta} \delta h_{\alpha\beta} \right)$. In the third step, we densitised $s^I$ (note that $s^I$ is always contracted such that variations on the density $\sqrt{h}$ drop out), partially integrated in the last summand and interchanged the indices $\alpha$ and $\beta$. In the fourth step, we replaced the second summand in square brackets using (\ref{eq:VarEuler}). 

Now we will have a closer look at the left hand side of (\ref{eq:a3c1}).
\begin{align}
	2\frac{E^{(2n)}}{\sqrt{h}} (\delta_{[1} \tilde{s}^I )(\delta_{2]} n_I) &= 2 E^{(2n)} (\delta_{[1} s^I)( \delta_{2]} n_I) + \frac{E^{(2n)}}{\sqrt{h}} \tilde{s}_I (\delta_{[1} \log h)( \delta_{2]} n^I) \nonumber 
	\\ &= 2 E^{(2n)} (\delta_{[1} s^I)( \delta_{2]} n_I) + \frac{E^{(2n)}}{\sqrt{h}} n^I (\delta_{[1}\tilde{s}_I )( \delta_{2]} \log h) \text{.} \label{eq:a3xxx}
\end{align}
Here, in the first step we varied $s^I$ and the density $\sqrt{h}$ independently. In the second step, we interchanged the variations and used $s_I \delta n^I = - n^I \delta s_I$ in the second summand. For the first summand, we find
\begin{align}
	& \;\;\;\;\; 2 E^{(2n)} (\delta_{[1} s^I)( \delta_{2]} n_I) \nonumber 
	\\ &= -\frac{2}{\sqrt{h}} \epsilon^{\alpha_1\beta_1...\alpha_n\beta_n} \epsilon^{\gamma_1\delta_1 ... \gamma_n\delta_n} R_{\alpha_1\beta_1\gamma_1\delta_1} ... R_{\alpha_n\beta_n\gamma_n\delta_n} (\delta_{[1} n^I)( \delta_{2]} s_I) \nonumber
	\\ &= -2\zeta  \epsilon^{\alpha_1\beta_1...\alpha_n\beta_n} \epsilon^{IJK_1L_1 ... K_nL_n} R^0_{\alpha_1\beta_1K_1 L_1} ... R^0_{\alpha_n\beta_nK_nL_n} n_I s_J (\delta_{[1} n^M)( \delta_{2]} s_M) \nonumber
	\\ &= -4\zeta \epsilon^{\alpha_1\beta_1...\alpha_n\beta_n} \epsilon^{IJK_1L_1 ... K_nL_n} R^0_{\alpha_1\beta_1K_1 L_1}  ...R^0_{\alpha_n\beta_nK_nL_n}  n_I s_{[J} (\delta_{[1} n^M) (\delta_{2]} s_{|M]}) \nonumber 
	\\ &=-4\zeta  \epsilon^{\alpha_1\beta_1...\alpha_n\beta_n}(\delta_{[1} n^{[M}) \epsilon^{I|J]K_1L_1 ... K_nL_n} R^0_{\alpha_1\beta_1K_1 L_1} ... R^0_{\alpha_n\beta_nK_nL_n}  n_I s_{J} (\delta_{2]} s_{M}) \nonumber
	\\ &= -2\zeta \epsilon^{\alpha_1\beta_1...\alpha_n\beta_n} \left((\delta_{[1} n^{I}) \epsilon^{K_1L_1 ... K_nL_nMJ} + 2n (\delta_{[1} n^{K_1}) \epsilon^{L_1 ... K_nL_nMJI}\right) \nonumber
	\\ & \quad \times R^0_{\alpha_1\beta_1K_1 L_1}  ... R^0_{\alpha_n\beta_nK_nL_n}  n_I s_{J} (\delta_{2]} s_{M}) \nonumber
	\\ &= -4\zeta n \epsilon^{\alpha_1\beta_1...\alpha_n\beta_n} \epsilon^{L_1 ... K_nL_nMJI} R^0_{\alpha_1\beta_1K_1 L_1}  (\delta_{[1} n^{K_1})R^0_{\alpha_2\beta_2K_2L_2}  ... R^0_{\alpha_n\beta_nK_nL_n}  n_I s_{J} (\delta_{2]} s_{M}) \nonumber 
	\\ &= 8\zeta n \epsilon^{\alpha_1\beta_1...\alpha_n\beta_n}\epsilon^{L_1 ... K_nL_nMJI} (D^{\Gamma^0}_{\alpha_1} D^{\Gamma^0}_{\beta_1} \delta_{[1} n_{L_1}) R^0_{\alpha_2\beta_2K_2L_2}  ... R^0_{\alpha_n\beta_nK_nL_n} n_I s_{J} (\delta_{2]} s_{M}) \nonumber 
	\\ &= -\frac{8 n}{h}  \epsilon^{\alpha_1\beta_1...\alpha_n\beta_n}\epsilon^{\gamma_1\delta_1 ... \gamma_n\delta_n} R^0_{\alpha_2\beta_2\gamma_2\delta_2} ... R^0_{\alpha_n\beta_n\gamma_n\delta_n} m_{\delta_1}\m^{J} (D^{\Gamma^0}_{\beta_1} \delta_{[1} n_{J}) m_{\gamma_1}\m^{L}(D^{\Gamma^0}_{\alpha_1} \delta_{2]} \tilde{s}_{L}) \text{,}
\end{align}
which shows that (\ref{eq:a3xxx}) coincides with (\ref{eq:VarGammaGamma}) iff $\delta \left(\frac{E^{(2n)}}{\sqrt{h}} \right)= 0$. Here, in the first step, we used the defining equation for $E^{(2n)}$ and in the second step we used 
\be
\frac{\zeta}{\sqrt{h}} \epsilon^{\gamma_1\delta_1 ... \gamma_n \delta_n} = \epsilon^{IJM_1N_1 ... M_nN_n} n_I s_J m^{\gamma_1}\m_{M_1} m^{\delta_1}\m_{N_1} ... m^{\gamma_n}\m_{M_n} m^{\delta_n}\m_{N_n} \label{eq:epsilonepsilon}
\ee
 and (\ref{eq:R0identity}). In the third step, we antisymmetrise in the lower pair of indices $J$ and $M$. Note that the additional term vanishes since $s^J \delta s_J = 0$ and the epsilon tensor enforces $\delta s_J$ to be projected into that direction. The fifth line is exactly the same as the fourth, we just moved $\delta n^M$ to the front and antisymmetrised the upper indices $J$ and $M$ instead of the lower ones. Now we again antisymmetrise the $D+2$ upper indices $M, I, J, K_1, L_1,..., K_n L_n$, which gives zero, and subtract the term we added for antisymmetrisation again. The first of these, the first summand in the round brackets in line 6, gives zero due to $n^I \delta n_I = 0$. The others all give the same term of the form $R^0_{\alpha \beta KL} \delta n^L = 2D^{\Gamma^0}_{[\alpha} D^{\Gamma^0}_{\beta]} \delta n_K$, which we used in the second to last line. One more integration by parts in the last line, again using (\ref{eq:epsilonepsilon}) and densitising $s^I$ gives the final result.

\subsection{Symplectic Structure for the SO$(4)$ based Beetle-Engle Connection}

\label{app:BeetleEngleConnection}

For $D=3$, we will show that one can bypass the restriction to spherically symmetric isolated horizons in complete analogy to the treatment of Beetle and Engle \cite{BeetleGenericIsolatedHorizons}, 
\begin{align}
	2\langle E^{(2)}\rangle (\delta_{[1} \tilde{s}^I) (\delta_{2]} n_I ) =   \epsilon^{IJKL} \epsilon^{\alpha \beta} \left(\delta_{[1} A_{\alpha IJ}\right) \left(\delta_{2]} A_{\beta KL}\right) \text{,} \label{eq:a3c2}
\end{align}
where $\langle E^{(2)} \rangle := \int_{S} d^2x E^{(2)}$ coincides, up to constant factors, with the Euler characteristic of the intersection of the Isolated Horizon with the spatial slices, and $A_{\alpha IJ}$ was defined in (\ref{eq:BeetleEngleConnectionSO(4)}). The assumption $\delta \frac{E^{(2)}}{\sqrt{h}} = 0$ is then replaced by $\delta \langle E^{(2)} \rangle = 0$, which however is already enforced by our choice of topology of the horizon.

To prove (\ref{eq:a3c2}), we start by noting that
\begin{align}
&  \epsilon^{IJKL} \epsilon^{\alpha\beta} \left(\delta_{[1} A_{\alpha IJ}\right) \left(\delta_{2]} A_{\beta KL}\right) \nonumber 
\\ &\qquad =  \epsilon^{IJKL} \epsilon^{\alpha \beta} \left[ \left(\delta_{[1} \Gamma^0_{\alpha IJ}\right) \left(\delta_{2]} \Gamma^0_{\beta KL}\right) + 2 \left(\delta_{[1} \Gamma^0_{\alpha IJ}\right) \left(\delta_{2]} K_{\beta KL}\right) + \left(\delta_{[1} K_{\alpha IJ}\right) \left(\delta_{2]} K_{\beta KL}\right) \right] \nonumber
 \\ &\qquad =: A + B + C\text{,} 
\end{align}
where we introduced the abbreviations $A$, $B$, $C$ for the three summands. The first summand in square brackets is, up to factors, the restriction to $D=3$ of what we just calculated above,
\begin{align}
A &=  \epsilon^{IJKL} \epsilon^{\alpha  \beta} \left(\delta_{[1} \Gamma^0_{\alpha IJ}\right) \left(\delta_{2]} \Gamma^0_{\beta KL}\right) \nonumber \\
&=  \frac{2E^{(2)}}{\sqrt{h} } (\delta_{[1} \tilde{s}^I )(\delta_{2]} n_I ) - 2 \left(\delta_{[1} \frac{E^{(2)}}{\sqrt{h}}\right) n^P \left(\delta_{2]} \tilde{s}_P\right) \text{.} \label{eq:a3AA}
\end{align}
Next, we need to calculate
\begin{align}
	\delta K_{\alpha IJ} &= \delta \left(2 m_{\alpha [I} m_{\beta |J]} h^{\beta \gamma} (D_{\gamma} \psi) \right) \nonumber
	\\ &= 2m_{\alpha[I} m^{\beta}\m_{J]} (D_{\beta} \delta \psi) + 4 (\delta m_{[\alpha|K}) m_{\beta][J}\bar{\bar{\eta}}^K\m_{I]}  h^{\beta\gamma} (D_{\gamma} \psi) + 2 m_{\alpha [I} m_{\beta|J]} (\delta h^{\beta \gamma}) (D_{\gamma} \psi) \nonumber 
	\\ &\quad + 4\zeta (\delta m_{[\alpha |K}) m_{\beta][J} n^K n_{I]}  h^{\beta \gamma} (D_{\gamma} \psi) +4 (\delta m_{[\alpha |K}) m_{\beta][J} s^K s_{I]}  h^{\beta \gamma} (D_{\gamma} \psi)  \text{,} \label{eq:a3deltaK}
\end{align}
where we again splitted the {\it (bar bar)} terms (second line) from the {\it (bar $n$)}, {\it (bar $s$)} terms (third line). Since no {\it ($n$ $s$)} terms appear, we find for $C$
\begin{align}
	C &= \epsilon^{IJKL} \epsilon^{\alpha \beta} \left(\delta_{[1} K_{\alpha IJ}\right) \left(\delta_{2]} K_{\beta KL}\right) \nonumber
	\\ &= 32\zeta \epsilon^{IJKL} \epsilon^{\alpha  \beta} (\delta_{[1} m_{[\alpha |M})m_{\gamma]J} n^M n_I h^{\gamma\epsilon} (D_{\epsilon} \psi) (\delta_{2]} m_{[\beta|N})m_{\delta]L} s^N s_K h^{\delta \zeta} (D_{\zeta} \psi) \nonumber
	\\ &= -32 \sqrt{h}  \epsilon^{\alpha \beta} (\delta_{[1} m_{[\alpha |M})\epsilon_{\gamma][\delta}  (\delta_{2]} m_{\beta]N}) n^M h^{\gamma\epsilon} (D_{\epsilon} \psi) s^N h^{\delta \zeta} (D_{\zeta} \psi) \nonumber
	\\ &=0 \text{,} \label{eq:a3CC}
\end{align}
where in the second step we used 
\be
\epsilon^{IJKL} n_I s_J m_{\alpha K} m_{\beta L} = \zeta \sqrt{h} \epsilon_{\alpha  \beta} \label{eq:epsilonepsilon2}
\ee
and the last equality is easily obtained when explicitly writing out all antisymmetrisations. For $B$, we find using (\ref{eq:a3delta1}) and (\ref{eq:a3deltaK})
\begin{align}
	B &= 2 \epsilon^{IJKL} \epsilon^{\alpha \beta}\left(\delta_{[1} \Gamma^0_{\alpha IJ}\right) \left(\delta_{2]} K_{\beta KL}\right) \nonumber
	\\ &= 2  \epsilon^{IJKL} \epsilon^{\alpha \beta}
		\begin{aligned}[t]
		& \left\{ \left[-2\zeta n_{I} s_{J} \left(D_{\alpha } n^M \delta_{[1} s_M\right)\right] \right.
			\begin{aligned}[t]
			 \left[2m_{\beta K} m^{\gamma}\m_{L} D_{\gamma} \delta_{2]} \psi + 4 (\delta_{2]} m_{[\beta |N}) m_{\gamma ]L}\bar{\bar{\eta}}^N\m_{K}  h^{\gamma \delta } D_{\delta} \psi \right. \quad& \nonumber
			\\  \left. \quad + 2 m_{\beta K} m_{\gamma L} (\delta_{2]} h^{\gamma \delta }) D_{\delta} \psi\right] & \nonumber 
			\end{aligned} \nonumber
		\\ &\left. \quad + \left[-2\zeta \bar{\bar{\eta}}^M\m_{[I} n_{J]} \left(D_{\alpha} \delta_{[1} n_M\right) \right] \left[4 (\delta_{2]} m_{[\beta |N}) m_{\gamma ]L} s^N s_{K}  h^{\gamma \delta} D_{\delta} \psi\right] \right. \nonumber 
		\\ &\left. \quad + \left[-2 \bar{\bar{\eta}}^M\m_{[I} s_{J]}\left(D_{\alpha} \delta_{[1} s_M\right)\right] \left[4\zeta (\delta_{2]} m_{[\beta |N}) m_{\gamma ]L} n^N n_{K}  h^{\gamma \delta} D_{\delta} \psi\right]  \right\} \nonumber
		\end{aligned}
	\\ &=-8\sqrt{h} \epsilon^{\alpha \beta } 
		\begin{aligned}[t]
		& \left\{ \left(D_{\alpha} n^M \delta_{[1} s_M\right) \left[\epsilon_{\beta \gamma } h^{\gamma \delta} D_{\delta} \delta_{2]} \psi + 2 (\delta_{2]} m_{[\beta |N}) \epsilon_{\epsilon |\gamma ]} m^{\epsilon N}  h^{\gamma \delta} D_{\delta} \psi +  \epsilon_{\beta \gamma }(\delta_{2]} h^{\gamma \delta}) D_{\delta} \psi\right]  \right. \nonumber 
		\\ &\left. \quad + 2 m^{\epsilon M} \left(D_{\alpha} \delta_{[1} n_M\right) (\delta_{2]} m_{[\beta |N}) \epsilon_{\epsilon|\gamma ]} s^N h^{\gamma \delta} D_{\delta} \psi \right. \nonumber 
		\\ &\left. \quad - 2m^{\epsilon M} \left(D_{\alpha} \delta_{[1} s_M\right) (\delta_{2]} m_{[\beta |N}) \epsilon_{\epsilon|\gamma ]} n^N h^{\gamma \delta} D_{\delta} \psi \right\} \nonumber
		\end{aligned}
	\\ &= -8\sqrt{h} 
		\begin{aligned}[t]
		& \left\{ \left(D_{\alpha} n^M \delta_{[1} s_M\right) \left[- D^{\alpha} \delta_{2]} \psi +  2(\delta_{2]} m_{\beta N}) m^{[\alpha |N}  D^{\beta ]} \psi - (\delta_{2]} m_{\beta N}) m^{\alpha N}  D^{\beta}  \psi - (\delta_{2]} h^{\alpha \delta}) D_{\delta} \psi\right]  \right. \nonumber 
		\\ &\left. \quad + 2  \left(D_{\alpha} \delta_{[1} n_M\right) (\delta_{2]} m_{\beta N}) s^N m^{[\alpha |M} D^{\beta ]} \psi - \left(D_{\alpha} \delta_{[1} n_M\right) (\delta_{2]} m_{\beta N}) s^N  m^{\alpha M} D^{\beta}  \psi \right. \nonumber 
		\\ &\left. \quad - 2\left(D_{\alpha} \delta_{[1} s_M\right) (\delta_{2]} m_{\beta N}) n^N  m^{[\alpha |M} D^{\beta ]} \psi +  \left(D_{\alpha} \delta_{[1} s_M\right) (\delta_{2]} m_{\beta N}) n^N m^{\alpha M} D^{\beta}  \psi \right\} \nonumber
		\end{aligned}
	\\ &= -8\sqrt{h} 
		\begin{aligned}[t]
		& \left\{ \left(D_{\alpha} n^M \delta_{[1} s_M\right) \left[- D^{\alpha} \delta_{2]} \psi - (\delta_{2]} m_{\beta N}) m^{\beta N}  D^{\alpha} \psi - (\delta_{2]} h^{\alpha \delta}) D_{\delta} \psi\right]  \right. \nonumber 
		\\ &\left. \quad +   \left(D_{\alpha} \delta_{[1} n_M\right) (\delta_{2]}  s^N) \bar{\bar{\eta}}_N\m^M D^{\alpha} \psi - \left(D_{\alpha} \delta_{[1} s_M\right) (\delta_{2]} n^N) \bar{\bar{\eta}}_N\m^M D^{\alpha} \psi \right\} \nonumber
		\end{aligned}
	\\ &=- 8\sqrt{h} 
		\begin{aligned}[t]
		& \left\{ \left(n^M \delta_{[1} s_M\right) \left[D_{\alpha} D^{\alpha} \delta_{2]} \psi + D_{\alpha}((\delta_{2]} \log \sqrt{h}) D^{\alpha} \psi) + D_{\alpha}((\delta_{2]} h^{\alpha \delta}) D_{\delta} \psi)\right]  \right. \nonumber 
		\\ &\left. \quad +  \left(D_{\alpha} \delta_{[1} n_M \delta_{2]}  s^N\right) \bar{\bar{\eta}}_N\m^M D^{\alpha} \psi \right\} \nonumber
		\end{aligned}
	\\ &= -8\sqrt{h} \left\{ \left(n^M \delta_{[1} s_M\right) \right.
			\begin{aligned}[t]
			& \left[ \Delta \delta_{2]} \psi + (D_{\alpha} \delta_{2]} \log \sqrt{h}) D^{\alpha} \psi + (\delta_{2]} \log \sqrt{h}) \Delta \psi \right. \quad \nonumber
			\\& \quad \left. \left. - (\delta_{2]} h_{\alpha \delta}) D^{\alpha} D^{\delta} \psi - (D^{\alpha}\delta_{2]} h_{\alpha \delta}) D^{\delta} \psi \right]  -   (\delta_{[1} n_M)(\delta_{2]}  s^M) \Delta \psi \right\}  \nonumber
			\end{aligned}
	\\ &= -8\sqrt{h} \left\{ \left(n^M \delta_{[1} s_M\right) (\delta_{2]}  \Delta \psi) - \frac{1}{\sqrt{h}}(\delta_{[1}n_M) \left[\sqrt{h}(\delta_{2]}  s^M) + s^M (\delta_{2]} \sqrt{h}) \right] \Delta \psi \right\} \nonumber
	\\ &= -8 \left\{ \left(n^M \delta_{[1} \tilde{s}_M\right) (\delta_{2]}  \Delta \psi) - (\delta_{[1}n_M) (\delta_{2]} \tilde{s}^M) \Delta \psi \right\} \text{,} \nonumber
	\\ \intertext{and since we assumed that $\Delta \psi = \frac{1}{4}\left(\frac{E^{(2)}}{\sqrt{h}} - \langle E^{(2)}\rangle\right)$ and $\delta \langle E^{(2)}\rangle = 0$, we find}
	&=-2 \left\{ \left(n^M \delta_{[1} \tilde{s}_M\right) \left(\delta_{2]} \frac{E^{(2)}}{\sqrt{h}}\right) + (\delta_{[1} \tilde{s}_M) (\delta_{2]} n^M)\left(\frac{E^{(2)}}{\sqrt{h}} - \langle E^{(2)}\rangle\right) \right\} \text{.} \label{eq:a3BB}
\end{align}
Here, in the second line, we inserted the expressions for $\delta \Gamma^0\m_{\alpha IJ}$ and $\delta K_{\alpha IJ}$ (\ref{eq:a3delta1}, \ref{eq:a3deltaK}). Note that since $\delta K_{\alpha IJ}$ does not contain {\it ($n$ $s$)} terms, the {\it (bar bar)} terms of $\delta \Gamma^0\m_{\alpha IJ}$ drop out. In the third step, we used (\ref{eq:epsilonepsilon2}) and $\bar{\bar{\eta}}_{IJ} = m_{\alpha I} m^{\beta}\m_{J}$, and in the fourth step, epsilon identities were used and antisymmetrisations in $(\beta, \gamma)$ were written out explicitly. When furthermore writing out the antisymmetrisations in $(\alpha,\beta)$, we find that several terms cancel (step 5) and additionally used $(\delta m_{\alpha I}) n^I = -(\delta n^I)m_{\alpha I}$, $(\delta m_{\alpha I}) s^I = -(\delta s^I)m_{\alpha I}$ and $m_{\alpha I} m^{\alpha}\m_{J} = \bar{\bar{\eta}}_{IJ}$. In the sixth step, the upper line is partially integrated and we used $(\delta m_{\alpha I})m^{\alpha I} = \frac{1}{2} (\delta h_{\alpha \beta}) h^{\alpha \beta} = \frac{1}{\sqrt{h}} \delta \sqrt{h}$, and the two summands of the lower line are combined into one term. The seventh step consists of writing out all individual terms appearing in the square brackets explicitly and partially integrating the last term. In step 8, we used
\be
	\delta \Delta \psi =  - (D^{\alpha} D^{\beta} \psi) \delta h_{\alpha \beta} +(\Delta \delta \psi)- (D^{\beta} \psi) (D^{\alpha} \delta h_{\alpha \beta}) +(D^{\gamma} \psi) (D_{\gamma} \delta \log \sqrt{h})
\ee
and the remaining steps are straightforward. 

Combining (\ref{eq:a3AA}),  (\ref{eq:a3BB}) and  (\ref{eq:a3CC}), we find immediately
\begin{align}
&  \epsilon^{IJKL} \epsilon^{\alpha\beta} \left(\delta_{[1} A_{\alpha IJ}\right) \left(\delta_{2]} A_{\beta KL}\right) \nonumber 
\\ &\qquad =  -\frac{2E^{(2)}}{\sqrt{h}} (\delta_{[1} n^I)( \delta_{2]} \tilde{s}_I)  + 2 \left(\delta_{[1} \frac{E^{(2)}}{\sqrt{h}}\right) n^P \left(\delta_{2]} \tilde{s}_P\right) \nonumber 
\\ &\qquad{} \quad -2 \left\{ \left(n^M \delta_{[1} \tilde{s}_M\right) (\delta_{2]} \frac{E^{(2)}}{\sqrt{h}}) + (\delta_{[1} \tilde{s}_M) (\delta_{2]} n^M)\left(\frac{E^{(2)}}{\sqrt{h}} - \langle E^{(2)}\rangle\right) \right\} \nonumber 
\\ &\qquad = 2 \langle E^{(2)} \rangle (\delta_{[1} \tilde{s}_M) (\delta_{2]} n^M)\text{.}
\end{align}

\newpage

\section{Higher-Dimensional Newman-Penrose Formalism}
\label{app:NPFormalism}
In this appendix, we will very briefly introduce the higher-dimensional Newman Penrose formalism as far as it is needed for the purpose of this paper. First, the Riemann tensor can be decomposed as follows
\be
R_{\mu \nu \rho \sigma}^{(D+1)} &=& C_{\mu \nu \rho \sigma}^{(D+1)} + \frac{2}{D-1} \left( R_{[\mu| \rho}^{(D+1)} g_{|\nu] \sigma} - R^{(D+1)}_{[\mu| \sigma} g_{|\nu] \rho} \right) - \frac{2}{D(D-1)} g_{[\mu|\rho} g_{|\nu] \sigma} R^{(D+1)} \nonumber \\
 &=& C_{\mu \nu \rho \sigma}^{(D+1)} + \frac{2}{D-1} \left( J_{[\mu| \rho}^{(D+1)} g_{|\nu] \sigma} - J^{(D+1)}_{[\mu| \sigma} g_{|\nu] \rho} \right) + \frac{2}{D(D+1)} g_{[\mu|\rho} g_{|\nu] \sigma} R^{(D+1)} \text{,}~~~
\ee
where $C_{\mu \nu \rho \sigma}^{(D+1)}$ denotes the $(D+1)$ Weyl tensor and $J_{\mu \nu}^{(D+1)} := R_{\mu \nu}^{(D+1)} - \frac{1}{D+1} g_{\mu \nu} R^{(D+1)}$ the trace-free Ricci tensor. In a given null frame $\{l, k, \{m_I\}\}$, $l^2 = k^2 = l \cdot m_I = k \cdot m_I = 0$, $l \cdot k = -1$, $m_I \cdot m_J = \bar{\eta}_{IJ}$, we will use the following notation (cf. \cite{ColeyClassificationOfThe}) for the components of the Weyl tensor
\begin{align}
\Psi_{0101} &:= C^{(D+1)}_{\mu \nu \rho \sigma} l^{\mu} k^{\nu} l^{\rho} k^{\sigma} \text{,} & \Psi_{010I} &:= C^{(D+1)}_{\mu \nu \rho \sigma} l^{\mu} k^{\nu} l^{\rho} m^{\sigma}_I \text{,} \nonumber \\
\Psi_{011I} &:= C^{(D+1)}_{\mu \nu \rho \sigma} l^{\mu} k^{\nu} k^{\rho} m^{\sigma}_I \text{,} & \Psi_{01IJ} &:= C^{(D+1)}_{\mu \nu \rho \sigma} l^{\mu} k^{\nu} m^{\rho}_I m^{\sigma}_J \text{,} \nonumber \\
\Psi_{0I0J} &:= C^{(D+1)}_{\mu \nu \rho \sigma} l^{\mu} m^{\nu}_I l^{\rho} m^{\sigma}_J \text{,} & \Psi_{0I1J} &:= C^{(D+1)}_{\mu \nu \rho \sigma} l^{\mu} m^{\nu}_I k^{\rho} m^{\sigma}_J \text{,} \nonumber \\
\Psi_{0IJK} &:= C^{(D+1)}_{\mu \nu \rho \sigma} l^{\mu} m^{\nu}_I m^{\rho}_J m^{\sigma}_K \text{,} & \Psi_{1I1J} &:= C^{(D+1)}_{\mu \nu \rho \sigma} k^{\mu} m^{\nu}_I k^{\rho} m^{\sigma}_J \text{,} \nonumber \\
\Psi_{1IJK} &:= C^{(D+1)}_{\mu \nu \rho \sigma} k^{\mu} m^{\nu}_I m^{\rho}_J m^{\sigma}_K \text{,} & \Psi_{IJKL} &:= C^{(D+1)}_{\mu \nu \rho \sigma} m^{\mu}_I m^{\nu}_J m^{\rho}_K m^{\sigma}_L \text{.}
\end{align}
We will use analogous notation for the $(D+1)$ Riemann tensor if convenient. From curvature tensor symmetries and tracelessness, the relations
\be
\Psi_{0I0}\m^I = \Psi_{1I1}\m^I =0,\, \Psi_{0[IJK]} = \Psi_{1[IJK]} = \Psi_{I[JKL]} = 0,\, \Psi_{0101} = - \Psi_{0I1}\m^I, \, \nonumber \\ \Psi_{010J} = - \Psi_{0IJ}\m^I, \,\Psi_{011J} = \Psi_{1IJ}\m^I, \, \Psi_{0I1J} = \frac{1}{2} \left(\Psi_{01IJ} + \Psi_{IKJ}\m^K\right)
\ee
can be derived \cite{ColeyClassificationOfThe}. For the components of the trace-free Ricci tensor $J^{(D+1)}_{\mu\nu}$, we introduce the notation
\begin{align}
\Phi_{00} & = J^{(D+1)}_{\mu\nu} l^{\mu} l^{\nu} \text{,} & \Phi_{01} & = J^{(D+1)}_{\mu\nu} l^{\mu} k^{\nu} \text{,} & \Phi_{0I} & = J^{(D+1)}_{\mu\nu} l^{\mu} m^{\nu}_I \text{,} \nonumber \\
\Phi_{11} & = J^{(D+1)}_{\mu\nu} k^{\mu} k^{\nu} \text{,} & \Phi_{1I} &= J^{(D+1)}_{\mu\nu} k^{\mu} m^{\nu}_I \text{,} & \Phi_{IJ} &= J^{(D+1)}_{\mu\nu} m^{\mu}_I m^{\nu}_J \text{,}
\end{align}
and, because of tracelessness, it holds that
\be
2 \Phi_{01} = \Phi_I\m^I \text{.}
\ee

\end{appendix}

\end{document}